\documentclass[twocolumn]{pasj01}
\SetRunningHead{Astronomical Society of Japan}{
$^{12}$CO (1--0) map of M83}
\Received{2017 October 23}
\Accepted{2018 May 3}
\Published{2018 June 27}


\newcommand\Tpeak[0]{T_{\mathrm{peak}}}
\newcommand\dV[0]{{\Delta}{V}}
\newcommand\Tedge[0]{T_{\mathrm{edge}}}
\newcommand\HAlpha[0]{\mathrm{H}\alpha}
\newcommand\HI[0]{\mathrm{H}\emissiontype{I}}
\newcommand\HII[0]{\mathrm{H}\emissiontype{II}}
\newcommand\NII[0]{\mathrm{N}\emissiontype{II}}
\newcommand\PaBeta[0]{\mathrm{Pa}\beta}
\newcommand\kmPerS[0]{\mathrm{km\ s}^{-1}}
\newcommand\KkmPerS[0]{\mathrm{K\ km\ s}^{-1}}
\newcommand\sigmax[0]{\sigma_{\mathrm{x}}}
\newcommand\sigmay[0]{\sigma_{\mathrm{y}}}
\newcommand\sigmav[0]{\sigma_{\mathrm{v}}}
\newcommand\sigmavz[0]{\sigma_{\mathrm{v}, 0}}

\newcommand\SigmaSFR[0]{\Sigma_{\mathrm{SFR}}}
\newcommand\SigmaGas[0]{\Sigma_{\mathrm{gas}}}
\newcommand\SigmaGMC[0]{\Sigma_{\mathrm{GMC}}}
\newcommand\SigmaMol[0]{\Sigma_{\mathrm{GMC}}}
\newcommand\rhoMol[0]{\rho_{\mathrm{GMC}}}

\newcommand\TauFF[0]{ \tau_{\mathrm{ff}}}
\newcommand\Fgas[0]{ {F_{\mathrm{gas}}} }
\newcommand\Frad[0]{ {F_{\mathrm{rad}}} }
\newcommand\Fgrav[0]{ {F_{\mathrm{grav}}} }
\newcommand\tAVG[0]{ t_{\mathrm{avg}}}
\newcommand\SFEff[0]{ \mathrm{SFE}_{\mathrm{ff}}}
\newcommand\SFEffi[0]{ \mathrm{SFE}_{\mathrm{ff,i}}}
\newcommand\SFEffa[0]{ \mathrm{SFE}_{\mathrm{ff}}}

\newcommand\THII[0]{ {T_{\mathrm{H}\emissiontype{II}}} }
\newcommand\RHII[0]{ {R_{\mathrm{H}\emissiontype{II}}} }
\newcommand\QLyc[0]{ {Q_{\mathrm{Lyc}}} }
\newcommand\rhos[0]{ {\rho_{\mathrm{s}}} }
\newcommand\alphaRec[0]{ {\alpha_\mathrm{rec}} }

\newcommand\Mvir[0]{{M}_{\mathrm{vir}}}

\newcommand\Qfb[0]{ {(\Frad + \Fgas)/\Fgrav} }
\newcommand\alphaVir[0]{{\alpha_{\mathrm{virial}}}}
\newcommand\Mmol[0]{{M}_{\mathrm{GMC}}}
\newcommand\LCO[0]{{L}_{\mathrm{CO}}}
\newcommand\MsunPerSqPC[0]{ {M}_{\odot}\ \mathrm{pc}^{-2} }
\newcommand\Msun[0]{ {M}_{\odot} }
\newcommand\MsunPerYear[0]{ {{M}_{\odot}\ \mathrm{yr}^{-1}} }

\begin{document}
\title{ALMA $^{12}$CO (J=1--0) imaging of the nearby galaxy M83: Variations in the efficiency of star formation in giant molecular clouds}
\author{
Akihiko Hirota, \altaffilmark{1, 2}
Fumi Egusa, \altaffilmark{1, 3}
Junichi Baba, \altaffilmark{4, 1}
Nario Kuno, \altaffilmark{5, 6}
Kazuyuki Muraoka, \altaffilmark{7}
Tomoka Tosaki, \altaffilmark{8}
Rie Miura, \altaffilmark{1}
Hiroyuki Nakanishi, \altaffilmark{9}
and Ryohei Kawabe \altaffilmark{1, 10, 11}
}

\altaffiltext{1}{National Astronomical Observatory of Japan, 2-21-1 Osawa, Mitaka, Tokyo 181-8588, Japan} \email{akihiko.hirota@nao.ac.jp}
\altaffiltext{2}{Joint ALMA Observatory, Alonso de C\'{o}rdova 3107, Vitacura, Santiago 763-0355, Chile}
\altaffiltext{3}{Institute of Space and Astronautical Science, Japan Aerospace Exploration Agency, Sagamihara, Kanagawa 252-5210, Japan}
\altaffiltext{4}{Earth-Life Science Institute, Tokyo Institute of Technology, 2-12-1 Ookayama, Meguro, Tokyo 152-8551, Japan}
\altaffiltext{5}{Division of Physics, Faculty of Pure and Applied Sciences, University of Tsukuba, 1-1-1 Tennodai, Tsukuba, Ibaraki 305-8571, Japan}
\altaffiltext{6}{Tomonaga Center for the History of the Universe, University of Tsukuba, 1-1-1 Tennodai, Tsukuba, Ibaraki 305-8571, Japan}
\altaffiltext{7}{Department of Physical Science, Graduate School of Science, Osaka Prefecture University, 1-1 Gakuen-cho, Naka-ku, Sakai 599-8531, Japan}
\altaffiltext{8}{Joetsu University of Education, Yamayashiki-machi, Joetsu, Niigata 943-8512, Japan}
\altaffiltext{9}{Graduate School of Science and Engineering, Kagoshima University, 1-21-35 Korimoto, Kagoshima, Kagoshima 890-0065, Japan}
\altaffiltext{10}{SOKENDAI (The Graduate University for Advanced Studies), 2-21-1 Osawa, Mitaka, Tokyo 181-8588, Japan}
\altaffiltext{11}{Department of Astronomy, School of Science, University of Tokyo, Bunkyo, Tokyo 113-0033, Japan}

\maketitle
\begin{abstract}
We present results of the $^{12}$CO (1--0) mosaic observations of the nearby barred-spiral galaxy M83 obtained with the Atacama Large Millimeter/submillimeter Array (ALMA). The total flux is recovered by combining the ALMA data with single-dish data obtained using the Nobeyama 45-m telescope.  The combined map covers a $\sim$13 kpc$^{2}$ field that includes the galactic center, eastern bar, and spiral arm with a resolution of \timeform{2''.03} $\times$ \timeform{1''.15} ($\sim45$ pc $\times$ $\sim25$ pc). With a resolution comparable to typical sizes of giant molecular clouds (GMCs), the CO distribution in the bar and arm is resolved into many clumpy peaks that form ridge-like structures. Remarkably, in the eastern arm, the CO peaks form two arc-shaped ridges that run along the arm and exhibit a distinct difference in the activity of star formation: the one on the leading side has numerous H\emissiontype{II} regions associated with it, whereas the other one on the trailing side has only a few.
\par
To see whether GMCs form stars with uniform star formation efficiency (SFE) per free-fall time ($\SFEff$), GMCs are identified from the data cube and then cross-matched with the catalog of $\HII$ regions to estimate the star formation rate for each of them.
179 GMCs with a median mass of 1.6 $\times$ 10$^{6}$ $\Msun$ are identified. The mass-weighted average $\SFEff$ of the GMCs is $\sim$9.4 $\times$ 10$^{-3}$, which is in agreement with models of turbulence regulated star formation. Meanwhile, we find that $\SFEff$ is not universal within the mapped region. In particular, one of the arm ridges shows a high $\SFEff$ with a mass-weighted value of $\sim2.7$ $\times$ 10$^{-2}$, which is higher by more than a factor of 5 compared to the inter-arm regions. This large regional variation in $\SFEff$ favors the recent interpretation that GMCs do not form stars at a constant rate within their lifetime.
\end{abstract}

\section{Introduction}

As most star formation takes place in giant molecular clouds (GMCs) and as star formation is one of the fundamental processes that drives the evolution of galaxies, it is essential to understand the processes that determine the star formation efficiency (SFE) in GMCs.
\par
Molecular clouds with sizes of 20--100 pc and masses of 10$^{4-6}$ $\Msun$ are often classified as GMCs (e.g., \cite{Sanders1985}; \cite{Solomon1987Larson}).
Motions inside GMCs are turbulent with resultant CO linewidths of several kilometers per second, which are 'supersonic' for typical temperatures in GMCs ($\sim$10 K).
Although early studies proposed that the turbulent linewidth could be a manifestation of the gravitational collapse of GMCs (e.g., \cite{GoldreichKwan1974}), in most places GMCs are assumed to be supported by turbulent pressure and magnetic fields against their self-gravity, and kept in near-virial equilibrium (e.g., \cite{Larson1981}; \cite{Solomon1987Larson}).
The observed balance between the virial mass and the cloud mass derived with a reasonable assumption regrading the CO-to-H$_2$ conversion factor, also known as Larson's second law, has been considered as one of the important manifestations of the nature of GMC that is close to the virial equilibrium.
\par
In the Milky Way (MW), GMCs form stars with a low efficiency in the sense that only 1\% of the GMC mass is converted into stars per free-fall time (\cite{ZuckermanEvans1974}; \cite{Krumholz2005TurbulenceSF}).
The low SFE was one of the earliest grounds for assuming that GMCs are in a state of near equilibrium rather than in a state of gravitational collapse because it was argued that if GMCs are in a state of collapse, more stars should be produced within a free-fall time.
The recognition of the low efficiency led to another discussion on the idea that some mechanisms should be regulating the star formation rate (SFR) in GMCs.
\par
In discussing mechanisms that regulate the SFE in GMCs, it is common practice to adopt a parameter called {\it star formation efficiency per free-fall time} ($\SFEff$).
The parameter is defined as $\SFEff$ $\equiv$ $\TauFF\mathrm{SFR}/\Mmol$, where $\Mmol$ and $\TauFF$ are the mass and free-fall time of a cloud, respectively.
As far as GMCs are concerned, the average value of $\SFEff$ is claimed to be uniform, approximately 0.01, in many galaxies, including the MW (\cite{Krumholz2012UniversalSF}).
\par
Turbulence is one of the mechanisms considered responsible for regulating $\SFEff$.
Turbulence could regulate star formation in a GMC by forming a small volume of high-density regions within it while keeping the bulk of the GMC in near virial balance (e.g., \cite{MacLow2004}).
Attempts have been made to construct theoretical descriptions that quantitatively reproduce an $\SFEff$ of $\sim$0.01 in a steady state (e.g., \cite{Krumholz2005TurbulenceSF}; \cite{Federrath2015}).
The efforts trying to achieve $\SFEff$ $\sim$0.01 in a steady state implies that GMCs are approximated as long-lived entities that form stars with constant efficiency.
\par

However, although the model of turbulence regulated star formation succeeded in explaining the low average value of $\SFEff$ to a certain degree (\cite{Federrath2015}), some studies imply that the assumption of quasi-steady virialized GMCs could be oversimplifying the nature of GMCs.
First, the lifetime of GMCs has been estimated to be 15--40 Myr, which is only a few $\TauFF$, by several studies; the GMC lifetime is suggested to be effectively limited by the disrupting roles of OB stars (\cite{BlitzShu1980}; \cite{Kawamura2009LMC}; \cite{Murray2011SFE}; \cite{Miura2012M33}) or large-scale shearing motions in inter-arm regions \citep{Meidt2015Lifetime}.
If this is true, then GMCs do not necessarily have to be long-lived entities that are kept in near virial balance.
Second, Larson's second law alone does not completely rule out the  possibility that GMCs are collapsing because even if a GMC is in a state of free-fall, the CO linewidth increases by only a factor of $\sqrt{2}$ compared to a GMC in virial balance (\cite{Larson1981}).
The difference is subtle and within the inevitable observational uncertainties.
If GMCs are not long-lived as suggested by the first argument, they are allowed to be at least partially collapsing as was argued by \citet{GoldreichKwan1974} (see \cite{BallesterosParedes2011A}).
\par
Further, recent studies point out that there is a wide spread, larger than two orders of magnitude, in $\SFEff$ observed for Galactic GMCs (\cite{Murray2011SFE}; \cite{Lee2016DynamicSF}; see also \cite{Mooney1988}).
Those authors argued that the large scatter in $\SFEff$ is hard explain by the models of turbulence-regulated star formation, which predict a weak dependence of $\SFEff$ on the properties of GMCs.
Instead, they proposed that $\SFEff$ should be a time-dependent variable that dynamically increases during the lifetime of GMCs, although the effects of observational bias mist be carefully examined (\cite{Feldmann2011}).
\par
Accepting that there is a wide scatter in $\SFEff$, it is natural to think of its spatial distribution within a galaxy.
If $\SFEff$ indeed increases dynamically during the lifetime of GMCs, there should be a variation in the observed $\SFEff$ across spiral arms because spiral arms have at least moderate impact in organizing the build-up and disruption of massive GMCs (\cite{Egusa2011M51}; \cite{Hirota2011IC342}; \cite{Colombo2014Env}).
On the other hand, if $\SFEff$ is stable, then the distribution of $\SFEff$ observed should not exhibit strong spatial variations.
Therefore, the investigation of the spatial distribution of $\SFEff$ in galaxies should provide a clue to understand how the rate of star formation is regulated in GMCs, a discussion that is closely connected with the nature of a GMC itself.
\par
M83 is an ideal target to investigate $\SFEff$ in GMCs because it is one of the nearest (4.5 Mpc, \cite{Thim2003M83Distance}) spiral galaxies that is seen face-on and hosts prominent galactic structures, namely a bar and spiral arms.
Metallicities in M83 are comparable to or even higher than in the MW \citep{Bresolin2016M83}, thus CO lines are effective in tracing molecular clouds.
Other basic parameters of M83 are listed in table \ref{TableGalaxyParameters}.
As this galaxy contains a large amount of molecular gas (3.2 $\times$ 10$^9$ $\Msun$, \cite{Crosthwaite2002M83}), numerous observational studies were made with both single-dish telescopes and interferometers (see references listed in \cite{Hirota2014}).
However, due to its low declination, interferometric observations made previous to the arrival of the Atacama Large Millimeter/submillimeter Array (ALMA) suffered from limited $u$-$v$ samplings, and the spatial resolutions achieved were insufficient to resolve individual GMCs.
Recently, \citet{Freeman2017} investigated the properties of GMCs in M83 using the ALMA data, but the data used lacked sensitivity to the total flux.
\par
We present the results of $^{12}$CO (1--0) mosaic observations of M83 taken with the ALMA. The interferometric data are combined with data taken with the Nobeyama 45-m telescope to recover the total flux.
Target fields of the observations are selected so that a variety of galactic structures are covered, including the bar, the spiral arm, inter-arm regions, and the galactic center.
We examine the observational properties of GMC including $\SFEff$ over galactic structures to see whether or not systematic variation in $\SFEff$ exists.
\par
We describe the CO observations in section 2 and present the CO distribution in section 3.
In section 4, we identify GMCs from the obtained CO data cube and examine their basic properties.
Scaling relations of the properties of the identified GMCs and their mass functions are also examined.
In section 5, by cross-matching the identified GMCs with $\HII$ regions, we derive $\SFEff$ for each GMC.
In section 6, discussions are made to interpreting the meaning of the observed variation in $\SFEff$ and also to see in which regions feedback from massive stars is large enough to disrupt GMCs.
In section 7, we present a summary.

\par

\begin{table} [htbp]
\caption{Adopted parameters of M83}
\begin{center}
\small
\begin{tabular} {lr}
\hline\hline
Parameter& Value\\
\hline
Morph. \footnotemark[1]& SAB(s)c \\
Center position	(J2000)\footnotemark[2]& \timeform{13h 37m 00s.72}\\
& $-$\timeform{29D 51' 57''.2}\\
Position angle$^{}$\footnotemark[3]& \timeform{225D}\\
Inclination angle$^{}$\footnotemark[3]& \timeform{24D}\\
Systemic velocity (LSR)\footnotemark[4]& 514 km\ s$^{-1}$\\
Distance\footnotemark[5]& 4.5 $\pm$ 0.3 Mpc\\
Linear scale & \timeform{1''} $\sim$ 22pc\\
SFR     \footnotemark[6] & 3.0 $\MsunPerYear$ \\
HI mass   \footnotemark[7]   &  7.9$\times$10$^9$ M$_{\odot}$\\
H$_2$ mass \footnotemark[8]  &  3.2$\times$10$^9$ M$_{\odot}$\\
E(B-V)\footnotemark[9]  & 0.070    \\
\hline
\end{tabular}
\end{center}
\label{TableGalaxyParameters}
\begin{tabnote}
\footnotemark[1] {\citet{deVaucouleurs1991RC3}.}
\footnotemark[2] {\citet{Thatte2000}.}
\footnotemark[3] {\citet{Comte1981M83Parameters}.}
\footnotemark[4] {\citet{Kuno2007Atlas}.}
\footnotemark[5] {\citet{Thim2003M83Distance}.}
\footnotemark[6] {\citet{Jarrett2013WISE}, adjusted to a distance of 4.5 Mpc.}
\footnotemark[7] {\citet{Heald2016M83}, adjusted to a distance of 4.5 Mpc.}
\footnotemark[8] {\citet{Crosthwaite2002M83}, adjusted to a distance of 4.5 Mpc.}
\footnotemark[9] {\citet{Schlegel1998DustMap}.}
\end{tabnote}
\end{table}
\section{Observations and Data}
\label{SecObservation}

\begin {figure*} []
\begin {center}
\FigureFile(160mm,160mm){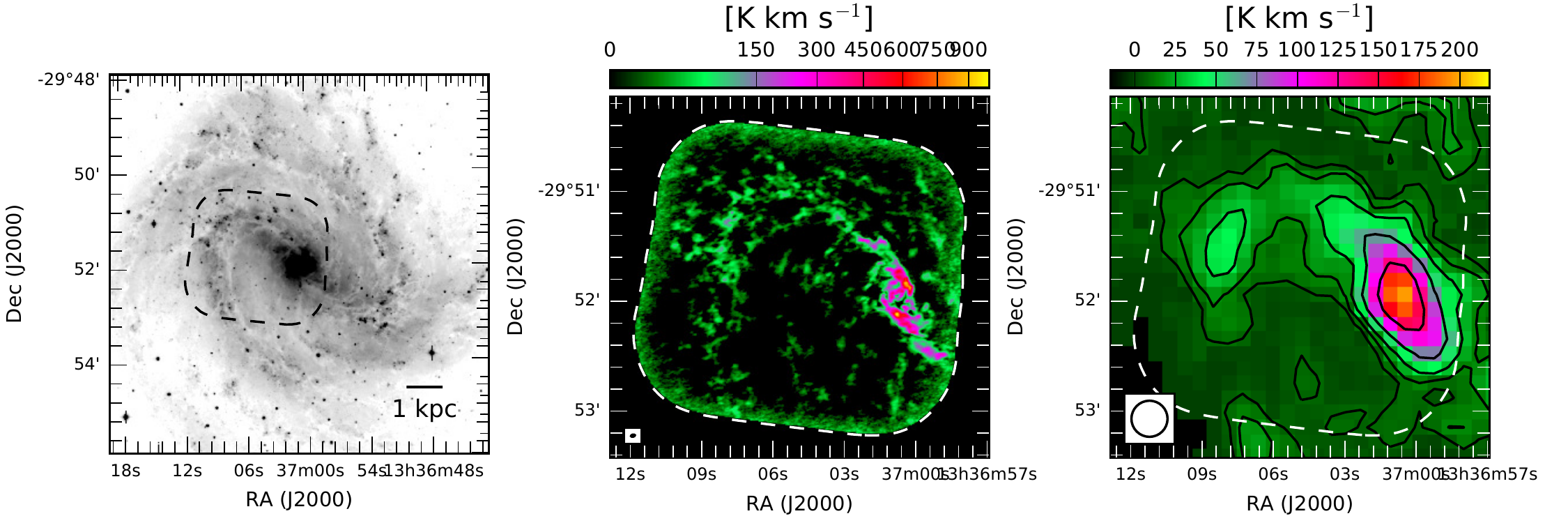}
\caption{(a) Optical $V$-band image of M83 taken from \citet{Larsen1999YMC}.
(b) Velocity integrated intensity $^{12}$CO (1--0) image of M83 generated from the ALMA data only. Due to the lack of sensitivity of zero-spacing data, extended emission are filtered out.
(c) Same as (b), but for the data taken with 45-m telescope. The contour levels are 8, 16, 32, 64, and 128 $\KkmPerS$.
In each of these three plots, the dashed line indicates the field of view (FOV) of the ALMA mosaic observations with a normalized gain level of 0.9.
}
\label{FigObsRegionEtc}
\end{center}
\end{figure*}
\subsection{ALMA observations}
\label{SubsecObsALMAOnly}
Aperture synthesis observations of the inner part of {M83} in $^{12}$CO(1--0) were carried out with the ALMA (program ID: 2011.0.00772.S).
Observations were made with about 16 antennas that have 12m diameter each.
The 64-input correlator was used to acquire cross-correlation data and configured to cover $\sim$1 GHz bandwidth around a center frequency of $\sim$114.92 GHz with a channel spacing of 0.244 MHz. As the Hanning window function was applied online, the effective frequency resolution was 0.488 MHz ($\sim$1.27 km s$^{-1}$).
\par
Figure \ref{FigObsRegionEtc}(a) shows the target field of the ALMA observations.
45 pointings were observed sequentially with an integration time of $\sim$6 s per pointing, and J1316-336 was observed once in approximately 20 min as a time dependent gain calibrator.
The absolute flux scale was calibrated using J1337-129 and Titan, and 3C279 was used for the passband calibration.
The system noise temperature was in the range from 100 to 150 K throughout the observations.
\par
The cross-correlated visibility data were processed with the standard ALMA calibration procedures using the CASA software package (\cite{McMullin2007CASA}; \cite{Petry2012CASA}) to correct for atmospheric and instrumental phase fluctuations, and to transfer the temperature from the primary flux calibrators.
These calibrations are made by applying calibration scripts bundled with the data set.
Inspection of the antenna-based gain solutions and also the calibrated data showed that there was no need to further edit of the data.
The calibrated visibility data were exported as UV-FITS and then converted to the MIRIAD \citep{SaultTeubenWright1995MIRIAD} data format for the subsequent reduction process.
\par
Before proceeding to the data combination, as a sanity check, the calibrated visibility data were Fourier-transformed with natural weighting, and the CLEAN method was applied to produce the deconvolved image.
The CLEAN beam was \timeform{2''.07} $\times$ \timeform{1''.13} with a position angle of $\sim-$\timeform{77.6D}.
The images were made for every two channels by averaging four neighboring channels to keep the resultant data cube oversampled in the velocity axis.
Therefore, each channel in the resultant cube has a width of $\sim$1.27 $\kmPerS$, but with an effective resolution of $\sim$2.54 $\kmPerS$.
After the deconvolution was performed, the spatial distribution of the rms noise was estimated from line free channels for each line of sight.
Within the mapped area, the median value of the rms was $\sim$7.4  mJy per beam with a velocity resolution of $\sim2.54$ $\kmPerS$.
However, we note that as the pointed mosaic observations made by the ALMA observing script assigned longer integration times to the southern half of the map, the spatial distribution of the noise is not uniform.
The 16th and 84th percentiles of the rms noise were 6.1 and 10.3 mJy per beam, respectively.
Figure \ref{FigObsRegionEtc}(b) shows the integrated intensity map of the ALMA image in $^{12}$CO (1--0).
\subsection{45-m Telescope observations and data combination}
As the ALMA observational data presented here lack sensitivities to the extended emission due to the central hole in the ($u,v$)-coverage ($<5k {\lambda}$), the ALMA data were combined with zero-spacing data obtained using the NRO 45-m telescope.
The 45-m observations were performed using the on-the-fly (OTF) observation mode during the winter seasons of 2008 and 2009.
The scanning rates were set between \timeform{40''} and \timeform{50''} s$^{-1}$, and data were sampled at a rate of 10 Hz, thereby providing a sufficient spatial sampling rate that met the Nyquist rate.
The T100 receiver was used as a front-end, and the SAM45 spectrometers configured to cover a 512 MHz bandwidth with 1024 channels used as back-ends.
Intensity calibration in the DSB antenna temperature ($T_{\mathrm{A}}^*$) scale was performed with the chopper-wheel method.
The Hanning window function was applied to the spectrometer outputs to achieve Nyquist sampling in the frequency axis.
To correct for the pointing offset, SiO (J=1--0) maser (42.821 and 43.122 GHz) sources were observed once every hour.
Any data with pointing errors larger than \timeform{6''} were excluded from the analysis.
\par
The OTF data were processed with the NOSTAR software package \citep{Sawada2008} to perform basic calibration procedures including baseline subtraction and application of main beam efficiency.
Imaging was also obtained with NOSTAR by gridding data points to a \timeform{6''} spacing grid and performing convolution with a spheroidal function (figure \ref{FigObsRegionEtc}c).
The total CO (1--0) luminosity detected within the field of view (FOV) of interferometric observations, defined as the region where the gain of the interferometric mosaic is above 0.9, is 4.3 $\times$ 10$^8$ $\pm$ 6.4 $\times$ 10$^6$ $\KkmPerS$ pc$^2$.
We note that the denoted error is obtained by scaling the rms noise level in line free channels, and thus the uncertainty of the absolute temperature scale is not included.
On the other hand, the luminosity detected by ALMA within the same area is 9.9 $\times$ 10$^7$ $\pm$ 5.7 $\times$ 10$^5$ $\KkmPerS$ pc$^2$.
Thus, only approximately 23\% of the CO flux was detected with the interferometric image synthesis.
\par
\subsection{Combined imaging}
\label{SubsecObsCombine}
THe the data sets taken with the different telescopes were combined by the procedure introduced in \citet{Kurono2009}.
The 45-m data cube was converted into pseudo visibility data sets using the same methodology as \citet{Hirota2014} to allow combined imaging and deconvolution with the data taken with ALMA.
The relative weighting between the two data sets was optimized so that combined image achieves a similar resolution as the ALMA only data and also so that it recovers the total flux detected by the 45-m observations.
The calibrated ALMA visibility data and pseudo-visibility data generated from the 45-m observations were Fourier-transformed and deconvolved with the CLEAN method together.
The parameters of the CLEAN beam are \timeform{2''.03} $\times$ \timeform{1''.15} (44.3 pc $\times$ 25.1 pc) with a position angle of $\sim-$\timeform{77.7D}.
As described in \S\ref{SubsecObsALMAOnly}, images were made for every two channels by averaging four neighboring channels; therefore the velocity resolution of the combined data is $\sim$2.54 $\kmPerS$.
The median value of the rms noise within the mapped area was $\sim$8.0 mJy per beam.
The combined data have a similar spatial distribution of noise as the ALMA-only data cube, with the 16th and 84th percentiles of the rms noise being 6.6 and 10.6 mJy per beam, respectively.

\subsection{Ancillary data sets}
\label{SubsecAncillayDataSets}
To compare with the CO data, data sets of the star formation tracers were also prepared.
\par
A catalog of $\HII$ regions was taken from \citet{Hirota2014} and was made by identifying $\HII$ regions from the $\HAlpha$ image retrieved from the data archive of the Survey for Ionization in Neutral-Gas Galaxies (SINGG; \cite{Meurer2006SINGG}) using the HIIphot software \citep{Thilker2000}.
The resolution of the $\HAlpha$ image is limited by a seeing of \timeform{1''.8}, which is comparable to the resolution of the CO data used here.
Correction for foreground extinction was made by adopting $A_{\mathrm{H}\alpha}$ = 0.18 mag, which is given by \citet{Meurer2006SINGG}.
Contamination of [$\NII$] lines at rest wavelengths of 6548 \AA \ and 6583 \AA \ were also corrected by using the estimated fractional contribution of [$\NII$] lines to the total bandpass, 0.11, which is given by \citet{Meurer2006SINGG}. Application of these corrections resulted in scaling the original image by a factor of $\sim$1.05.
\par
An image of $\PaBeta$ was produced from the narrow- and the broad-band filter images retrieved from the data archive of the HST/WFC3 ERS program (GO-11360).
Continuum subtraction was achieved by subtracting the scaled broad-band image (F110W) from the narrow-band image (F128N).

\section{GMC scale molecular gas distribution in M83}
\label{SecDistribution}

\begin{figure*} []
\begin{center}
\FigureFile(178mm,178mm){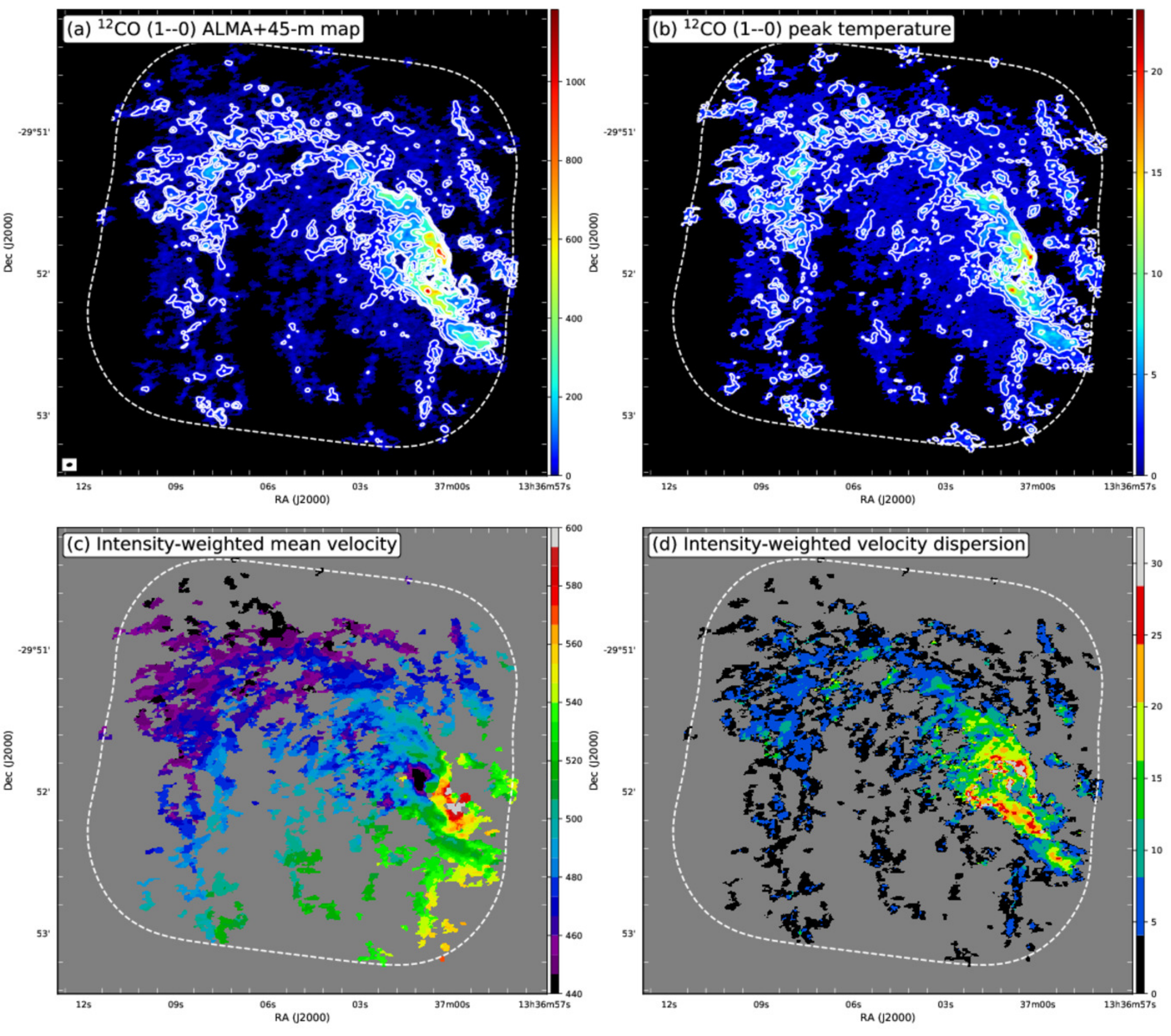}
\caption{
(a) Velocity integrated intensity map of $^{12}$CO (1--0) made from the combined data.
Contour levels are 24, 48, 96, 192, 384, and 768 $\KkmPerS$, respectively.
The dashed line indicates the gain pattern of the mosaic observations at the normalized level of 0.9.
The gain pattern is also indicated in each of the subsequent plots.
Due to the application of the mask referred to in the main text, the rms noise of the integrated intensity image is not uniform.
Typical rms values are $\sim$1 and $\sim$2 K km s$^{-1}$ in the inter-arm and arm regions, respectively.
(b) Peak temperature map for the combined $^{12}$CO (1--0) data.
Contour levels are 1.92, 3.84, and 7.68 K, respectively (1$\sigma$ = 0.32 K).
(c) Intensity-weighted mean velocity map of CO emission.
(d) Intensity-weighted velocity dispersion map of CO emission.
}
\label{FigMulti1}
\end{center}
\end{figure*}

\begin{figure*} []
\begin{center}
\FigureFile(178mm,178mm){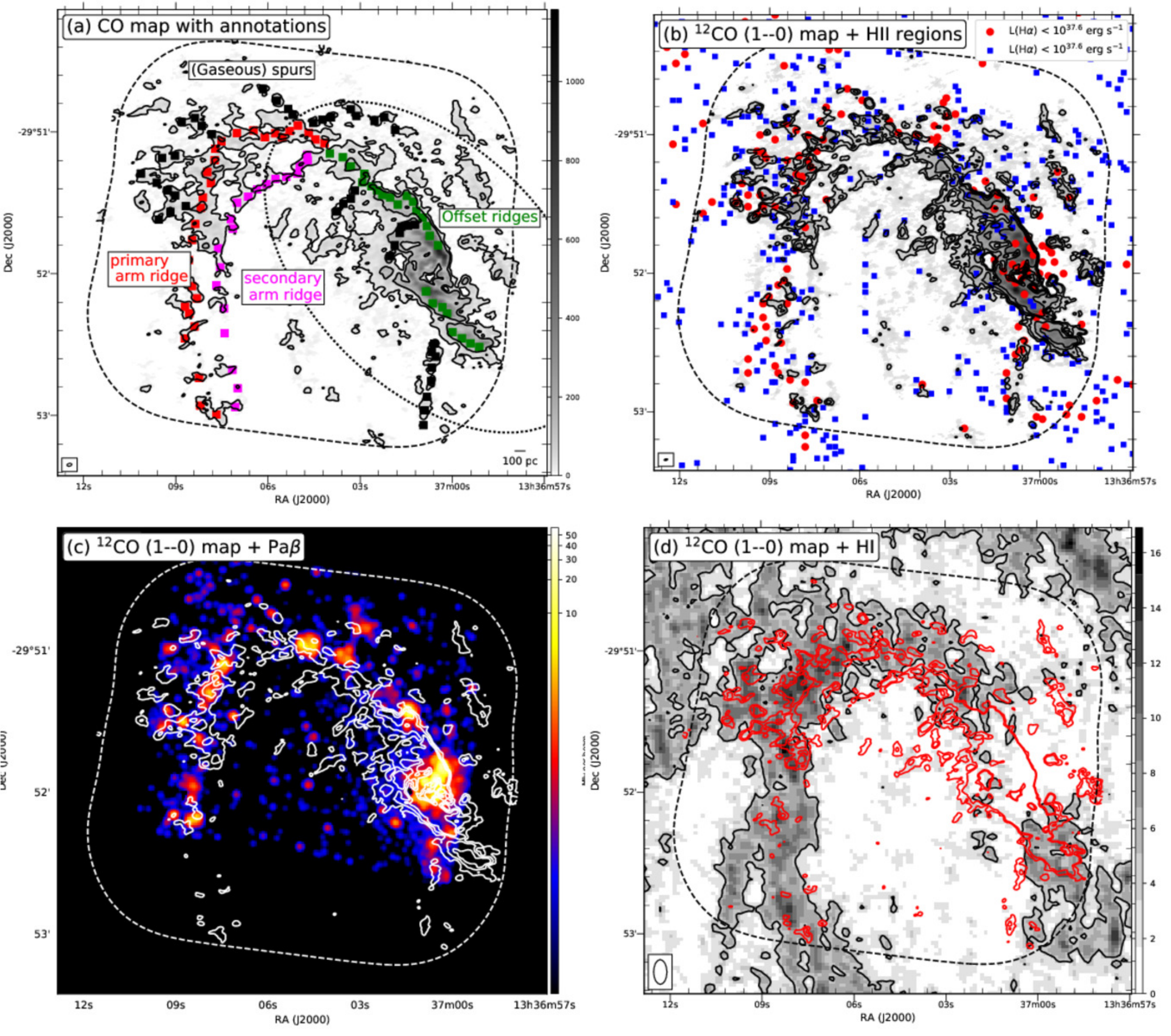}
\caption{
(a) CO map overlaid with annotations to indicate the names of the places referred to in the text.
(b) CO map compared with the distribution of $\HII$ regions.
Red and blue markers indicate the location of $\HII$ regions with the uncorrected $\HAlpha$ luminosity greater and lower than 10$^{37.6}$ erg s$^{-1}$, respectively.
Contour levels are the same as in figure \ref{FigMulti1}(a).
(c) CO contours compared with $\PaBeta$ image in arbitrary units.
The continuum-subtracted $\PaBeta$ image is smoothed to a resolution of \timeform{1''.8} to align with that of the $\HAlpha$ image used to identify the $\HII$ regions that are shown in (b).
THe contour levels are 33, 100, and 200 $\KkmPerS$.
(d) CO contours in red compared with $\HI$ map in units of $\MsunPerSqPC$.
The black contour is drawn at a level of 4 $\MsunPerSqPC$ for $\HI$.
The contour levels of the CO map are 30 and 60 $\KkmPerS$.
}
\label{FigMulti2}
\end{center}
\end{figure*}
\subsection{CO distribution}
\label{SubsecDistribution}
Figures\ref{FigMulti1}(a) and \ref{FigMulti1}(b) show maps of the integrated intensity and the peak temperature for $^{12}$CO (1--0), respectively.
Both of the CO maps were prepared from the combined CO (1--0) data cube presented in the previous section.
The entire velocity range of the CO emission within the mapped region is approximately 350--630 $\kmPerS$ and is much wider than the linewidths of individual molecular clouds which are several kilometers per second.
Therefore, when generating the two-dimensional maps, a mask was applied to the original data cube to reject pixels with low signal-to-noise ratios (SNRs).
The mask was made by including all pixels with SNRs over 4 and also including pixels with SNRs over 2 and that are morphologically connected to the pixels with SNRs over 4.
From the masked CO data cube, the first- and second-moment maps were also constructed (figures \ref{FigMulti1}(c) and figure \ref{FigMulti1}(d)).
For reference, figure \ref{FigMulti2}(a) also shows the contour map of the distribution of CO integrated intensity with annotations overlaid to indicate names of places that will be referred to hereafter.
\par
\par
The mapped region contains prominent galactic structures, including the stellar bar and the spiral arm that extends from the eastern end of the bar.
The stellar bar in M83 lies at a position angle of $\sim$\timeform{45D} and has a semi-major radius of approximately \timeform{84''} (\cite{Hirota2014}).
The approximate extent of the stellar bar is indicated in figure \ref{FigMulti2} by the dotted line.
The bar-to-arm transition region located around the eastern end of the bar was mapped by several studies in CO lines (e.g., \cite{Wiklind1990M83}; \cite{RandLordHidgon1999M83}).
\par
The  spatial resolution (44.3 pc $\times$ 25.1 pc) being comparable to the typical sizes of GMCs ($>$20 pc; \cite{Sanders1985}), the CO emission is resolved into many complex features.
The spiral arm is resolved into many clumpy peaks that are arranged in arc-shaped ridges, and the bar is also arranged in ridges but with a more continuous distribution.
The CO distribution in the arm and bar is narrow, in the sense that the widths are comparable to the beam size.
\par
Along the leading sides of the bar, two continuous ridges of CO emission extend almost symmetrically with respect to the galactic center.
The two molecular ridges correspond to 'offset ridges' commonly seen in molecular-rich barred-spiral galaxies (annotated in figure \ref{FigMulti2}a).
The offset ridges are found to be highly abundant in molecular gas.
By assuming a CO-to-H$_2$ conversion factor of 2.0 $\times$ 10$^{20}$ cm$^{-2}$ (K km s$^{-1}$)$^{-1}$ (e.g., \cite{Bolatto2013ConversionFactor}), a correction factor of 1.36 for the existence of helium, and an inclination angle of \timeform{24D}, the molecular gas surface density is found to be $\sim$200 $\MsunPerSqPC$ ($\sim$50 $\KkmPerS$) at the eastern edge of the ridge, and it even increases to $\sim$800 $\MsunPerSqPC$ ($\sim$200 $\KkmPerS$) or higher near the galactic center.
These values are comparable to or even higher than the well-quoted typical surface density of Galactic GMCs ($\sim$170 $\MsunPerSqPC$; \cite{Solomon1987Larson}).
\par
Near the eastern end of the bar, the molecular bar seen as a continuous ridge bifurcates into two molecular ridges that run through the leading and trailing sides of the spiral arm.
The ridge on the leading side is brighter in CO compared to the one on the trailing side.
Hereafter, we will refer to the ridge on the leading side as the 'primary arm ridge' and the one on the trailing side as the 'secondary arm ridge' (see figure \ref{FigMulti2}a).
Along the molecular ridges in the spiral arm, signatures of streaming motion can be recognized from the first-moment velocity map (figure \ref{FigMulti1}c) with a velocity shift of $\sim$20 $\kmPerS$.
The existence of the velocity shift is in agreement with the finding made by \citet{RandLordHidgon1999M83} and suggests that gas clouds in these two spiral arm ridges are subject to shock compression.
\par
In addition to the ridges mentioned above, there exist secondary structures that extend with large opening angles from the primary arm ridge and the bar seen in CO (annotated in figure \ref{FigMulti2}a).
The secondary features extending from the primary arm ridge are on the leading side; they most likely correspond to the structures often seen in spiral arms and are referred to as 'spurs' or 'feathers' (e.g., \cite{ElmegreenDM1980Spurs}; \cite{LaVigne2006}, hereafter, referred to as spurs).
We note that the four outermost spurs indicated in figure \ref{FigMulti2}(a) roughly coincide in position with the ones also identified by \citet{LaVigne2006} from an optical image.
Unlike the spurs on the leading side of the primary arm ridge, some spurs that extend from the bar are on trailing side.
The trailing spurs extending from a bar are often seen in other barred-spiral galaxies (\cite{Sheth2000NGC5383}; \cite{Zurita2008}).
The detailed mechanisms that produce the leading side spurs are still under active discussion, but most theories agree that gas clumps formed in an arm are sheared while moving toward the leading side of the arm (e.g., \cite{Wada2004Wiggle}; \cite{DobbsBonnell2006Spurs}; \cite{Renaud2013Subpc}).
On the other hand, the formation mechanism of the spurs on the trailing sides of a bar has not been investigated in detail.
Except for the trailing spurs in the bar, most of the spurs on the leading side are closely associated with $\HII$ regions.
Active star formation in spurs is in agreement with the suggestion that spurs, possibly produced from bound gas clumps, are preferred sites for star formation (e.g., \cite{Schinnerer2017}).

\subsection{Comparison with tracers of high mass star formation}
Figures \ref{FigMulti2}(b) and \ref{FigMulti2}(c) compare the CO map with the distribution of the $\HII$ regions and a map of $\PaBeta$ emission, respectively.
Comparing the distribution of CO emission with the tracers of massive star formation, the SFE in molecular clouds does not appear to be uniform within the region observed.
A large number of $\HII$ regions are found in limited areas such as the primary arm ridge, spurs that extend toward the leading side of the primary ridge, and around the galactic center.
On the other hand, it is evident that fewer $\HII$ regions are found along the secondary arm ridge and in the inter-arm region located on the trailing side of the bar.
Later in \S\ref{SecStarFormationInGMCs}, we will quantitatively investigate the SFE for individual GMCs.
Further in \S\ref{SecDiscussion}, we will discuss whether the apparent regional variation in star formation activity is caused by intrinsic variations in the SFE in GMCs.
\subsection{Comparison with the distribution of atomic hydrogen}
Figure \ref{FigMulti2}(d) compares the CO map with a map of the atomic hydrogen ($\HI$) retrieved from the archive of The HI Nearby Galaxy Survey (THINGS; \cite{Walter2008THINGS}), which is an imaging survey of $\HI$ in nearby galaxies made with the Very Large Array (VLA).
The archival image was made with robust weighting of the visibilities and the size of the synthesized beam is \timeform{10''.4} $\times$ \timeform{5''.6}.
By assuming optically thin emission of $\HI$, we converted the unit of the $\HI$ image into surface density which resulted in a sensitivity of $\sim$0.3 $\MsunPerSqPC$.
\par
Over the FOV of the CO observation, the surface density of $\HI$ is at most 12 M$_{\odot}$ pc$^{-2}$, and is well below that of molecular gas.
As $\HI$ is suggested to have a smooth distribution even at small spatial scales (finer than 100 pc; \cite{Leroy2013ISMClumpiness}), we assume that the surface density of $\HI$ is still well below that of molecular gas even when compared at the scale of GMCs.
\par
Another concern might be the missing flux of the $\HI$ image.
As the THINGS $\HI$ image is not sensitive to structures larger than \timeform{15'} ($\sim$20 kpc) due to the lack of short spacing baselines, the contribution of the missing flux might be a concern.
From a comparison of the $\HI$ flux detected by VLA observations with single dish telescope observations, \citet{Crosthwaite2002M83} estimated that large-scale features greater than \timeform{15'} in M83 could have a mean surface $\HI$ density of 2$\pm$0.5 $\MsunPerSqPC$.
Although the data used by \citet{Crosthwaite2002M83} and those presented as part of THINGS survey are not identical, the total $\HI$ flux detected in the two data sets do not differ greatly from each other (480 Jy $\kmPerS$ vs. 360 Jy $\kmPerS$).
Therefore, we hereafter assume that the contribution of $\HI$ is not significant compared to the molecular gas, and do not take the $\HI$ gas content into account in the following analyses.
\par
On the basis of the spatial distribution of CO and $\HI$ in spiral arms, several authors suggested that in the molecular dominated part of the galactic gas disks, $\HI$ is preferentially found in star-forming regions as the product of photo-dissociation (M51 by \cite{RandKulkarniRice1992}; the MW by \cite{Koda2016}).
For the particular case of the eastern spiral arm of M83, \citet{RandLordHidgon1999M83} compared their interferometric CO map with the $\HI$ map of \citet{TilanusAllen1993}.
They noted that it is difficult to identify systematic spatial offsets between $\HI$ and CO peaks which were taken as evidence for $\HI$ being the photo-dissociation product in M51 (\cite{RandKulkarniRice1992}).
However, we note that the previous CO map of M83 obtained by \citet{RandLordHidgon1999M83} has missed the secondary arm ridge and the arm traced by the $\HI$ emission is more aligned with the primary arm ridge compared to the secondary one, although the beam size of the $\HI$ image is considerably larger than that of the CO image.
As the SFE is suggested to be lower in the secondary arm ridge compared to the primary arm ridge, the notion of $\HI$ as the photo-dissociation product seems to still hold in the eastern arm of M83.

\section{Properties of GMCs}
\label{SectionGMCProp}
\subsection{Cloud identification}
\label{SubsecCloudIdentification}
GMCs are identified from the three-dimensional $^{12}$CO (1--0) data.
Before performing the cloud identification, the original data cube is smoothed spatially by convolving with a Gaussian beam to obtain an axisymmetric beam profile of \timeform{2.1''} ($\sim46$ pc).  The 16th, 50th, and 84th percentiles of the rms noise are 7.3, 10.6, and 13.7 mJy per beam with a velocity resolution of $\sim2.54$ $\kmPerS$.
We used the {\it astrodendro}\footnote{http://www.dendrograms.org} software package to identify clouds from the data cube.
Although the software can identify hierarchically nested sets of topologically closed surfaces, we have adopted only closed surfaces with the finest spatial scales, which are commonly referred to as {\it leaves}.
The reason for using only {\it leaves} is that a resolution of $\sim$46 pc does not resolve the typical size of GMCs, because of which {\it leaves} should not represent substructures inside GMCs.
Instead, each of them should correspond to an individual GMC or a complex of GMCs that are closely packed within a resolution limit.
\par
Parameters for the cloud decomposition algorithm are configured such that each {\it leaf} has a peak SNR over 3, and the peak SNR of each {\it leaf} is higher by at least 3 from the merging level to neighboring {\it leaves}.
In addition, to avoid picking up small features that could be mere noise fluctuations, we discarded {\it leaf} candidates where the number of pixels occupied in the three-dimensional data cube is below 3 $\times$ $\left(A_{\mathrm{beam}} / A_{\mathrm{pixel}}\right)$, where $A_{\mathrm{beam}}$ and $A_{\mathrm{pixel}}$ are the areas of the beam and a pixel.
We also tried a threshold value of 2 $\times$ $\left(A_{\mathrm{beam}} / A_{\mathrm{pixel}}\right)$, which appears more plausible considering the factor-of-2 oversampling used in the velocity direction in the data cube (\S\ref{SubsecObsCombine}).
However, we discarded it because while the number of identified {\it leaves} increased by approximately 6\% , some of the newly detected ${\it leaves}$ are smaller than the beam size and thus are too small for the size parameters to be correctly determined later in \S\ref{SubsecDeriveBasicParameters}.
\par
With the adopted set of parameters, GMCs with a peak SNR down to 6 which corresponds to $\sim$1.4 K, are expected to be detected.
Assuming that a minimum detectable GMC has a Gaussian profile with a radius and linewidth of 46 pc and 5 $\kmPerS$, respectively, and also adopting the 'standard' CO-to-H$_2$ conversion factor (see \S\ref{SubsecDefinitionQuantities}), the mass threshold for the GMC identification is estimated to be 2.3 $\times$ 10$^5$ $\Msun$.
We will see later in \S\ref{SubsecMassSpectrum} that the mass threshold estimated here is consistent in regions where molecular clouds are not closely packed.
On the other hand, in regions with high average molecular gas density, such as the bar and the central region, the detection limit is elevated to around 10$^{6}$ $\Msun$; this point will also be reviewed in \S\ref{SubsecMassSpectrum}.
\par
All the identified {\it leaves} are checked by examining the two-dimensional distribution obtained by projecting each on the $x$-$y$ plane.
After manually flagging three {\it leaves} with too dispersed appearance, 179 {\it leaves} remained.
We regard each {\it leaf} as the 'kernel' of a GMC because the surface boundary determined by each {\it leaf} does not extend to the 0 K level and thus traces just a limited volume around the peak of each GMC.
\subsection{Derivation of basic cloud parameters}
\label{SubsecDeriveBasicParameters}
The centroid position and the velocity of each GMC are derived by calculating the intensity-weighted first moment within each GMC kernel.
The maximum temperatures within each GMC kernel are taken as peak temperatures ($\Tpeak$), and we also record the minimum temperature as the edge temperature ($\Tedge$).
\par
Two-dimensional root-mean-square sizes and the velocity dispersion of each GMC ($\sigma_\mathrm{x}$, $\sigma_\mathrm{y}$, and $\sigma_\mathrm{v}$, where $\sigma_{\mathrm{x}}$ $>$ $\sigma_{\mathrm{y}}$) are derived by calculating second moments in three axes within each kernel mask.
The second-moment value calculated within each kernel mask is extrapolated to the 0 K level by following the procedure proposed by \citet{Rosolowsky2006CPROPS}, which calculates a moment value with various temperature thresholds in the range between $\Tedge$ and $\Tpeak$ and linearly extrapolates it to the 0 K level.
The extrapolated second moments are further corrected by subtracting the resolution element size in quadrature to account for the finite resolution.
\par
\begin{figure} []
\begin{center}
\FigureFile(88mm,88mm){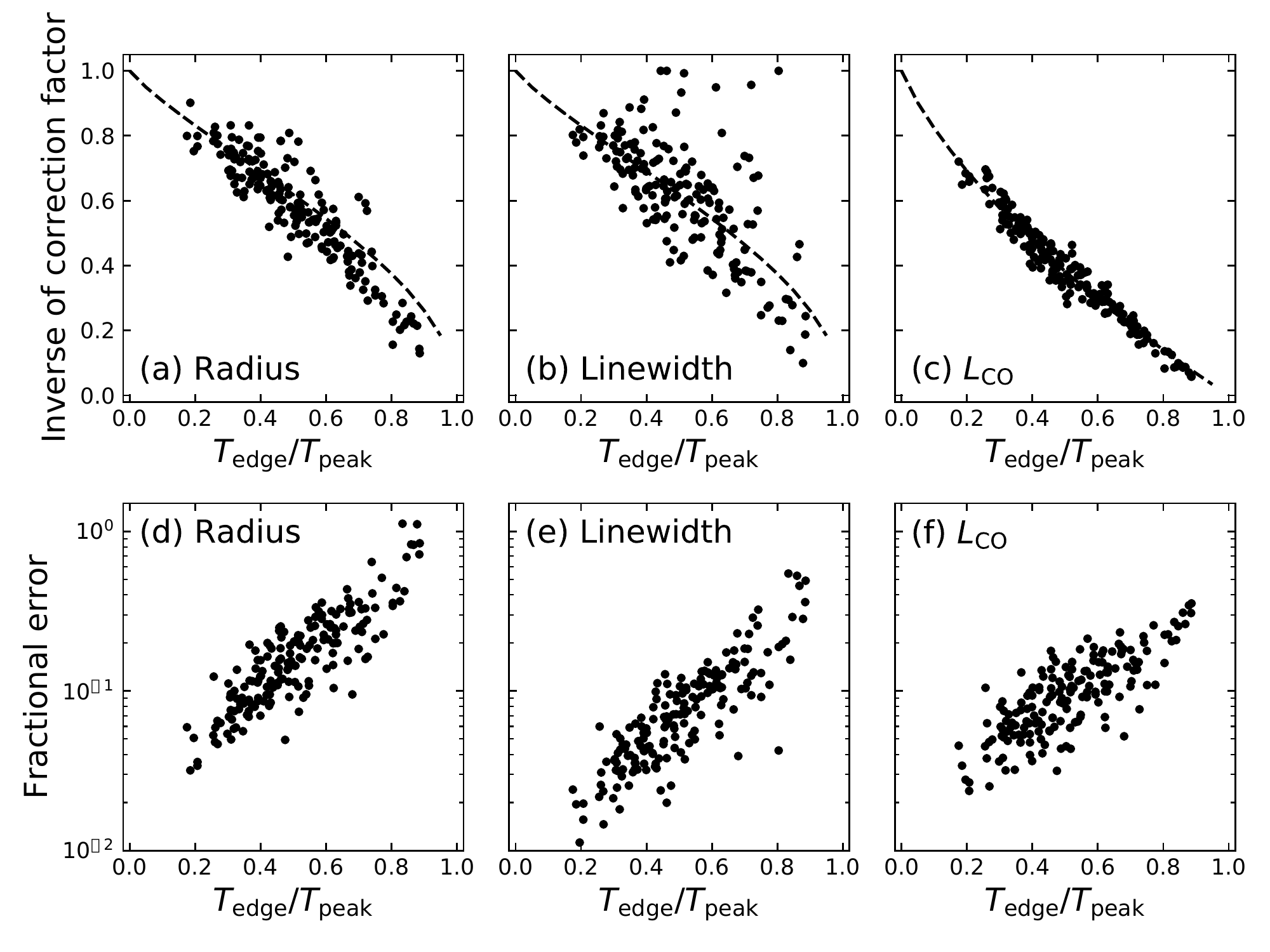}
\caption{
(a-c) Ratio of the extrapolated value to unextrapolated value for size, velocity dispersion, and CO luminosity, respectively, as a function of $\Tedge/\Tpeak$.
(d-f) Fractional uncertainty for each quantity in (a--c) as a function of $\Tedge/\Tpeak$.
Note that the ordinate in each plot is in logarithmic scale.}
\label{FigExtrapolationFactors}
\end{center}
\end{figure}
The CO luminosity of the cloud is first calculated within each GMC kernel and then corrected by applying the same extrapolation method as for the size parameters.
\par
The three plots in the top row of figure \ref{FigExtrapolationFactors} show the ratio of the extrapolated value to the unextrapolated value for size ($\propto$ $\sqrt{\sigmax\sigmay}$), velocity dispersion, and CO luminosity, respectively, as a function of the ratio of $\Tpeak$ to $\Tedge$.
The dashed line in each plot indicates the expected behavior with a simple Gaussian profile.
We see in each plot that most of the data points for the identified GMCs agree with the expected line for a Gaussian profile.
\par
Uncertainties for the size parameters and CO luminosity are determined by performing bootstrap error estimates.
As shown by the three plots in the bottom row of figure \ref{FigExtrapolationFactors}, the uncertainty assigned to each GMC quantity depends on $\Tpeak$ / $\Tedge$.
As we will see later in \S\ref{SubsecBoxplot1}, some GMCs in the center and the bar have a low contrast to their ambient regions---i.e., $\Tedge$ / $\Tpeak$ is high--- and thus those GMCs are assigned a larger uncertainty for cloud parameters compared to the GMCs in uncrowded regions.
\subsubsection{Adopted definition of physical quantities}
\label{SubsecDefinitionQuantities}
After determining the rms size values and the CO luminosity for each cloud, those values are mapped to physical quantities.
The linewidth of the cloud ($\Delta{V}$) is simply taken as $\Delta{V} = \sqrt{8 \log{2}} \sigma_{\mathrm{v}}$.
The effective size of the cloud ($R$) is defined as
\begin{equation}
R = \frac{3.4}{\sqrt{\pi}} \sqrt{ \sigma_{\mathrm{x}} \sigma_{\mathrm{y}} },
\end{equation}
where $3.4 / \sqrt\pi$ is an empirical factor that is determined by \citet{Solomon1987Larson}.
\par
The molecular gas mass for each GMC ($\Mmol$) is calculated from the CO luminosity ($\LCO$) by applying the CO-to-H$_{2}$ conversion factor of 2.0 $\times$ 10$^{20}$ cm$^{-2}$ (K km s$^{-1}$)$^{-1}$, which is considered to be valid for the inner disk of massive spiral galaxies (\cite{Bolatto2013ConversionFactor}) along with a correction factor of 1.36 that accounts for the contribution of helium and other elements:
\begin{equation}
\label{EqLCOtoMsun}
\left(\frac{\Mmol}{\Msun}\right) = 4.4 \left(\frac{\LCO}{\mathrm{K\ km\ s}^{-1}\ \mathrm{pc}^{2}}\right).
\end{equation}
THe virial mass is calculated with the following equation which neglects the magnetic energy and external pressure,
\begin{equation}
\Mvir =  \frac{1}{a_1}\frac{5 R \sigmav^2}{G},
\end{equation}
where $G$ is the gravitational constant, and $a_{1}$ is a factor that encapsulates the effects of the density distribution.
The factor, $a_{1}$, can be written as $a_1$ = (1 - $n$/3) / (1 - 2$n$/5) for a cloud that follows the power law density profile of $\rho(r)$ $\propto$ $r^{-n}$ (\cite{BertoldiMcKee1992}).
Hereafter, we employ $n$=1 \citep{Solomon1987Larson}.
\par
The virial parameter (\cite{BertoldiMcKee1992}), which is a measure of the ratio of the kinetic to gravitational energy of a GMC, is defined as
\begin{equation}
\alphaVir \equiv \Mvir / \Mmol.
\end{equation}
The Average gas surface density of GMC ($\SigmaMol$) is defined as
\begin{equation}
\SigmaMol = \Mmol / (\pi R^2).
\end{equation}
The volume density is calculated with the following equation taken from \citet{Leroy2015NGC253}:
\begin{equation}
\rhoMol = 1.26 \Mmol / \left(\left( 4 / 3 \right) \pi R^3\right)
\end{equation}
The free-fall time ($\TauFF$) of GMC is calculated as follows:
\begin{equation}
\TauFF = \sqrt{\frac{3\pi}{32G \rhoMol}}.
\end{equation}
\subsection{Distribution of identified GMCs}
\begin{figure*} [htbp]
\begin{center}
\FigureFile(138mm,138mm){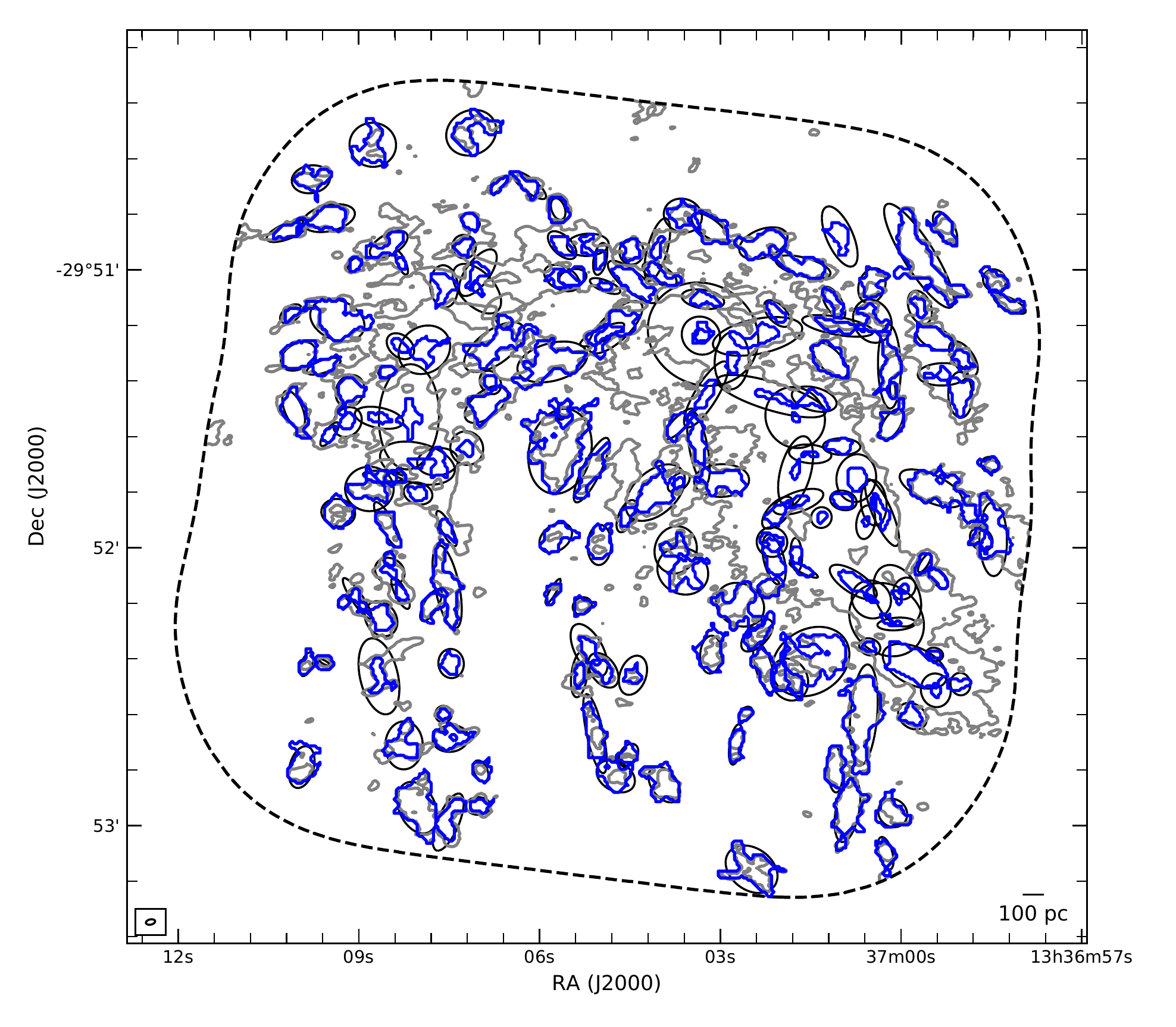}
\caption{
Distribution of the identified GMCs overlaid on the CO map indicated with a contour line with 13 K km s$^{-1}$ level (gray line).
Each ellipse indicates the position and the effective major and minor radii of a GMC (i.e. $3.4 / \sqrt{\pi}$ $\sigmax$ and $3.4 / \sqrt{\pi}$ $\sigmay$).
The solid blue line indicates the boundary of the kernel for each GMC projected on the $x-y$ plane.
}
\label{FigCloudDistribution1}
\end{center}
\end{figure*}
Figure \ref{FigCloudDistribution1} shows the distribution of the identified GMCs.
The spatial extent of each GMC is displayed by an ellipse with major and minor radii of $3.4 / \sqrt{\pi}$ $\sigmax$ and $3.4 / \sqrt{\pi}$ $\sigmay$, respectively.
The sum of the GMC mass derived by applying the extrapolation method and the CO conversion factor described in the previous section is $\sim$5.6 $\times$ 10$^{8}$ $\Msun$, which is approximately 33\%  of the total molecular gas mass within the target region ($\sim$1.7 $\times$ 10$^{9}$ $\Msun$). The fractional molecular gas mass identified as GMC is comparable to the value found in M51 ($\sim$54\%, \cite{Colombo2014Env}) with a similar spatial resolution.
\subsection{Definition of environmental masks}
\begin{figure} [htbp]
\begin{center}
\FigureFile(88mm,88mm){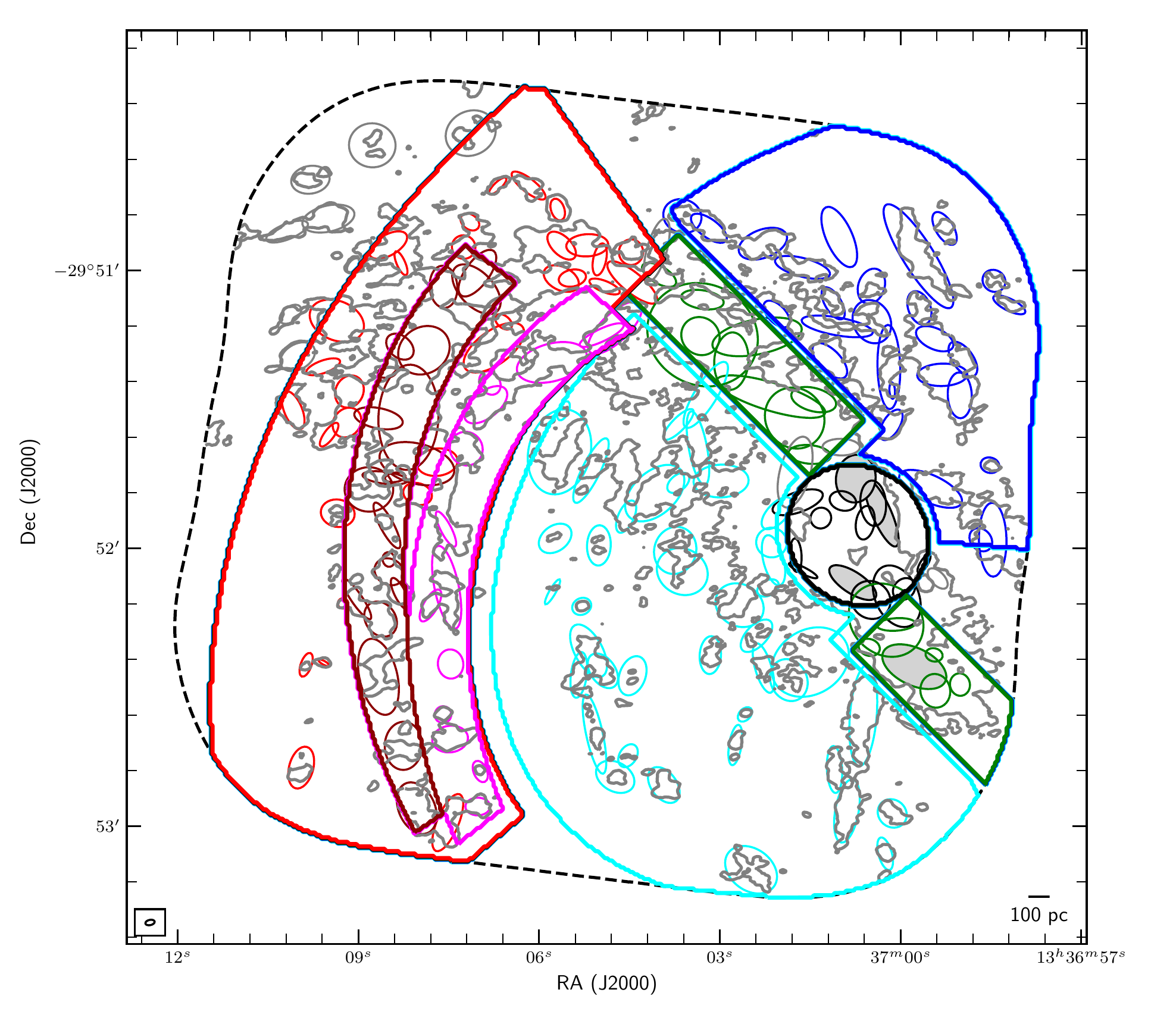}
\caption{Definition of the environmental mask indicated with heavy solid lines, overlaid on the distribution of GMCs (ellipses in thin line) and on the CO map (gray contour). The central region, the bar, inter-arm (trailing side), inter-arm (leading side), the arm, the primary arm ridge, and the secondary arm ridge are defined and indicated with black, green, cyan, blue, red, dark-red, and magenta, respectively.
The four gray-shaded ellipses indicate the four most massive clouds, which appear as outliers in the plots of mass spectra (will be discussed later in \S\ref{SubsecMassSpectrum}).
}
\label{FigEnvDef}
\end{center}
\end{figure}

To investigate the regional variation in the properties of GMCs, we define regional masks indicated in figure \ref{FigEnvDef}.
The area within the galactocentric radius of \timeform{16''} ($\sim$350 pc) is defined as the central region to include a gas ring with a radius of $\sim$9$^{''}$ (\cite{Elmegreen1998M83DoubleRing}) and condensations of molecular gas formed at the crossing points of the ring and the bar offset set ridges (e.g., \cite{Sakamoto2004}).
Two rectangles that cover the offset ridges of the molecular bar are defined as the bar region.
Two inter-arm regions are defined at the leading and trailing sides of the bar, respectively.
The arm region is defined to cover the two spiral arm ridges and associated spurs that extend toward the leading side.
Within the arm region, we also define two subregions that trace the spiral arm ridges described in \S\ref{SubsecDistribution}, i.e., the primary and secondary arm ridges.
\subsection{Environmental variation in GMC properties}
\label{SubsecVariationBoxplot}

\begin{figure} []
\begin{center}
\FigureFile(88mm,88mm){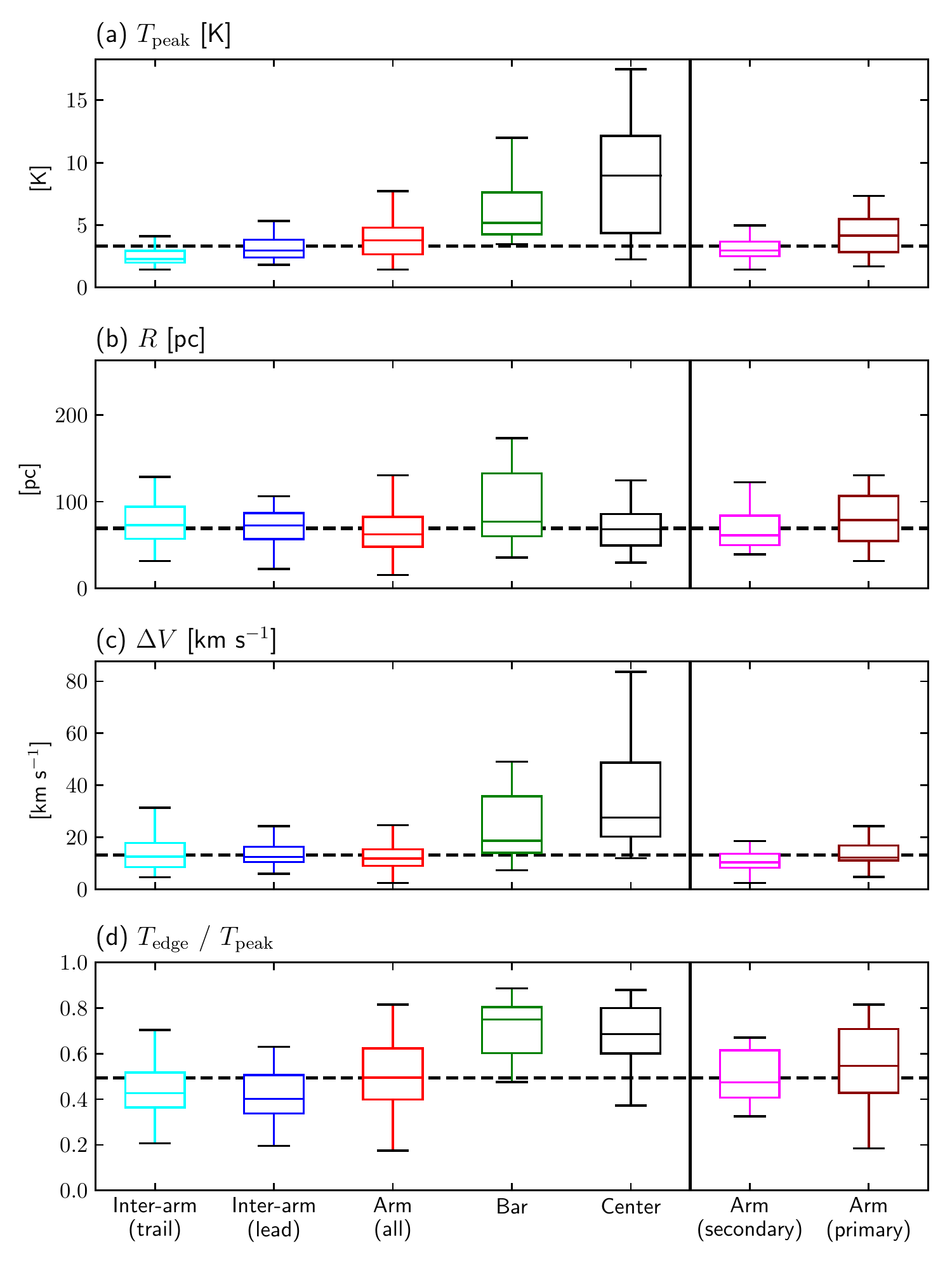}
\caption{Box-and-whisker plots for
(a) peak temperature ($\Tpeak$),
(b) effective radius ($R$),
(c) linewidth ($\dV$), and
(d) edge-to-peak temperature ratio ($\Tedge$/$\Tpeak$).
Upper and lower edges of each box indicate 25th ($\mathrm{Q1}$) and 75th ($\mathrm{Q3}$) percentiles of the number distribution.
Upper whisker extends up to the last data point less than $\mathrm{Q3} + 1.5\mathrm{IQR}$, where $\mathrm{IQR}$ is the interquartile range ($=\mathrm{Q3}-\mathrm{Q1}$).
Lower whisker extends down to the last data point greater than $\mathrm{Q1} - 1.5 \mathrm{IQR}$.
Data points outside the whiskers are indicated with pluses.
Median value is indicated with a horizontal line within each box.
Dashed horizontal line indicates the median value for all samples.
}
\label{FigBoxplot1}
\end{center}
\end{figure}
\begin{figure} []
\begin{center}
\FigureFile(88mm,88mm){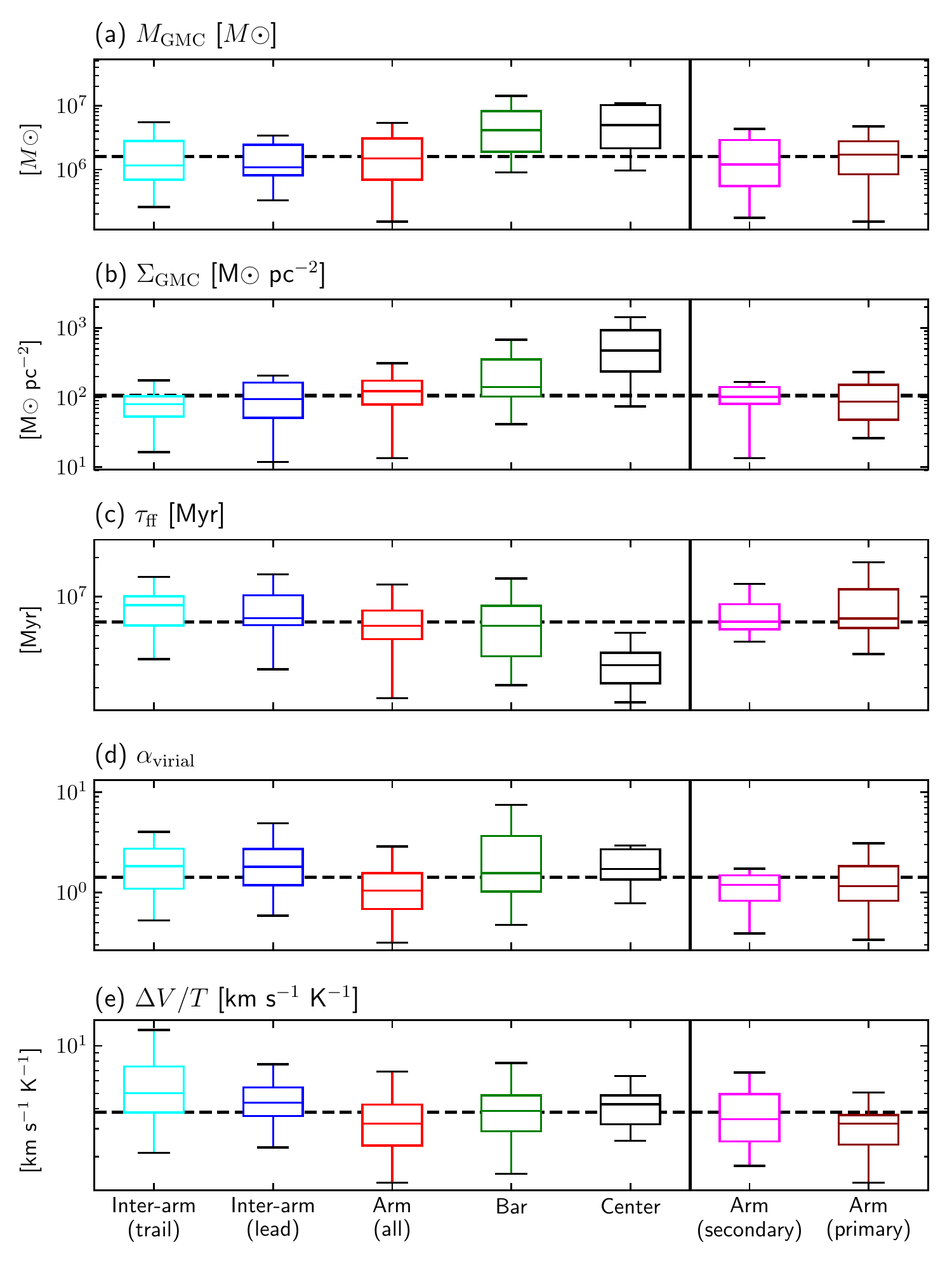}
\caption{Same as figure \ref{FigBoxplot1}, but for
(a) cloud mass ($\Mmol$),
(b) average surface density ($\SigmaMol$),
(c) free-fall time ($\TauFF$),
(d) virial parameter ($\alpha$), and
(e) $\dV$/$\Tpeak$, respectively.}
\label{FigBoxplot2}
\end{center}
\end{figure}
In this subsection, we investigate the environmental variation in GMC properties by examining the statistical distribution of each quantity.
We employ a box-and-whisker plot to display the quartiles of each quantity, as \citet{Colombo2014Env} used for M51.
Table \ref{TableProps} also tabulates the 25th, 50th, and 75th percentiles of relevant GMC properties, including the peak temperature ($\Tpeak$), edge temperature ($\Tedge$), size ($R$), linewidth ($\dV$), cloud mass ($\Mmol$), gas surface density ($\SigmaMol$), virial parameter ($\alphaVir$), and free-fall time ($\TauFF$) of the GMCs in each regional mask.
\begin{table*}
\tbl{Properties of identified GMCs}{
\begin{tabular}{lrrrrrrrrr}
\hline
\hline
&N&$T_{\mathrm{peak}}$&$T_{\mathrm{edge}}$&$R$&${\Delta}V$&$M_{\mathrm{GMC}}$&$\Sigma_{\mathrm{GMC}}$&$\alpha_{\mathrm{virial}}$&$\tau_{\mathrm{ff}}$\\
&&$(\mathrm{K})$&$(\mathrm{K})$&$(\mathrm{pc})$&$(\mathrm{km}\ s^{-1})$&$(10^6\ \mathrm{M}_{\odot})$&$(\mathrm{M}_{\odot}\ \mathrm{pc}^{-2})$&$$&$(\mathrm{Myr})$\\
\hline
all&179&$3.3_{-0.8}^{+1.4}$&$1.6_{-0.5}^{+0.6}$&$69_{-16}^{+20}$&$13.3_{-3.9}^{+5.1}$&$1.6_{-0.8}^{+1.7}$&$106_{-40}^{+71}$&$1.4_{-0.5}^{+0.9}$&$6.4_{-1.6}^{+2.9}$\\
\hline
Inter-arm (trail)&42&$2.3_{-0.3}^{+0.7}$&$1.0_{-0.2}^{+0.3}$&$72_{-15}^{+21}$&$12.6_{-4.0}^{+5.3}$&$1.2_{-0.5}^{+1.6}$&$80_{-26}^{+24}$&$1.8_{-0.7}^{+0.9}$&$8.6_{-2.6}^{+1.5}$\\
Inter-arm (lead)&27&$3.0_{-0.6}^{+0.8}$&$1.2_{-0.1}^{+0.1}$&$72_{-15}^{+14}$&$12.6_{-1.9}^{+3.9}$&$1.1_{-0.3}^{+1.4}$&$95_{-43}^{+68}$&$1.8_{-0.6}^{+0.9}$&$6.8_{-0.8}^{+3.4}$\\
\hline
Arm&66&$3.8_{-1.1}^{+1.0}$&$1.7_{-0.6}^{+0.8}$&$62_{-14}^{+20}$&$11.9_{-2.9}^{+3.5}$&$1.5_{-0.8}^{+1.6}$&$123_{-43}^{+50}$&$1.0_{-0.4}^{+0.5}$&$6.0_{-1.2}^{+1.8}$\\
Arm ridge (secondary)&15&$3.2_{-0.7}^{+0.6}$&$1.6_{-0.5}^{+0.1}$&$64_{-13}^{+21}$&$9.8_{-1.3}^{+3.5}$&$1.3_{-0.7}^{+1.7}$&$113_{-31}^{+24}$&$1.1_{-0.5}^{+0.3}$&$6.7_{-1.0}^{+1.8}$\\
Arm ridge (primary)&21&$4.1_{-1.3}^{+1.3}$&$2.1_{-1.0}^{+1.6}$&$74_{-19}^{+29}$&$12.3_{-3.9}^{+4.1}$&$1.7_{-0.8}^{+1.1}$&$101_{-47}^{+52}$&$1.1_{-0.3}^{+0.5}$&$6.7_{-1.0}^{+4.7}$\\
\hline
Bar&17&$5.2_{-0.9}^{+2.4}$&$3.9_{-0.7}^{+2.1}$&$76_{-16}^{+55}$&$18.7_{-4.5}^{+17.1}$&$4.2_{-2.3}^{+4.1}$&$142_{-38}^{+209}$&$1.6_{-0.5}^{+2.1}$&$6.0_{-2.5}^{+2.5}$\\
\hline
Center&13&$9.8_{-5.6}^{+2.7}$&$5.6_{-2.9}^{+3.0}$&$77_{-25}^{+10}$&$29.5_{-9.9}^{+20.3}$&$5.9_{-3.8}^{+5.0}$&$411_{-184}^{+538}$&$1.7_{-0.4}^{+1.0}$&$3.0_{-0.6}^{+0.9}$\\
\hline
\end{tabular}
}
\label{TableProps}
\begin{tabnote}
Each GMC property is noted as $M_{D25}^{D75}$, where $M$, $D25$, and $D75$ are median, the distance to the 25th percentile from the median, and the distance to the 75th percentile from the median of the number distribution, respectively.
\end{tabnote}
\end{table*}
\subsubsection{Variation in basic properties}
\label{SubsecBoxplot1}
Many of the GMC properties are dimensionally proportional to combinations of three parameters, namely, the temperature, the size, and the linewidth.
Therefore, we start by examining the distribution of these quantities before investigating other properties.
\par
Figures \ref{FigBoxplot1} (a)-(c) show the environmental variation in the peak temperature ($\Tpeak$), effective size ($R$), and linewidth ($\dV$), respectively.
The median peak temperature of the identified clouds is approximately 3.3 K above the background emission, suggesting that most of the clouds in our sample are marginally resolved at the 46 pc spatial resolution.
There is environmental variation in the peak temperature.
By examining the range of the 25th and 75th percentiles for $\Tpeak$ in each region, the peak temperature is found to exhibit regional variations and is highest in the center (4.2--12.5 K), followed by the bar (4.3--7.6 K), arm (2.7--4.8 K), and inter-arm regions (2.0--3.0 K and 2.4--3.8 K for trailing and leading sides, respectively).
Even within the arm region, there is a variation in the peak temperature: $\Tpeak$ in the primary arm ridge extends to a higher regime (2.8--5.4 K) compared to the secondary one (2.5--3.8 K).
In addition to $\Tpeak$, the linewidth also follows a similar trend, although the difference between the inter-arm and the arm is moderate.
\par
On the other hand, the effective size of the clouds does not clearly follow the same trend.
The only notable variation is that clouds in the bar region are larger than the others and there is little variation in the size parameters among other regions.
The lack of strong variation in the size parameter is not unforeseeable, because the cloud partition algorithm tends to divide cloud emission into structures with rather uniform size scale that is comparable to the beam size \citep{Hughes2013GMCs}.
\par
Figure \ref{FigBoxplot1}(d) shows the ratio of the edge-to-peak temperatures of clouds ($\Tedge$ / $\Tpeak$).
The higher this value is for a GMC, the less distinct the cloud becomes from its surroundings.
The median value for the entire sample is approximately 0.49, and thus most of the identified GMCs are not well isolated from their environment.
In particular, GMCs in the bar and the center exhibit quite elevated values (0.6--0.8 for 25th to 75th percentile range).
As this quantity primarily determines the uncertainties assigned to each cloud parameter (as seen in figure \ref{FigExtrapolationFactors}), GMC properties in the center and the bar regions are more uncertain compared to the GMCs in other regions.
\subsubsection{Variation in derived properties}
\label{SubsecBoxplot2}
Here, we look into the statistical distribution of the cloud mass ($\Mmol$), gas surface density ($\SigmaMol$), free-fall time ($\TauFF$), and virial parameter ($\alphaVir$).
Box-and-whisker plots for these parameters are presented in figure \ref{FigBoxplot2}.
These four quantities have a dimensional dependence on the combinations of three quantities, temperature ($T$), linewidth ($\dV$), and cloud size ($R$) as follows:
\begin{eqnarray*}
\SigmaMol &\propto& {T} \dV, \\
\Mmol     &\propto& {T} \dV {R}^2, \\
{\TauFF}  &\propto& \left({T} \dV / {R}\right)^{-1/2}, \\
\alphaVir &\propto& \dV / ({T} {R}).
\end{eqnarray*}
The median value of $\SigmaMol$ for all samples is $\sim$106 $\MsunPerSqPC$ and is close to the values found in the Galactic GMCs (80--120 $\MsunPerSqPC$; \cite{Heyer2009}).
$\SigmaMol$ is highest in the central region with lower and upper quartiles of $\sim$230 and $\sim$950 $\MsunPerSqPC$, respectively.
The bar region also exhibits elevated values of $\SigmaMol$ with lower and upper quartiles of $\sim$100 and $\sim$350 $\MsunPerSqPC$, respectively.
The environmental variation expressed by the fact that $\SigmaMol$ is higher in the center and the bar is easily understood considering the dimensional dependence of $\SigmaMol$ on $T{\dV}$.
\par
The cloud mass, $\Mmol$, shows a similar trend as $\SigmaMol$.
This is also comprehensive because $\Mmol$ has an extra dependence on $R^2$ compared to $\SigmaMol$, and $R$ does not exhibit strong regional variation.
\par
The median value of the free-fall time, $\TauFF$, for all clouds is 6.4 Myr.
It is shortest in the central region with lower and upper quartiles of 2.4 and 3.9 Myr, respectively.
As $\TauFF$ is proportional to the inverse square of $T{\dV}$, the shorter $\TauFF$ in the central region is reasonable.
\par
The virial parameter, $\alphaVir$, has a median value of $\sim$1.4 for all samples.
It tends to be smaller in the arm (median value is $\sim$1.0) compared to other regions including the inter-arm, bar, and center (median value is $\sim$1.8).
As it is dimensionally proportional to ${\dV}/(TR)$ and as $R$ does not exhibit strong regional variation except for the bar, an environmental variation of $\dV/T$ is expected to be mainly responsible for the variation in $\alphaVir$.
Figure \ref{FigBoxplot2}(e) shows the distribution of $\dV/\Tpeak$.
At least for the difference between the inter-arm regions and the arm region, the distribution of $\dV/\Tpeak$ is in agreement with the expectation.

\subsection{Scaling relations}
\label{SubsecScalingRelations}
Early studies of Galactic molecular clouds have revealed the existence of scaling relations between the cloud properties, which are often referred to as Larson's three laws (\cite{Larson1981}; \cite{Solomon1987Larson}):
(1) velocity dispersion of the cloud has a power-law dependence on the cloud size, $\sigmav \propto R^{\beta}$, with an index $\beta$ of approximately 0.5;
(2) the virial mass of the cloud is close to the luminosity mass ($\Mvir$ $\simeq$ $\Mmol$);
and (3) the average volume density of the cloud is nearly inversely proportional to the cloud size ($\rhoMol$ $\propto$ $R^{-1}$), or in other words, the average surface density is almost constant among GMCs.
In reality, only two of these three laws are independent because if two of them are provided, then the remaining one can be algebraically derived from the two laws.
Therefore, we examine the first two of Larson's laws with the GMCs in M83.
\par
\subsubsection{Linewidth--size relation}
\label{SubsecLarsonFirst}
\begin{figure*} []
\begin{center}
\FigureFile(158mm,158mm){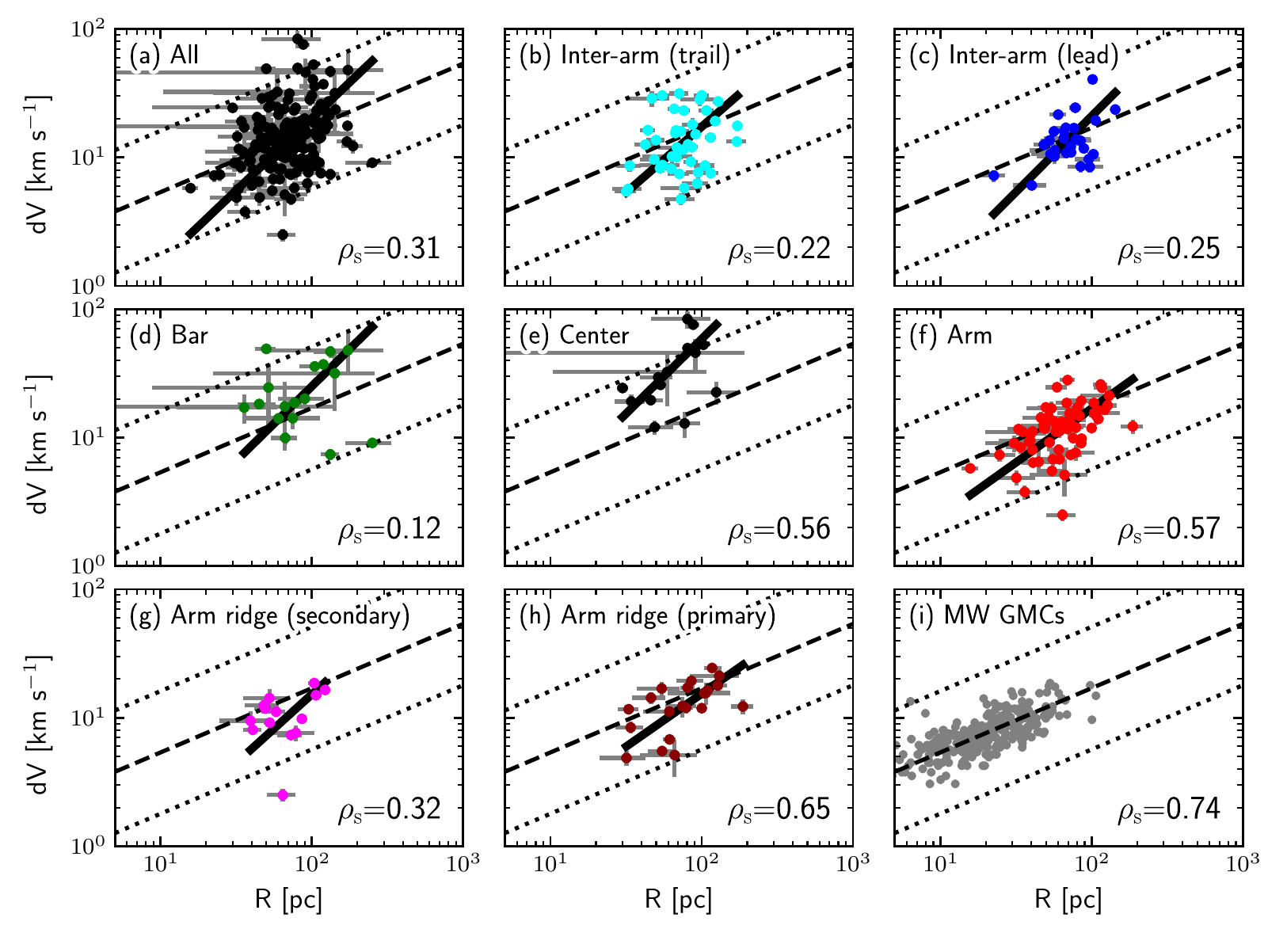}
\caption {(a) Linewidth--size relationship for all the GMC samples in M83.
(b-h) Same as (a), but for each subregion defined in figure \ref{FigEnvDef}.
(i) Same as (a), but for Galactic clouds \citep{Solomon1987Larson} and M51 clouds \citep{Colombo2014Env} indicated with gray and purple markers, respectively.
In each plot, the dashed line indicates the fit to Galactic GMCs given by \citet{Solomon1987Larson}, $\dV$ $\sim$ 1.70 $R^{0.5}$, and the two dotted lines show $\times$3 and 1/3 of it, respectively.
Spearman's rank correlation coefficient ($\rhos$) is noted at the bottom right corner of each plot.
Black solid lines indicate the power-law fitting results for the GMCs in M83.
}
\label{FigLinewidthSize}
\end{center}
\end{figure*}
Figure \ref{FigLinewidthSize}(a) shows the linewidth--size relation for all GMC samples, and (b)-(h) show the same relation for each subregion.
For reference, figure \ref{FigLinewidthSize}(i) plots Galactic GMC samples from \citet{Solomon1987Larson}.
The linewidth--size relation given by \citet{Solomon1987Larson} is plotted with a dashed line in each plot.
In each plot, Spearman's rank correlation coefficient ($\rhos$), a nonparametric measure of a rank correlation coefficient, is given to indicate the degree of correlation.
We regard a value of $\rhos$ greater than 0.7 as a sign of significant correlation.
Data points in each plot are fitted with a power-law relation by minimizing the effective variance \citep{Orear1982}, and the fitted result is overplotted.
\par
If all the GMC samples are taken together, it is difficult to identify a sign of the existence of correlation (figure \ref{FigLinewidthSize}a).
The calculated $\rhos$ is approximately 0.3, and this is quite low compared to the value found from the Galactic GMCs ($\rhos$ $\sim$ 0.7).
Even with division into subregions, the apparent lack of correlation is also the same, supported by the low value of $\rhos$ (figure \ref{FigLinewidthSize}b-h).
We will shortly discuss why we observed a weaker correlation for the linewidth--size relation in M83 compared to the MW samples of \citet{Solomon1987Larson} in \S\ref{SubsecPoorCorrelation}.
\subsubsection{Relationship between virial mass and CO luminosity}
\label{SubsecLarsonSecond}
\begin{figure*} []
\begin{center}
\FigureFile(158mm,158mm){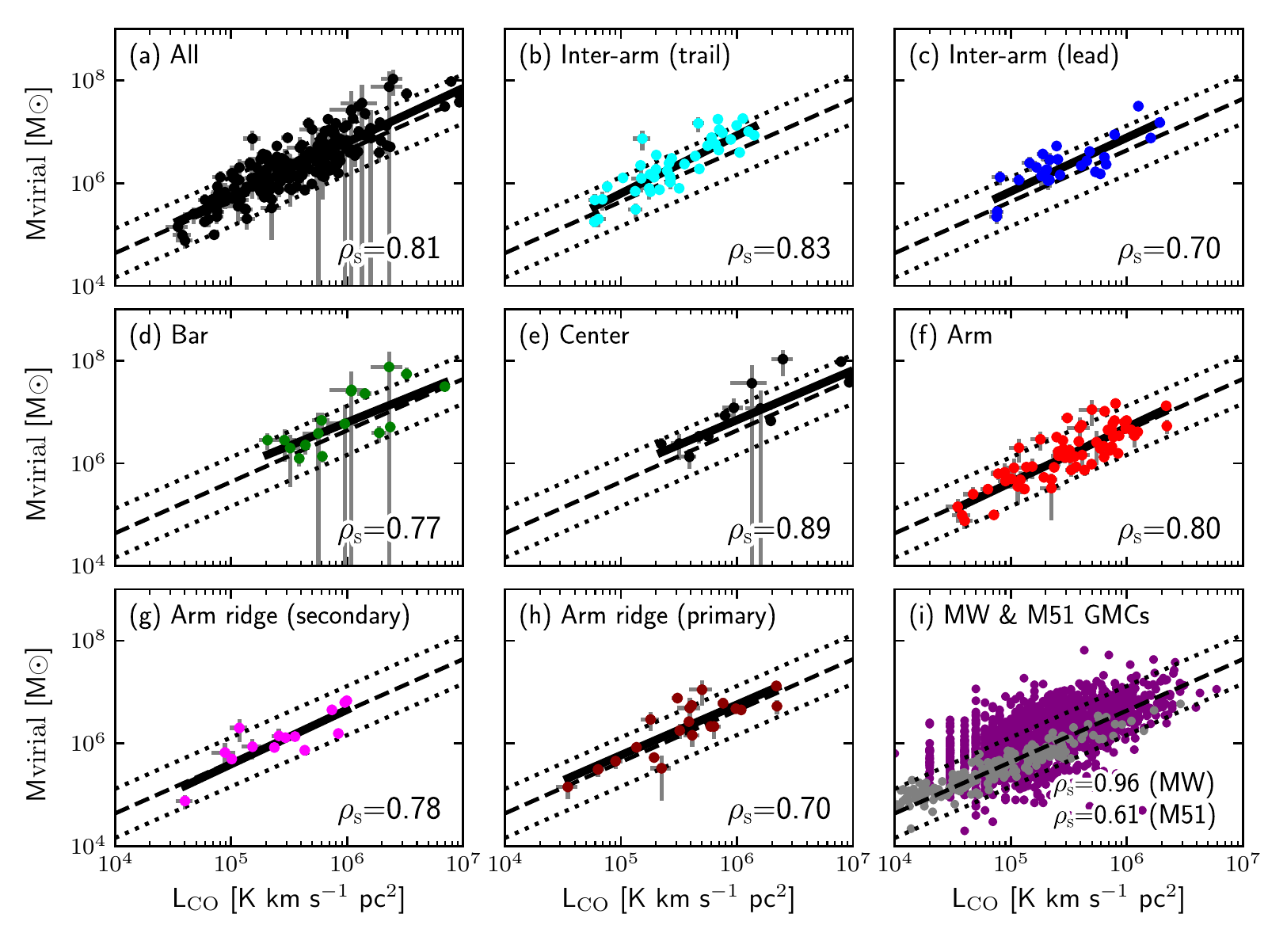}
\caption{Same as figure \ref{FigLinewidthSize}, but for the relationship between virial mass and CO luminosity.
The Dashed line in each plot indicates that the line for the virial mass is equal to the cloud mass derived with equation \ref{EqLCOtoMsun}, and two dotted lines indicate $\times$3 and 1/3 of it, respectively.}
\label{FigMvirLCO}
\end{center}
\end{figure*}
Figure \ref{FigMvirLCO} shows the relationship between the virial mass and CO luminosity for the GMCs in M83.
In each plot, the line representing the equality of the virial mass and the luminosity mass calculated from equation \ref{EqLCOtoMsun} is overplotted.
\par
Contrary to the low correlation found for the linewidth--size relation, the GMCs in M83 are mostly in agreement with Larson's second law.
This is consistent with the narrow range of $\alphaVir$, lower and upper quartiles being 0.9 and 2.3 for all samples, seen in the previous subsection.
Also, the power-law fitting resulted in almost linear relations for all of the subregions considered here.
Correlation coefficients, $\rhos$, fall in the range between 0.70 and 0.89 and support the apparent existence of the significant correlation.
\subsubsection{Why the observed linewidth--size relation exhibits weak correlation?}
\label{SubsecPoorCorrelation}
\begin{figure*} []
\begin{center}
\FigureFile(158mm,158mm){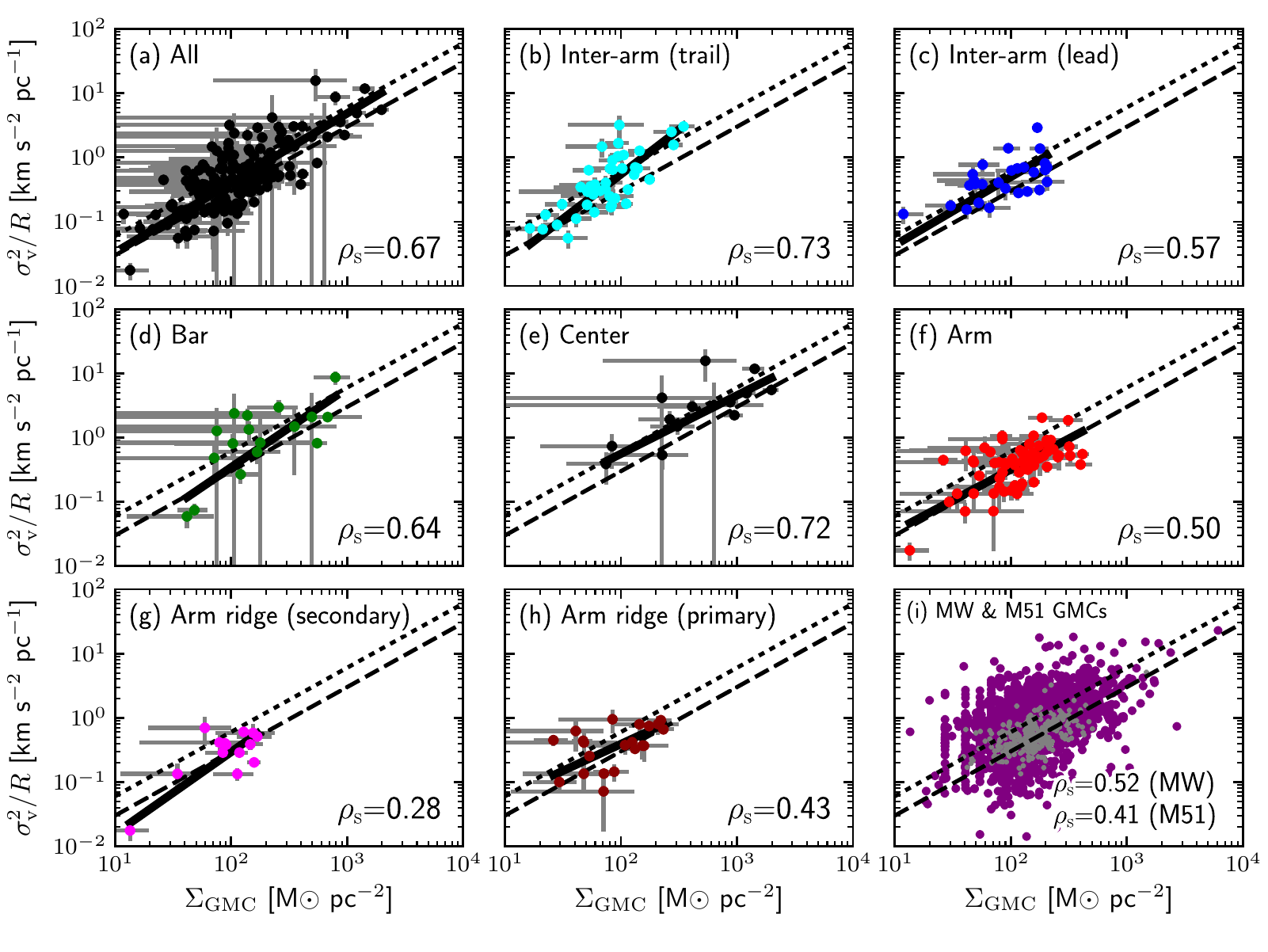}
\caption{Same as figure \ref{FigLinewidthSize}, but for the square of the linewidth--size relationship coefficient (${\sigmavz}^2 = \sigmav^2 / R$) as a function of the mean surface density for the GMCs in M83.
In each plot, dashed and dotted lines indicate the line for $\alphaVir=1$ and $\alphaVir=2$.
}
\label{FigSizelinewidthCoeffSigma}
\end{center}
\end{figure*}
Among the three Larson's laws, the linewidth--size relation has attracted particular attention because the similarity of its form with the Kolmogorov cascade, which happens for incompressible turbulence, is considered to be the manifestation of the hierarchical nature of GMC which is governed by the turbulent motion (\cite{Larson1981}).
The tight correlation between the linewidth and size in Galactic clouds found by \citet{Solomon1987Larson} and the similarity of the internal structure function of GMCs studied by \citet{HeyerBrunt2004} are taken as evidence for the universality of the linewidth--size relation.
However, later with the updated dataset, \citet{Heyer2009} pointed out that the coefficient of the linewidth--size relation ($\sigmavz$) is in reality not uniform, but it depends on the surface density ($\SigmaGas$) of the cloud as follows,
\begin{equation}
\sigmavz (= \sigmav/R^{1/2}) \propto \SigmaGas^{1/2}.
\end{equation}
\citet{Heyer2009} also pointed out that this $\sigmavz$--$\SigmaGas$ relation is mathematically equivalent to Larson's second law, which is also equivalent to $\Mvir=\Mmol$ if the assumed CO-to-H$_2$ conversion factor is correct.
\par
The weak correlation observed for the linewidth--size relation with a $\rhos$ of approximately 0.3 (\S\ref{SubsecLarsonFirst}) should partly be attributed to the fact that the GMC samples adopted here have a limited dynamic range, in particular for the size parameter (\S\ref{SubsecBoxplot1}).
In addition to the limited dynamic range, the relatively large uncertainty associated with the cloud properties seen in figure \ref{FigExtrapolationFactors} could be responsible for the weak correlation of the linewidth--size relation.
\par
However, the firm correlation observed between the virial mass and CO luminosity (\S\ref{SubsecLarsonSecond}) implies that the variation in the surface density is yet another factor that contributes to the weak correlation of the linewidth--size relation, as pointed out by \citet{Heyer2009}.
Figure \ref{FigSizelinewidthCoeffSigma} plots the square of the linewidth--size relation coefficient as a function of the surface density for the GMCs in M83.
The dashed line indicates the following relation,
\begin{equation}
\label{EqSizelinecoeff1}
{\sigmavz}^2 = \left( \frac{{\sigmav}^2}{R} \right) = \frac{a_1}{5} \pi G \SigmaMol,
\end{equation}
which is equivalent to $\Mvir$ = $\Mmol$ ($\alphaVir$ = 1, i.e., virial equilibrium) if the assumed CO-to-H$_2$ conversion factor is correct.
As a further reference, the dotted line in each plot corresponds to the relation $\alphaVir = 2$, which is expected for the case of simple gravitational collapse (\cite{BallesterosParedes2011A}; see also \cite{Larson1981}).
\par
Plotting the all samples together (figure\ref{FigSizelinewidthCoeffSigma}a), we see that the linewidth--size relation coefficient does depend on the gas surface density with a modest sign of correlation ($\rhos$ $\sim$0.67), and data points are clustered around $\alphaVir$ = 1.
Divided into subregions, correlation coefficients become lower in some regions, but this seems simply due to the reduced dynamic range of the gas surface density.
The lower correlation coefficient obtained for the $\sigmavz$ - $\SigmaMol$ relation compared to the $\Mvir$--$\Mmol$ relation is expected because the former coefficient is derived from the latter by dividing both sides by $R^2$ and associated constant factors that make the dynamic range of the $\sigmavz$--$\SigmaMol$ relation smaller.
\par
Galactic GMCs from \citet{Solomon1987Larson} are also clustered around the line $\alphaVir=1$ (figure \ref{FigSizelinewidthCoeffSigma}i), although $\rhos$ is low ($\sim0.52$).
The lower correlation coefficient for the MW GMCs would also be due to the limited variation in $\SigmaMol$ sampled by the survey of \citet{Solomon1987Larson}.
\par
Comparing figures \ref{FigSizelinewidthCoeffSigma}(a) and \ref{FigSizelinewidthCoeffSigma}(i), it is apparent that the GMCs in M83 cover a wider range of $\SigmaMol$ than the samples from \citet{Solomon1987Larson}.
Therefore, we conclude that the dependence of the linewidth--size coefficient on the surface density at least partly explains why the linewidth--size relation for the GMCs in M83 exhibits weak correlation.
The limited dynamic range of the size parameter should also play a role.
It is notable that \citet{Hughes2013GMCs} also observed a weak correlation for the linewidth--size relation in extragalactic GMCs with a nominal procedure of cloud decomposition which tends to identify structures close to a resolution limit.
However, with an alternative decomposition procedure that tends to achieve uniform surface density threshold, they obtained a stronger correlation.
The stronger correlation obtained with the uniform surface density threshold is in agreement with the fact that $\sigmavz$ scales as $\SigmaMol$.
\subsection{Mass spectrum of GMCs}
\label{SubsecMassSpectrum}
The mass spectrum of GMCs is known to follow a power-law relation, ${dN}/{dM} \propto M^{\gamma}$.
In the inner disk of the MW, the index $\gamma$ is claimed to be approximately -1.5 by several studies (\cite{Solomon1987Larson}; \cite{WilliamsMcKee1997}; \cite{Rosolowsky2005MassSpectra}).
On the other hand, in the outer disk of the MW, the slope of the GMC mass spectrum is suggested to be steeper compared to the inner disk with $\gamma$ being smaller than -1.5 (\cite{Heyer2001OuterGalaxy}; \cite{Rosolowsky2005MassSpectra}).
This difference of $\gamma$ found between the inner and outer parts of MW may imply that the mass spectrum of GMCs is not uniform and could vary over different environments.
Indeed, nonuniform values of $\gamma$ are reported by extragalactic studies:
citing some examples, in M33, LMC, M51, NGC 300, NGC 4526, and NGC 1068, $\gamma$ is found to be
-2.0 $\pm$ 0.2 (\cite{Rosolowsky2007M33}; also -2.0 $\pm$ 0.1 by \cite{Gratier2012M33}),
-1.75 $\pm$ 0.06 (\cite{Fukui2008}; although $>$ -2 by \cite{Wong2011Magma}),
-2.29 $\pm$ 0.09 (\cite{Colombo2014Env}),
-2.7 $\pm$ 0.5 (\cite{Faesi2014NGC300APEX}),
-2.39 $\pm$ 0.03 (\cite{Utomo2015}),
and -1.25 $\pm$ 0.07 (\cite{Tosaki2017}), respectively.
In addition to these galaxy-to-galaxy variations in $\gamma$, some of these extragalactic studies also reported regional variations in $\gamma$ within a galaxy (\cite{Rosolowsky2007M33}; \cite{Gratier2012M33}; \cite{Colombo2014Env}; \cite{Utomo2015}).
\par
Another important characteristic of GMC mass spectra is that although most of them could be fitted with a power-law relation at certain mass ranges, some of them are underpopulated at higher masses (e.g., \cite{Fukui2008}).
To take the deviation from the power-law relation at higher masses into account, it is a common practice to express the mass spectrum with a truncated power law, which is written as follows in integral form \citep{WilliamsMcKee1997}:
\begin{equation}
N(> {M}) = -\frac{N_{\mathrm{u}}}{\gamma + 1} \left[ \left(\frac{M}{M_\mathrm{u}}\right)^{(\gamma + 1)} - 1 \right],
\label{EqMF}
\end{equation}
where $M_{\mathrm{u}}$ is the upper cutoff mass of GMCs.
For Galactic GMCs in the inner MW, \citet{WilliamsMcKee1997} fitted the mass spectrum with this equation and found $\gamma$=-1.6, $M_\mathrm{u} = 6 \times 10^6 $ $\Msun$, and $N_\mathrm{u}$ = 63, respectively.
The total cloud mass integrated over a mass spectrum within the range from the lowest cloud mass $M_{\mathrm{l}}$ to the highest cloud mass $M_{\mathrm{u}}$ is given as
\begin{equation}
M_{\mathrm{total}}(\ge M_{\mathrm{l}}) = \frac{N_{\mathrm{u}} M_{\mathrm{u}}}{\gamma + 2} \left[1 - \left(\frac{M_{\mathrm{l}}}{M_{\mathrm{u}}}\right)^{(\gamma +2)}\right].
\label{EqMFSummation}
\end{equation}
If $\gamma$ is larger than -2, then massive clouds account for a large proportion of the total cloud mass, and the total cloud mass can be approximately given as
$ \frac{N_{\mathrm{u}} M_{\mathrm{u}}}{\gamma + 2}$.
Substituting the parameters for the inner MW quoted above, the total gas mass in the inner MW could be estimated as $\sim$1 $\times$ 10$^9$ $\Msun$ (\cite{WilliamsMcKee1997}), which is in agreement with other estimates (\cite{Heyer2015Review}).
\par
The shapes of GMC mass spectra are considered to reflect the properties of the environments from where the GMCs are sampled.
In a region where formation of massive GMCs is enhanced, either by collisional agglomeration of preexisting smaller clouds or by the roles of instabilities, the slope of GMC mass spectra would be shallower (e.g., \cite{Dobbs2008GMCFormation}).
On the other hand, destructive processes such as stellar feedback (\cite{Wilson1990M33}) and large-scale shear motion (e.g., \cite{Meidt2015Lifetime}) work to decrease the number of massive GMCs and thus steepen the slope of mass spectra.
The variation in $\gamma$ observed could be due to the environmental variation in the balance between the formation and destruction processes of GMCs (\cite{Inutsuka2015}; \cite{Kobayashi2017MF}).
\par
In this subsection, we will derive the mass spectra of the identified GMCs in M83 and fit them with the truncated power law.
The slopes of the fitted mass spectra, $\gamma$, will be compared with other extra-galactic studies.

\par

\par

\par
\begin{figure*} []
\begin{center}
\FigureFile(168mm,168mm){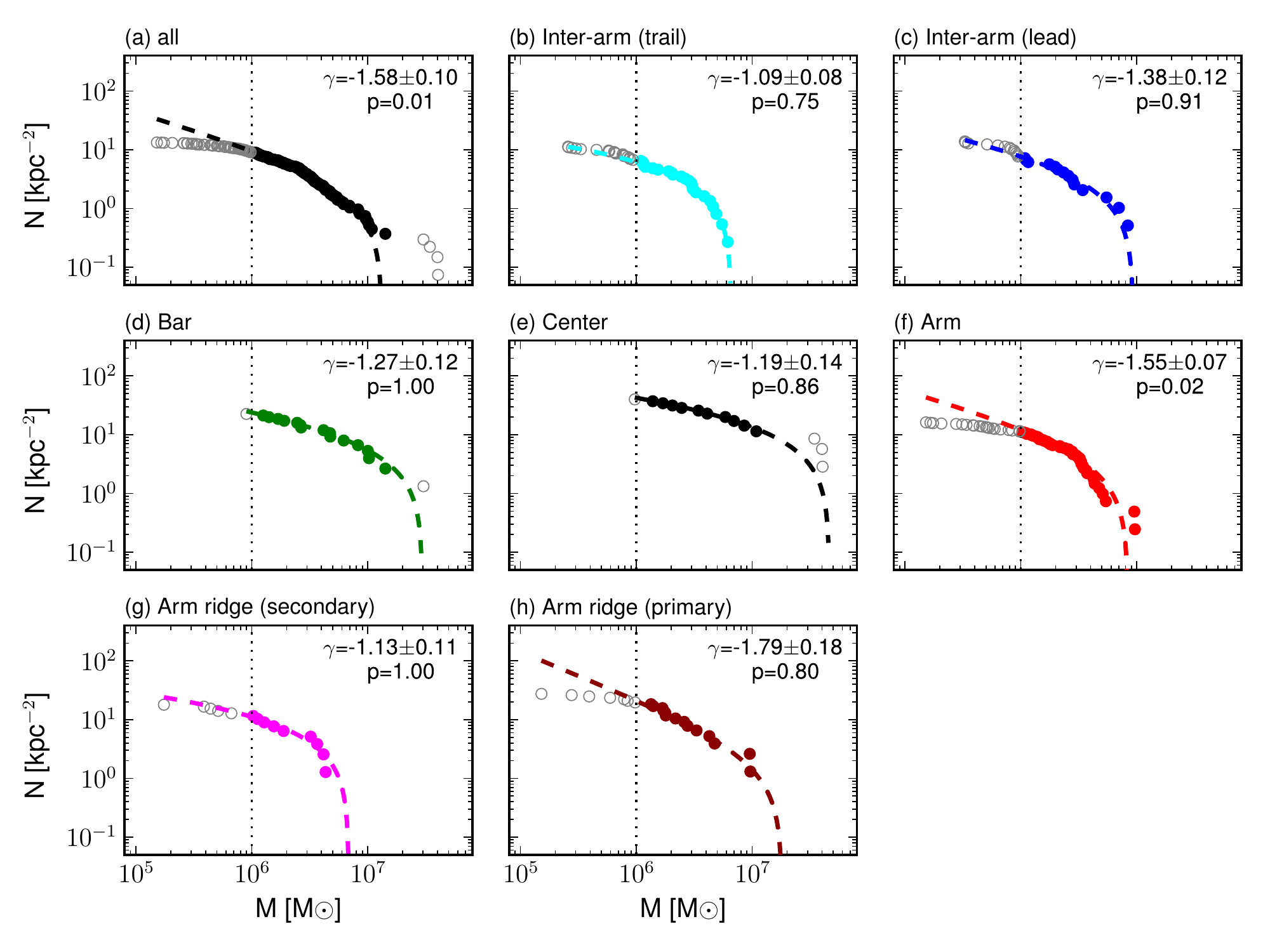}
\caption{(a) Cumulative mass function for all of the GMC samples in M83.
(b-h) Same as (a), but for the GMCs in each subregion defined in figure \ref{FigEnvDef} (b-h).
In each plot, filled and open circles indicate the GMCs that are used and not used for fitting the spectra with equation (\ref{EqMF}), respectively and the dashed line indicates the fitted function.
The dotted vertical line indicates the lower mass limit, 10$^6$ $\Msun$, for the fitting of mass spectra.
GMCs with masses below this limit were not used for the fitting.
The slope of the fitted mass spectrum, $\gamma$, and the $p$-value for KS test are noted at the top right corner of each plot.
}
\label{FigMassFunctions}
\end{center}
\end{figure*}
Figure \ref{FigMassFunctions} shows the cumulative mass function of the GMCs in M83, for all samples (a) and for each subregion (b-h).
The ordinates of the plots are normalized by the area of each region from which GMCs are sampled.
Although the mass detection limit is expected to be approximately 2.3 $\times$ 10$^5$ $\Msun$ in uncrowded regions (\S\ref{SubsecCloudIdentification}), the minimum masses of the sampled GMCs in the bar and center are approximately 10$^6$ $\Msun$.
On the other hand, in regions other than the bar and center, smaller GMCs with masses down to a few multiples of 10$^5$ $\Msun$ are sampled, in agreement with the expected detection limit.
The elevated detection limits in the bar and center are most likely due to the high background surface density of molecular gas in those regions.
As can be seen in figure \ref{FigMulti1}(a), the CO intensity in the bar and center is mostly over the contour line of 48 $\KkmPerS$, which corresponds to the molecular gas surface density of $\sim$200 $\MsunPerSqPC$ with the same assumptions made in \S\ref{SubsecDistribution}.
Therefore, GMCs in the bar and center regions are surrounded by 'ambient' gas with a high surface density, comparable to the typical values of Galactic GMCs (\cite{Solomon1987Larson}).
Only clouds with significant over-densities above the background level of $\sim$200 $\MsunPerSqPC$ can be identified as GMCs.
On the other hand, if there is a cloud with a mass of 10$^{6}$ $\Msun$ and with a size similar to the spatial distribution (FWHM of $\sim$46 pc), then the average surface density of cloud is $\sim$420 $\MsunPerSqPC$.
As this value is only a factor of 2 higher than the background, it is natural that clouds smaller than 10$^{6}$ $\Msun$ are not detected in the bar and center regions.
\par
We fitted the cumulative mass function for each region with equation (\ref{EqMF}), using the GMC samples with a mass between 10$^6$ $\Msun$ and 3 $\times$ 10$^7$ $\Msun$.
The lower end corresponds to the approximate minimum GMC mass in the bar and the center.
The higher end is taken to exclude four clouds with cloud masses of 3.0, 3.4, 4.0, and 4.4 $\times$ 10$^7$ $\Msun$ because their masses are larger than those of other clouds by more than 0.3 dexes and appear as outliers on figure \ref{FigMassFunctions}(a).
\par
A bootstrap method with 100 draws was used to estimate the confidence interval of the fitted parameters by taking the uncertainties of each GMC mass.
A draw for the bootstrap estimation is made by adding random mass noise to each GMC and fitting equation (\ref{EqMF}) to the cumulative mass function obtained from the data with noise added.
The mass noise is derived by assuming a Gaussian distribution with a standard deviation equivalent to the mass uncertainty derived for each GMC.
The most likely value and the confidence interval for each parameter were obtained by taking the median and the median absolute deviation (MAD) of 100 bootstrap realizations for each parameter.
After obtaining the most likely values for each parameter, the two-sided Kolmogorov--Smirnov (KS) test was performed to evaluate the goodness of fit.
For the KS test, GMCs outside of the fitting range were also included.
The resultant fitting parameters along with the $p$-value for the KS test are summarized in table \ref{TableMFParams}.
In addition, table \ref{TableMFParams} also lists the total cloud mass integrated over the fitted mass spectrum obtained using equation (\ref{EqMFSummation}) with an $M_{\mathrm{l}}$ of 10$^{3}$ $\Msun$ and the sum of the GMC mass for each region.
\par
\begin{table*}[htbp]
\tbl{Parameters of the GMC mass distribution}{
\begin{tabular}{lrrrrrrr}
\hline
\hline
Name&Area \footnotemark[(1)]&$\gamma$ \footnotemark[(2)]&$N_{u}$ \footnotemark[(3)]&$M_{u}$ \footnotemark[(4)]&$p$-value \footnotemark[(5)]&$M_\mathrm{cloud, total}$ \footnotemark[(6)]&$\Sigma$ $M_\mathrm{GMC}$ \footnotemark[(7)]\\
&$(\mathrm{kpc}^{-2})$&&&$(10^6 M\odot)$&&$(10^7 M\odot)$&$(10^7 M\odot)$\\
\hline
all&13.5&-1.58 $\pm$ 0.10&21.1 $\pm$ 6.3&13.3 $\pm$ 2.1&0.01&65.2&56.0\\
\hline
Inter-arm (trail)&3.7&-1.09 $\pm$ 0.08&11.4 $\pm$ 1.0&6.6 $\pm$ 0.3&0.75&8.3&7.6\\
Inter-arm (lead)&1.9&-1.38 $\pm$ 0.12&4.2 $\pm$ 0.7&9.3 $\pm$ 0.5&0.91&6.3&5.3\\
\hline
Arm&4.1&-1.55 $\pm$ 0.07&11.9 $\pm$ 1.4&8.4 $\pm$ 0.3&0.02&21.6&13.8\\
Arm ridge (secondary)&0.8&-1.13 $\pm$ 0.11&3.9 $\pm$ 0.6&6.9 $\pm$ 0.6&1.00&3.1&2.4\\
Arm ridge (primary)&0.8&-1.79 $\pm$ 0.18&1.4 $\pm$ 1.0&18.0 $\pm$ 5.6&0.80&10.9&5.3\\
\hline
Bar&0.8&-1.27 $\pm$ 0.12&3.2 $\pm$ 1.2&29.7 $\pm$ 8.1&1.00&13.0&10.8\\
\hline
Center&0.4&-1.19 $\pm$ 0.14&2.5 $\pm$ 1.2&46.3 $\pm$ 20.7&0.86&14.4&16.3\\
\hline
\end{tabular}
}
\label{TableMFParams}
\begin{tabnote}
\footnotemark[(1)] Area of the region.
\footnotemark[(2-4)] Parameters of the fitted truncated power-law function.
\footnotemark[(5)] The $p$-value of the KS test.
\footnotemark[(6)] The total cloud mass obtained by integrating the fitted mass spectrum within the range 10$^{3}$ $\Msun$ to $M_{\mathrm{u}}$.
\footnotemark[(7)] The sum of the GMC masses within the region.
\end{tabnote}
\end{table*}
The fitted mass spectrum for all samples (figure \ref{FigMassFunctions}a) has a slope of $\sim$-1.6 $\pm$ 0.1, which is in agreement with the value found in the inner disk of the galaxy ($\gamma$=-1.6; \cite{WilliamsMcKee1997}).
The KS test reported a low $p$-value of $\sim$0.01. This is mostly due to the disagreement between the fitted function and the actual data below 10$^6$ $\Msun$, which we suspect to be caused by the elevated detection limit in crowded areas.
\par
Divided into subregions, the slopes of the mass spectrum show a modest regional variation that ranges between -1.1 and -1.8.
The slope in the arm region ($\sim$-1.6) is steeper than both of the inter-arm subregions ($\sim$-1.1 and $\sim$-1.4).
The two arm ridges located inside the arm also exhibit different slopes from each other.
The primary arm ridge has the steepest slope among all of the subregions ($\sim$-1.8).
On the other hand, the slope for the secondary arm is $\sim$-1.2 and is similar to that in the inter-arm regions.
\par
All of the slopes found here are greater than -2, which implies that massive GMCs dominate a significant fraction of the total cloud mass, as in the MW.
With a mass spectrum having a shallow slope, the contribution of smaller clouds is expected to be less important. This point can be confirmed from the closeness between the total mass integrated over the fitted mass spectrum and the summation of the identified GMC mass in regions with shallow $\gamma$, such as the inter-arm, bar, and center regions (7th and 8th columns in Table \ref{TableMFParams}).
\subsubsection{Comparison with other extragalactic measurements of mass spectra}
In general, a galaxy-to-galaxy comparison of GMC mass spectra has to be made with great care because differences in the observational setup and in the cloud identification algorithm affect the observed shapes of mass spectra (e.g., \cite{Wong2011Magma}).
A cloud identified as a single entity with a particular spatial resolution could be divided into several smaller clouds if a finer spatial resolution is utilized.
The sensitivity limit also affects the shape of the GMC mass spectrum because if a physical upper limit of the GMC mass exists, the slope of the mass spectrum steepens at a higher mass and thus with a higher mass limit, the observed slope tends to be steeper (e.g., \cite{Fukui2008}).
For example, while the index $\gamma$ is found to be $\sim$-1.75 in LMC with a resolution of 50 pc (\cite{Fukui2008}), with a finer spatial solution, \citet{Wong2011Magma} reported a slope of $\gamma$ $\le$ -2.
\par
Despite the difficulty of galaxy-to-galaxy comparison of GMC mass spectra, or more precisely work-to-work comparison, several studies agreed in pointing out that the slope of GMC mass spectra tends to steepen when the galactocentric radius within a galaxy increases (e.g., \cite{Rosolowsky2007M33}; \cite{Gratier2012M33}; \cite{Colombo2014Env}).
In general, the fraction of molecular gas is lower in the outer part of a galactic disk compared to the inner disk. Thus, the steeper slope in the outer part could be interpreted as a consequence of the difficulty in building massive GMCs in regions where the amount of molecular gas is lower (\cite{Dobbs2008GMCFormation}).
\par
A few galaxies have been investigated for environmental variation in GMC mass spectra with a spatial resolution comparable to the one used here ($\sim$46 pc).
For example, two independent studies reported the radial variation of the index in M33, which changes from $\gamma$ $>$ -2 in the inner part, which is within a galactocentric radius of 2.2 kpc, to $\gamma$ $<$ -2 in the outer part (-1.8 vs -2.1 in \cite{Rosolowsky2007M33}, -1.6 vs -2.3 in \cite{Gratier2012M33}).
The variation in $\gamma$ apparently correlates with the radial variation in the molecular gas fraction in M33, because the fraction declines from 60\% to 20\% from the center to the outer disk (e.g., \cite{Rosolowsky2007M33}).
Another example is the environmental variation in GMC mass spectra observed in M51 (\cite{Colombo2014Env}).
Within a galactocentric radius of approximately 1.2 kpc ($\sim$\timeform{30''}) in M51, $\gamma$ is reported to be -1.3 -- -1.6 (regions labeled as the nuclear bar and molecular ring).
On the other hand, outside of 1.2 kpc, the reported values of $\gamma$ range between -1.8 and -2.6.
Therefore, M51 also exhibits a decreasing trend in $\gamma$ with increasing galactocentric radius.
As in M33, the decline in $\gamma$ is possibly correlated with the radial variation in the molecular gas fraction because the molecular gas fraction within 1.2 kpc is 90\% or higher, while outside of 1.2 kpc, it slightly declines to 70\%--80\% (read from the map in \cite{Koda2009}).
\par
The values of $\gamma$ found in M83 here range between -1.1 and -1.8, and are shallower than most values found in M33 and M51.
The target regions studied here in M83, which are located within a galactocentric radius of 2.6 kpc, are predominantly molecular (\S\ref{SubsecDistribution}).
Therefore, the shallower slopes found here seem to agree with a tendency for the slopes of GMC mass spectra to be shallower in the molecular-dominated part of the galactic disks.
We also note that from the examples cited above, similarly shallow slopes are reported only in the central $\sim$1.2 kpc of M51 (-1.3 and -1.6 in \cite{Colombo2014Env}).
\subsubsection{Possible interpretation of the variation in the mass spectrum slope}
\citet{Inutsuka2015} and \citet{Kobayashi2017MF} pointed out that if the continuity equation of molecular clouds in mass space is assumed, the slope of the mass spectrum, $\gamma$, could be described as
\begin{equation}
\label{EqTfTd}
\gamma = -(1 + T_{\mathrm{f}} / T_{\mathrm{d}}),
\end{equation}
where $T_{\mathrm{f}}$ and $T_{\mathrm{d}}$ are the time scales for the formation and destruction processes of molecular clouds, respectively.
Although the formation and destruction timescales are not well constrained at this moment, the essence of equation \ref{EqTfTd},
which is the dependence of $\gamma$ on the formation and destructive processes,
appears qualitatively in agreement with suggestions made by previous studies.
For example, \citet{Dobbs2008GMCFormation} suggested that in regions where formation of massive GMCs is enhanced, the mass spectrum slope becomes shallower.
On the other hand, \cite{Wilson1990M33} suggested that in regions where stellar feedback are effective enough to depopulate massive GMCs, the slope becomes shallower (\cite{Wilson1990M33}).
\par
As we have seen in the previous subsection, the slopes of the GMC mass spectra obtained here range from -1.1 to -1.8, which are rather shallow compared to the values obtained in other studies that have investigated the regional variation in GMC mass spectra within a galaxy (\cite{Rosolowsky2007M33}; \cite{Gratier2012M33}; \cite{Colombo2014Env}).
As the gas disk within the observed region is already predominantly molecular, GMCs do not have to be formed from atomic gas but instead are formed directly from molecular gas (\cite{Dobbs2008GMCFormation}).
In light of the idea that the balance between the formation and destruction timescales of GMCs determines the shape of the GMC mass spectrum (\cite{Inutsuka2015}; \cite{Kobayashi2017MF}; also \cite{Wilson1990M33}), the shallow slopes found here in M83 might be a reflection of effective formation of massive GMCs in the molecular-dominated gas disk with a short $T_{\mathrm{f}}$.
\par
We also speculate that the steeper slope found in the primary arm ridge (-1.8) compared to the other regions in M83 studied here is a result of the enhanced influence of stellar feedback, which effectively shortens $T_{\mathrm{d}}$.
We will see later in \S\ref{SectionFeedback} that in the primary arm ridge, GMCs exhibit an elevated SFE and many of them appear to be disrupted by stellar feedback.
If the destructive role of stellar feedback is more efficient in the primary arm ridge compared to other regions, then $T_{\mathrm{d}}$ would be effectively shorter and thus the slope of the mass spectrum steeper.

\section{Star formation in GMCs}
\label{SecStarFormationInGMCs}
As referred to in the introduction, GMC scale examination of the SFE is important to address how star formation in GMCs and galaxies is regulated to achieve Gyr-long depletion time.
In this section, we examine the SFR of the GMCs in M83 presented in the previous section.
We will derive the SFR for individual $\HII$ regions (\S\ref{SubsecSFRDerivation}) and then cross-match $\HII$ regions with the GMCs (\S\ref{SubsecCrossMatch}). Using the cross-matched data sets, we will examine their SFE  (\S\ref{SubsecSFE}, \S\ref{SubsecRadialKSLaw}, and \S\ref{SubsecSFEff}).
\subsection{Derivation of SFR for individual $\HII$ regions}
\label{SubsecSFRDerivation}
\subsubsection{Derivation of $\HAlpha$ luminosity}
The catalog of the $\HII$ region presented in \S\ref{SecDistribution} is used as a starting point for the derivation of the SFR.
As stated in \S\ref{SecDistribution}, the $\HAlpha$ luminosity for each $\HII$ region has been corrected for the contamination of [$\NII$] lines and foreground extinction using the correction factors presented by \citet{Meurer2006SINGG}.
We further corrected for the internal extinction of the $\HII$ regions by utilizing the image of the $\PaBeta$ line emission presented in \S\ref{SecDistribution}.
A median filter with a size of $\sim$\timeform{8''} was applied to the $\PaBeta$ image to remove the background emission and the residual of continuum subtraction.
Subsequently, the filtered image was smoothed to align the resolution with the $\HAlpha$ image, and the $\PaBeta$ flux value was calculated by using the spatial mask defined by HIIphot \citep{Thilker2000}.
\par
The extinction curve of \citet{Cardelli1989} and the intrinsic ratio of $I_{\HAlpha} / I_{\PaBeta}$ = 17.1 \citep{DopitaSutherland2003} were assumed in deriving $A_{\HAlpha}$ from the measured $\PaBeta$ and $\HAlpha$ flux for each $\HII$ region.
For the $\HII$ regions that are within the FOV of the $\PaBeta$, the median and MAD of the derived A$_{\HAlpha}$ are found to be 1.9 mag
and 1.1 mag, respectively.
For the $\HII$ regions that are outside the FOV of the $\PaBeta$ image,
we assign A$_{\HAlpha}$=1.9 mag to them with an error bar of 1.1 mag.
\subsubsection{Derivation of SFR}
\label{SubsecHaToSFR}
To derive SFR from the $\HAlpha$ luminosity, we use the SFR calibration provided
by \citet{Calzetti2007}:
\begin{equation}
\label{EqSFRCalibration}
SFR\ \left(\MsunPerYear\right) = 5.3 \times 10^{-42}\ L(\HAlpha)\ \left(erg\ s^{-1}\right).
\end{equation}
We note that this calibration factor is calculated using the Starburst99 model \citep{Leitherer1999} by assuming a constant SFR with the Starburst99 default stellar initial mass function (IMF), which resembles the IMF of \citet{Kroupa2001}.
The assumption of a constant SFR requires steady state balance between the number of ionizing stars that are forming and dying: at least $\sim$6 Myr of continuous star formation at a constant rate is needed to achieve this, and thus it might be unrealistic to expect this assumption to hold for individual $\HII$ regions.
In fact, by modeling the observed $\HAlpha$ and UV luminosities for $\HII$ regions in NGC 300, \citet{Faesi2014NGC300APEX} derived calibration factors that produce on average a factor-of-2-higher SFR than the \citet{Calzetti2007} calibration does.
As we have not constrained the star formation history of individual $\HII$ regions yet, we just adopt the conventional extra-galactic calibration of \citet{Calzetti2007} here but note that the SFR derived here could be underestimated by approximated a factor of 2.
\subsection{Cross-matching}
\label{SubsecCrossMatch}
\begin{figure} []
\begin{center}
\FigureFile(88mm,88mm){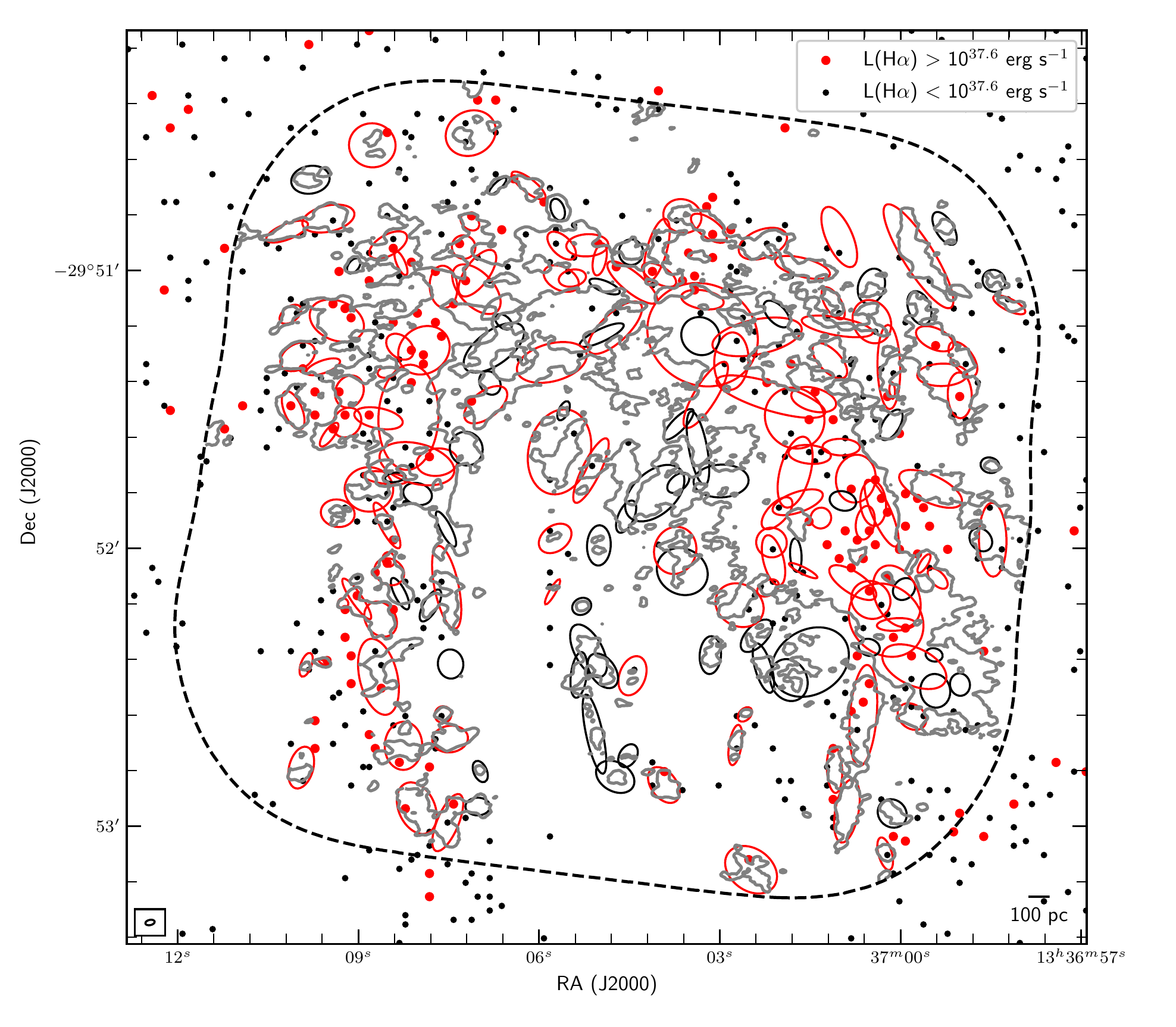}
\caption {Distribution of $\HII$ regions shown together with the distribution of GMCs overlaid on the integrated CO image with a contour line at 13 $\KkmPerS$. Black and red ellipses indicate the position of the GMCs without and with associated $\HII$ region(s). The size of the ellipses indicates effective major and minor radii for the GMCs, which are the same as the ones shown in figure \ref{FigCloudDistribution1}.
}
\label{FigCloudHIIAssociation}
\end{center}
\end{figure}
To identify host GMCs for $\HII$ regions, we cross-match the $\HII$ region catalog with the GMC samples presented in \S\ref{SectionGMCProp}.
As $\HII$ regions can drift away from host GMCs and can also expel the surrounding molecular material by stellar feedback, they do not necessarily reside within the densest part of host GMCs.
To account for this effect, we adopted ellipses that define effective major and minor radii for GMCs as their boundaries, which are the same as the ones shown in figure \ref{FigCloudDistribution1}.
If the central position of an $\HII$ region is within the boundary of a GMC, then the $\HII$ region is judged to be associated with the GMC.
Figure \ref{FigCloudHIIAssociation} shows the spatial relationship between the $\HII$ regions and the GMCs.
\par
Out of a total of 179 GMCs, 122 GMCs are found to be associated with one or more HII $\HII$ regions.
Sixty-six GMCs are associated with a single $\HII$ region each, and 56 GMCs are associated with more than one $\HII$ region each.
If an $\HII$ region has a one-to-one relationship with a GMC, then the SFR for the $\HII$ region is just directly registered to the GMC.
For an $\HII$ region that is associated with multiple GMCs, the SFR of the HII region is divided among the associated GMCs, weighted by the cloud mass.
\par
There also exist 57 GMCs that lack associated HII regions.
For these GMCs without HII regions, we set an upper limit on the SFR.
From the distribution of the uncorrected $\HAlpha$ luminosity, we determined the completeness limit of $\HII$ region detection to be approximately 10$^{37.1}$ erg s$^{-1}$.
By adopting the median value for the internal extinction, $A_{\HAlpha}$=1.9, and using equation (\ref{EqSFRCalibration}), the completeness limit for the detection of $\HII$ regions is translated into the limiting SFR of 3.5 $\times$ 10$^{-4}$ $\MsunPerYear$.
We just take this SFR value as an upper limit for the SFR of GMCs that lack associated $\HII$ regions.
The assumption involved here is that even some $\HII$ regions with an $\HAlpha$ luminosity comparable to 10$^{37.1}$ erg s$^{-1}$ could fail the HIIphot identification and not find a place in the $\HII$ region catalog used here, such bright but undetected $\HII$ regions are rare, and thus a GMC is associated with at most one of those undetected $\HII$ regions.
We believe that this is not an overoptimistic assumption because over half of the $\HII$ regions associated with GMCs have a one-to-one relation and as most of the GMCs that lack associated $\HII$ regions reside in inter-arm regions, where $\HII$ regions are sparse (figure \ref{FigCloudHIIAssociation}).
We also note that this upper limit is rather conservative, because approximately one-third of the HII regions have uncorrected $\HAlpha$ luminosity under the adopted completeness limit of 10$^{37.1}$ erg s$^{-1}$, down to $\sim$10$^{36.0}$ erg s$^{-1}$.
\par
The upper limit on the SFR adopted here also satisfies the requirement that our method for the SFR derivation requires the presence of at least one massive star so that $\HII$ region can be identified.
The mean lifetime of ionizing stars averaged over the stellar IMF is $\sim$4 Myr.
If stars form with a constant rate of 3.5 $\times$ 10$^{-4}$ $\MsunPerYear$, which is the adopted upper limit on the SFR, for 4 Myr, $\sim$1400 $\Msun$ of stellar mass is produced.
On the other hand, $\sim$1000 $\Msun$ of stellar mass is enough to sample the stellar IMF in the sense that at least one O star is produced (\cite{Calzetti2010}; \cite{Koda2012}).

\begin{figure*} []
\begin{center}
\FigureFile(158mm,158mm){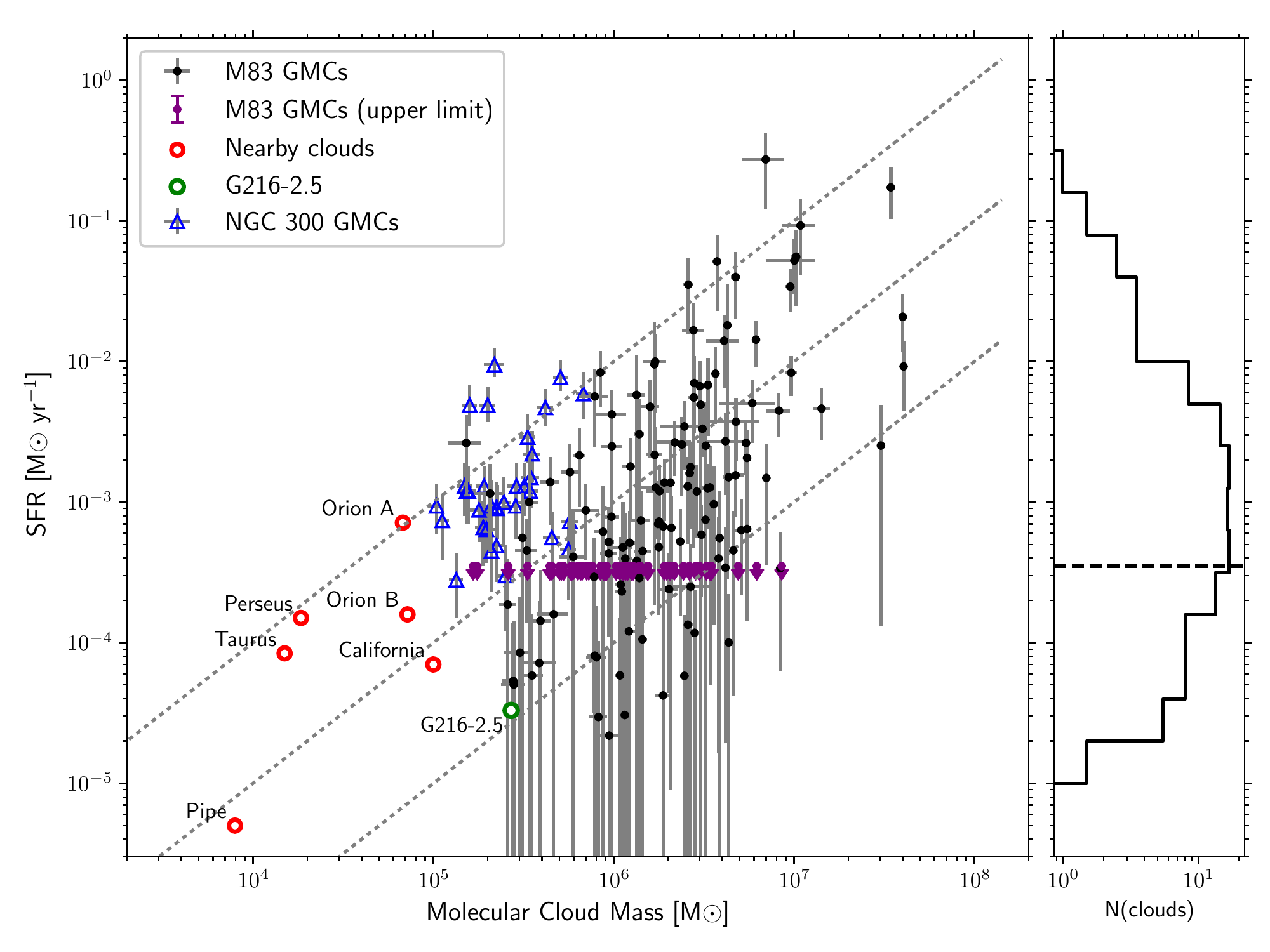}
\caption {
(Left) Star formation rate (SFR) to cloud mass relation for the GMCs in M83.
Black dots with error bars represent the GMCs with an associated $\HII$ region.
The GMCs without an associated $\HII$ region are indicated with the upper limit on SFR (purple).
Red open circles indicates nearby star-forming regions taken from \citet{Lada2010}.
The green circle indicate the data for G216-2.5.
Blue triangles indicate $\HII$ regions in NGC 300 that are detected in CO with $\sim$250 pc resolution (\cite{Faesi2014NGC300APEX}).
Dotted lines indicate lines of constant star formation efficiency (SFE) for 0.1, 1, and 10 Gyr$^{-1}$, respectively.
(Right) Histogram of the SFR assigned to individual GMCs in M83. The dashed horizontal line indicates the limit on SFR assigned to GMCs that are not associated with HII regions.
}
\label{FigSFRMass}
\end{center}
\end{figure*}

\subsection{GMC scale view of SFE}
\label{SubsecSFE}
Figure \ref{FigSFRMass} shows the relationship between the SFR and the molecular gas mass for the GMCs in M83. GMCs that lack associated $\HII$ regions are indicated with an upper limit on the SFR.
\par
To compare with the GMCs in M83, we incorporated some literature values into the plot.
\citet{Lada2010} estimated the SFR and the gas mass for nearby star-forming clouds that are located closer than 500 pc from the Sun, based on the number of young stellar objects (YSOs) and the $K$-band extinction, respectively.
Cloud samples in \citet{Lada2010} with mass greater than 5 $\times$ 10$^3$ $\Msun$ are included in the plot.
In addition, we have included G216-2.5, a Galactic cloud considered as an archetype for a low-activity cloud (also known as Maddalena's cloud; \cite{Maddalena1985}).
\citet{Lee1996G21625} and \citet{Megeath2009} detected a few YSOs around and within this cloud, respectively. On the basis of their YSO counts, \citet{Imara2015G21625} estimated the SFR of this cloud to be $\sim$33 $\Msun$ Myr$^{-1}$.
The $^{12}$CO luminosity of G216-2.5 is 6.1 $\times$ 10$^4$ K km s$^{-1}$ \citep{Lee1994G21625}.
Thus, using equation (\ref{EqLCOtoMsun}), we estimated the cloud mass to be 2.7 $\times$ 10$^{5}$ $\Msun$.
We note that there exist several other estimations for this cloud. \citet{Imara2015G21625} estimated the mass to be $1.2$ $\times$ 10$^5$ $\Msun$ on the basis of the far-infrared luminosity, which is close to another mass estimate made by \citet{Lee1994G21625} on the basis of the $^{13}$CO luminosity with an assumption of local thermodynamic equilibrium ($1.1$ $\times$ 10$^5$ $\Msun$).
In addition to the Galactic clouds, $\HII$ regions in NGC 300, which are detected in CO with 250 pc resolution (\cite{Faesi2014NGC300APEX}), are also included in the plot.
\par
A comparison of the SFR between the the GMCs in M83 and the Galactic samples quoted here has to be treated with care.
The SFR for GMCs in M83 is derived by tracing the $\HAlpha$ emission from $\HII$ regions, which are almost solely powered by massive stars.
As the nominal lifetime of massive stars is $\sim$4 Myr, the SFR derived from the $\HAlpha$ emission is a time-averaged value with a window size of $\sim$4 Myr.
On the other hand, the SFR for Galactic GMCs is derived with a shorter window size because it is based on the number counts of YSOs, which formed within the last 2$\pm$ 1 Myr (e.g., \cite{Evans2009c2d}).
The YSO counting method has a further advantage that it samples much smaller stars compared to the $\HAlpha$ method.
Thus it is less affected by stochastic samplings of the stellar IMF.
From the comparison of different SFR calibration methods applied to Galactic clouds, \citet{ChomiukPovich2011} concluded that the SFR derived with YSO counting methods are approximately a factor of 2 to 3 higher than either of the SFR derived from the 24$\mu$m flux or the Lyman continuum production rate.
\par
Despite the uncertainty about the relative consistency between the different SFR calibration methods, the GMCs in M83 share similar characteristics as the Galactic clouds in regard to SFE, defined as the ratio between the SFR and the cloud mass.
The data points for the GMCs in M83 fall in the SFE range between 0.1 and 10 Gyr$^{-1}$, indicating that the peak-to-peak SFE variation is approximately two orders of magnitudes.
This scatter is comparable to the values found in Galactic GMCs (\cite{Mooney1988}; \cite{Murray2011SFE}; \cite{Lee2016DynamicSF}).
Specifically, the median value of the SFE is $\sim$0.5 Gyr$^{-1}$ for the M83 GMC samples, and the scatter is 0.52 dex in MAD.
We note that the range of SFEs observed in the M83 GMCs is comparable to that found in the Galactic cloud samples plotted in figure \ref{FigSFRMass}, which places G216-2.5 and Orion A at the both ends.
\subsection{Radially averaged SFR to mass relation for GMCs}
\label{SubsecRadialKSLaw}
\begin{figure*} []
\begin{center}
\FigureFile(142mm,142mm){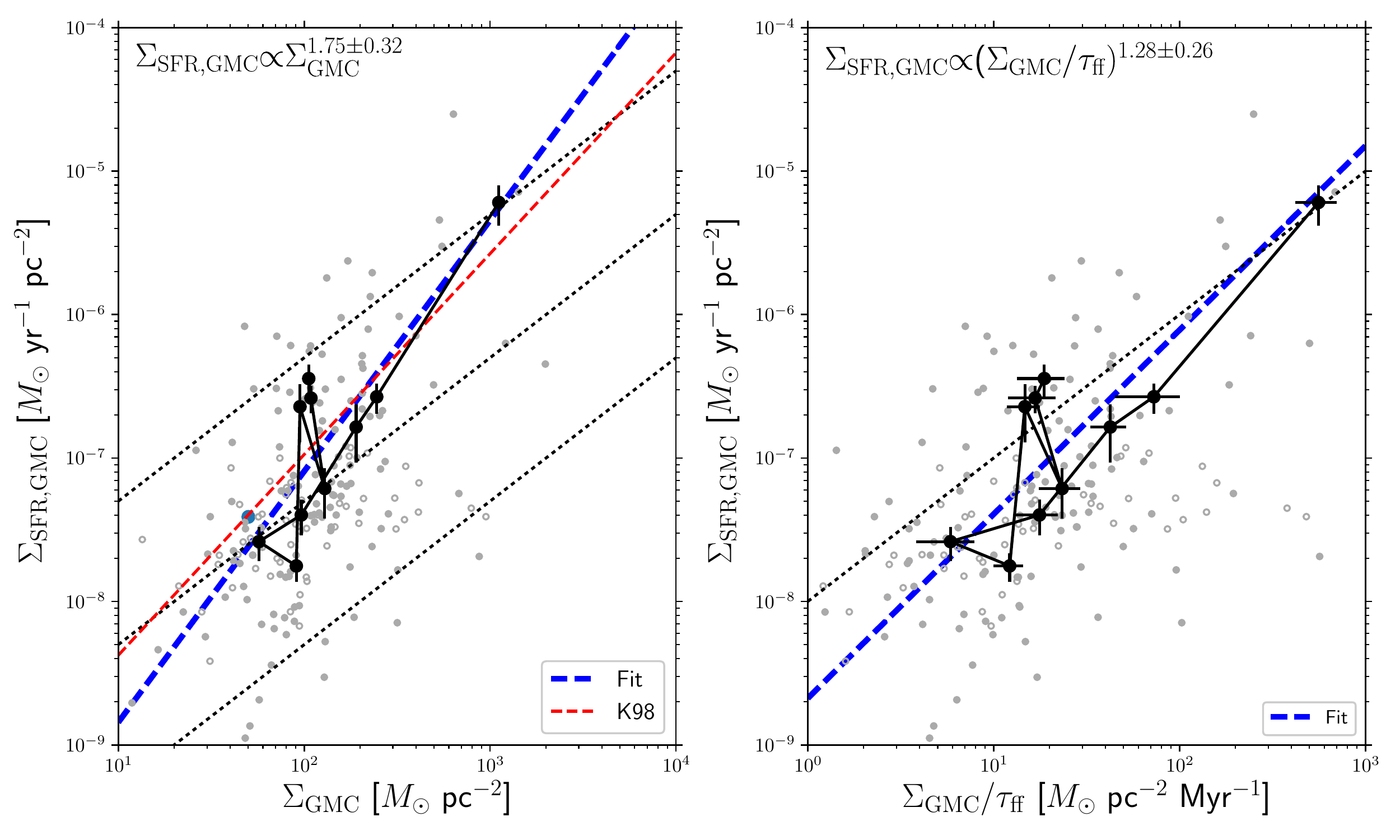}
\caption{(Left)
Relation between SFR per unit area and gas surface density for GMCs.
Gray filled and open circles indicate the data for individual GMCs with and without associated $\HII$ regions, respectively.
Black circles with error bars indicate the radially averaged data, calculated with radial bins specified with galactocentric radius ranges of 0--12, 12--24, 24--36, ..., and 108--120 arcseconds, respectively.
Neighboring radial data points are connected to each other with solid lines.
In each plot, the data point with the highest surface mass density indicates the innermost radial bin.
The blue heavy dashed line indicates the fitted function.
The red dashed line indicates the relation given by \citet{Kennicutt1998SchmidtLaw} but scaled by a factor of 0.67 to account for the difference of the assumed stellar IMF.
Dotted lines are for constant depletion times of 0.2, 2, and 20 Gyr, respectively.
(Right)
Same as the left one, but for SFR per unit area and gas surface density divided by the free-fall time for GMCs.
The dotted line is for a constant $\SFEff$ of 0.01.
}
\label {FigRadialSFLaws}
\end{center}
\end{figure*}

In this subsection, we inspect the relation between the surface density of the SFR and the GMC mass on a radial average basis.
First, the SFR surface density per unit area, $\SigmaSFR$, of each GMC is derived by dividing the SFR by the effective area of GMC, $\pi$ R$_{\mathrm{eff}}^2$.
Next, radial bins with a width of \timeform{12''} ($\sim$264 pc) are generated, and the radially averaged values of $\SigmaSFR$ and $\SigmaGMC$ are derived.
Figure \ref{FigRadialSFLaws} (left) plots the relation between $\SigmaSFR$ and $\SigmaGMC$.
The solid line in the plot indicates the radially averaged values of $\SigmaSFR$ and $\SigmaGMC$.
In calculating the radial average, the SFR for the GMCs without associated $\HII$ regions are calculated with upper limits on the SFR.
Therefore, the radially averaged values in this plot have to be treated as upper limits.
\par
A least-square fitting is made to the radially averaged data points by minimizing the effective variance \citep{Orear1982}.
The fitted slope is 1.71 $\pm$ 0.31, which is well above unity.
This super-linear relation does not contradict the well-quoted nonlinear index of 1.4 $\pm$ 0.15 \citep{Kennicutt1998SchmidtLaw} obtained by fitting data of nearby galaxies.
For reference, the relation of \citet{Kennicutt1998SchmidtLaw} is overplotted with a modification of the coefficient to adjust for the difference in the assumed stellar IMF.
The radially averaged data points are within a factor of three variations from the relation of \citet{Kennicutt1998SchmidtLaw} and appear to be in reasonable agreement with it.
\par
Next, we divide the quantity on the abscissa, $\SigmaGMC$, by the free-fall time of the GMC.
If the SFR and SFE are characterized by a constant or a nearly constant $\SFEff$, then the observed data points should exhibit a linear relation.
Figure \ref{FigRadialSFLaws} (right) shows the relation between the radially averaged $\SigmaSFR$ and $\SigmaGMC/\TauFF$.
We also fit a proportional relationship, $\SigmaSFR$ $\propto$ $\SigmaGMC/\TauFF$, to the radial data points to estimate a proportionality coefficient, which is $\SFEff$.
The resultant value for $\SFEff$ is 5.8$^{+1.8}_{-1.4}$ $\times$ 10$^{-3}$ and is not far from the well quoted value of 0.01 (e.g., \cite{Krumholz2012UniversalSF}).
\par
The results obtained with radially averaged data points are not so sensitive to the way $\HII$ regions are associated with GMCs (\S\ref{SubsecCrossMatch}).
If the threshold radius for associating $\HII$ regions with GMCs is changed, it is possible that an $\HII$ region associated with a GMC with a particular threshold will be assigned to other neighboring GMC with a different threshold.
However, as $\SigmaSFR$ and $\SigmaGMC$ are averaged within radial bins, the shift of SFR from a GMC to another GMC is not expected to make a severe impact.
Indeed, if the association radius is increased and decreased by 20\% (by 44\% in terms of the area), the index changes are 1.63 $\pm$ 0.30 and 1.86 $\pm$ 0.32 for the $\SigmaSFR$-$\SigmaGMC$ relation and 1.33 $\pm$ 0.26 and 1.20 $\pm$ 0.25 for the $\SigmaSFR$-$\SigmaGMC$/$\TauFF$: these values are consistent with each other.
\par
Another source of uncertainty is the different detection limits of GMCs in different environments (\S\ref{SubsecMassSpectrum}).
In figure \ref{FigRadialSFLaws} (left), the single data point with the highest $\SigmaGMC$ has a strong influence in determining the index of the fitted function, and this also applies for figure \ref{FigRadialSFLaws} (right).
The area of the radial bin for that point overlaps with the central region in which the minimum of the detected GMC masses is a factor of few higher than in uncrowded regions, such as the inter-arm regions.
Thus, one might suspect the lack of smaller clouds in the central region may have a significant impact on the fitted parameters.
However, GMC mass spectra investigated in \S\ref{SubsecMassSpectrum} exhibited slopes shallower than $\gamma$ $<$ -2.
In particular, the slope $\gamma$ in the central region is quite shallow with $\gamma$ $\sim$-1.2.
With the fitted parameters in the central region and using equation (\ref{EqMFSummation}), clouds with masses between 10$^{3}$ and 10$^{6}$ $\Msun$ are expected to comprise less than 5\% of the total cloud mass above 10$^{3}$ $\Msun$.
Therefore, it seems reasonable to suppose that the impact of the varying detection GMC limits is not significant.

\par
Thus far, as far as the averaged data points are concerned, or in other words, as seen on a global scale, the star-forming properties of the GMCs in M83 are in agreement with the model that star formation takes place with a nearly constant efficiency per free-fall time. However, from figure\ref{FigRadialSFLaws} (right), the GMC-to-GMC variation in $\SFEff$ appears to be large.
We will investigate the environmental dependence of $\SFEff$ in the next subsection (\S\ref{SubsecSFEff}) and will discuss it in \S\ref{SubsecConstantSFEffOrNot} and \S\ref{SubsecImplications}.

\subsection{$\SFEff$ for individual GMCs}
\label{SubsecSFEff}
\begin{figure*} []
\begin{center}
\FigureFile(128mm,128mm){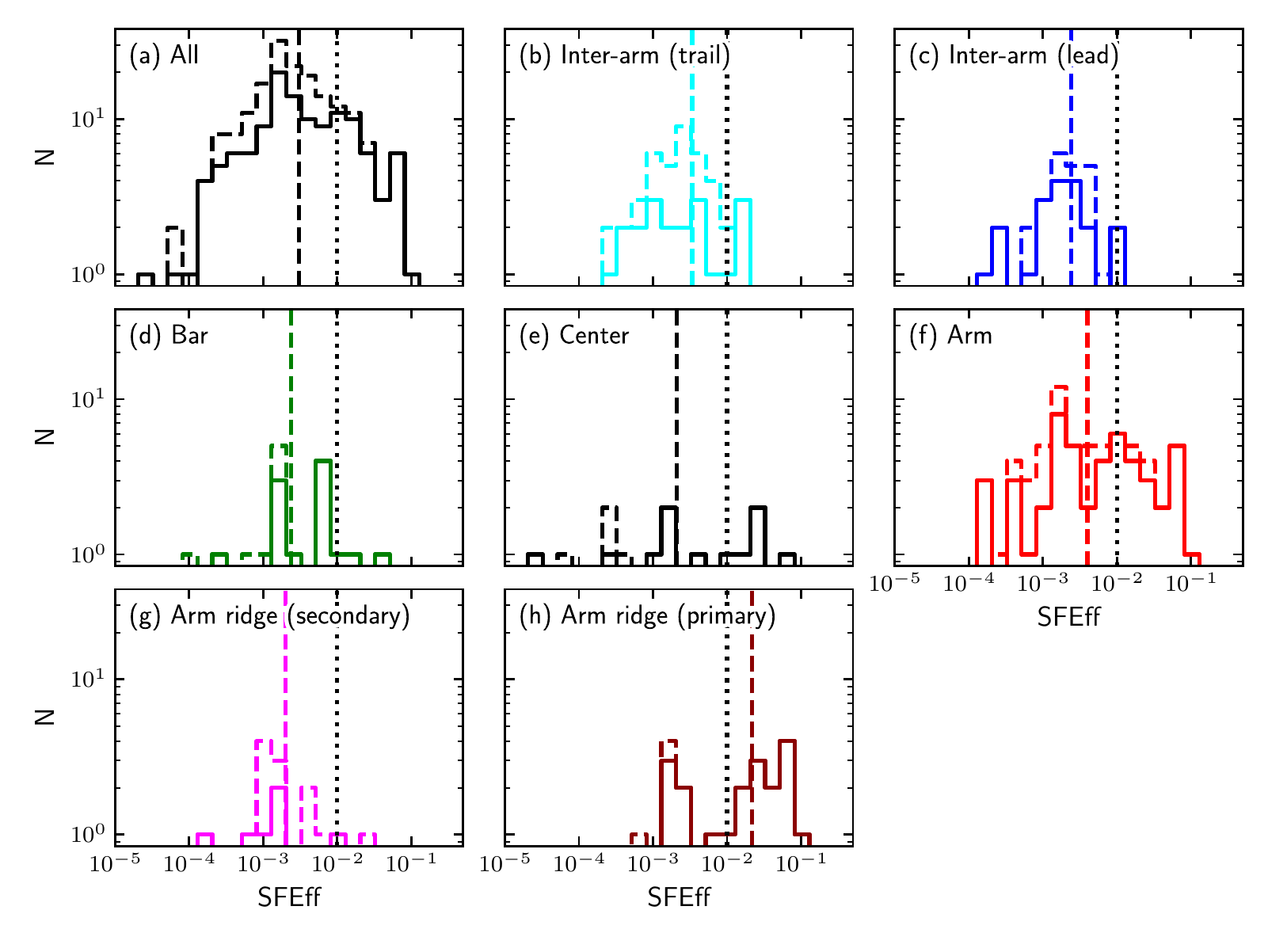}
\caption {
Histogram of $\SFEff$ plotted for each region.
In each plot, the solid histogram shows the number distribution of $\SFEff$ for the GMCs that have associated $\HII$ regions.
The dashed histogram in each plot includes both GMCs with and without associated $\HII$ regions.
For the GMCs without associated $\HII$ regions, upper limits on $\SFEff$ are used to calculate the dashed histogram.
The dotted vertical line indicates an $\SFEff$ of 0.01, which is the mass-weighted mean value for all of the GMCs in M83.
The dashed vertical line in each plot indicates the median value of $\SFEff$ for the samples contained in the plot.
Note that only in the primary arm ridge is the median $\SFEff$ higher than the global mass-weighted average value of 0.01.
}
\label{FigSFEffHistogram}
\end{center}
\end{figure*}
Now we proceed to examine $\SFEff$ for individual GMCs.
We derive $\SFEff$ for each GMC as
\begin{equation}
\mathrm{SFE}_{\mathrm{ff}} = \frac{\mathrm{SFR}}{\Mmol} \TauFF.
\end{equation}
Quantities on the right hand side, SFR, $\Mmol$, and $\TauFF$, are taken from the
values determined for each GMC.
\par
We must note that accurate determination of $\SFEff$ for individual GMCs is far from trivial because the stochastic sampling effect of the stellar IMF and the uncertainty of the age introduce errors when deriving the SFR.
Further, as the SFRs used here are derived from $\HAlpha$ luminosities, they are essentially time-averaged values with a window size of approximately 3--4 Myr
because the production rate of the Lyman continuum from a stellar cluster quickly decays after the deaths of massive stars (\cite{Murray2011SFE}).
This duration is close to the time scale for an individual GMC: the free-fall time is approximately 6--10 Myr for GMCs.
\citet{Feldmann2011} pointed out that observationally derived $\SFEff$ values for individual clouds have to be treated as 'apparent' values, because if a significant fraction of the gas mass is lost during the averaging time imposed by the SFR tracer, the observed value of $\SFEff$ should be biased upward.
Despite these uncertainties, we first just review the observed distribution of $\SFEff$ in this subsection.
Later in the next section, we will discuss the effects that can contribute to increasing the apparent variation in $\SFEff$.
\par
Figure \ref{FigSFEffHistogram} shows histograms of the derived $\SFEff$ for the GMCs for each regional mask.
In table \ref{TableSFEff}, we list the mass-weighted mean ($\langle{\SFEff}\rangle$), median, and MAD of $\SFEff$ for each regional mask.
\begin{table}[htbp]
\tbl{Mean, median, and MAD of $\SFEff$}{
\begin{tabular}{llll}
\hline
\hline
& $\langle{\SFEff}\rangle$ & Median & MAD \\
& $\left(\%\right)$ & $\left(\%\right)$ & (dex)\\
\hline
all & 0.93 & 0.30 & 0.60 \\
\hline
Inter-arm (up) & 0.44 & 0.34 & 0.47 \\
Inter-arm (down) & 0.26 & 0.24 & 0.41 \\
\hline
Arm & 1.48 & 0.40 & 0.79 \\
... Arm ridge (secondary) & 0.36 & 0.20 & 0.46 \\
... Arm ridge (primary) & 2.68 & 2.17 & 0.81 \\
\hline
Bar & 0.92 & 0.24 & 0.78 \\
\hline
Center & 1.00 & 0.21 & 1.16 \\
\hline
\end{tabular}
}
\label{TableSFEff}
\end{table}
The mass-weighted mean of $\SFEff$, $\langle\SFEff\rangle$, for all GMC samples is $\sim$9.3 $\times$ 10$^{-3}$.
This value is close to the value of $\sim$0.01 assumed in many theories based on turbulence-regulated star formation (e.g., \cite{Krumholz2005TurbulenceSF}; \cite{Krumholz2012UniversalSF}).
By changing the association radius by $\pm$20\% as was done in the previous subsection, the mass-weighted mean changes to 0.79--1.06 $\times$ 10$^{-2}$ and is still in agreement with $\sim$0.01.
The agreement of the mass-weighted mean of $\SFEff$ with 0.01 appears to reconfirm the fact the models of turbulence regulated star formation have succeeded in describing the global averaged property of star formation.
\par
Although $\langle\SFEff\rangle$ shows good agreement with the theoretical expectation, we also see a huge scatter in the distribution of $\SFEff$ if all samples are taken together (figure \ref{FigSFEffHistogram}a).
The observed peak-to-peak variation in $\SFEff$ is approximately three orders of magnitude and is consistent with other studies that also reported huge cloud-to-cloud variations of $\SFEff$ (\cite{Murray2011SFE}; \cite{Lee2016DynamicSF}).
As foreseeable from the large scatter, the median value of $\SFEff$ is lower $\langle\SFEff\rangle$, and is 3.0 $\times$ 10$^{-3}$.
We note that \citet{Leroy2017M51SFEff} reported similar value of 3--3.6 $\times$ 10$^{-3}$ in M51 using infrared emission as SFR tracer at resolutions of $\sim$370 pc and $\sim$1100 pc.
\par
In addition, the variation in $\SFEff$ exhibits a strong regional dependence (figure \ref{FigSFEffHistogram}b-h).
For almost all of the clouds in the inter-arm regions, $\SFEff$ is below the global average of 0.01.
Most GMCs in the bar also show $\SFEff$ values below 0.01, although there is one GMC that exceeds $\SFEff$=0.01.
The center and the arm contain several GMCs that exceed $\SFEff$=0.01.
Within the arm, however, the two arm ridges exhibit a clear difference from each other concerning the distribution of $\SFEff$: GMCs with $\SFEff$ values higher than 0.01 are concentrated in the primary arm ridge, while in the secondary arm ridge, almost all of the GMCs are below $\SFEff$=0.01.
\section{Discussion}
\label{SecDiscussion}
\subsection{Is there intrinsic environmental variation of the SFE?}
\label{SubsecConstantSFEffOrNot}
In the previous subsection, we find that there are approximately three orders of magnitude peak-to-peak spread in the values of observationally derived $\SFEff$.
Not only is the scatter of the distribution large, but there are also regional variations in $\SFEff$.
In particular, the two ridges within the arm exhibited a clear difference from each other concerning the distribution of $\SFEff$.
Before discussing the meaning of the regional variation in $\SFEff$, we first discuss possibilities that the large scatter and regional variations in $\SFEff$ are actually caused by observational effects that produce 'apparent' variations in $\SFEff$ even if $\SFEff$ is intrinsically a constant parameter.
We consider three mechanisms here: (1) stochastic sampling of the stellar IMF, (2) uncertainty of stellar ages that affects the SFR calibration factor, and (3) mass consumption of GMCs due to stellar feedback.
\par
\begin{figure*} []
\begin{center}
\FigureFile(158mm,158mm){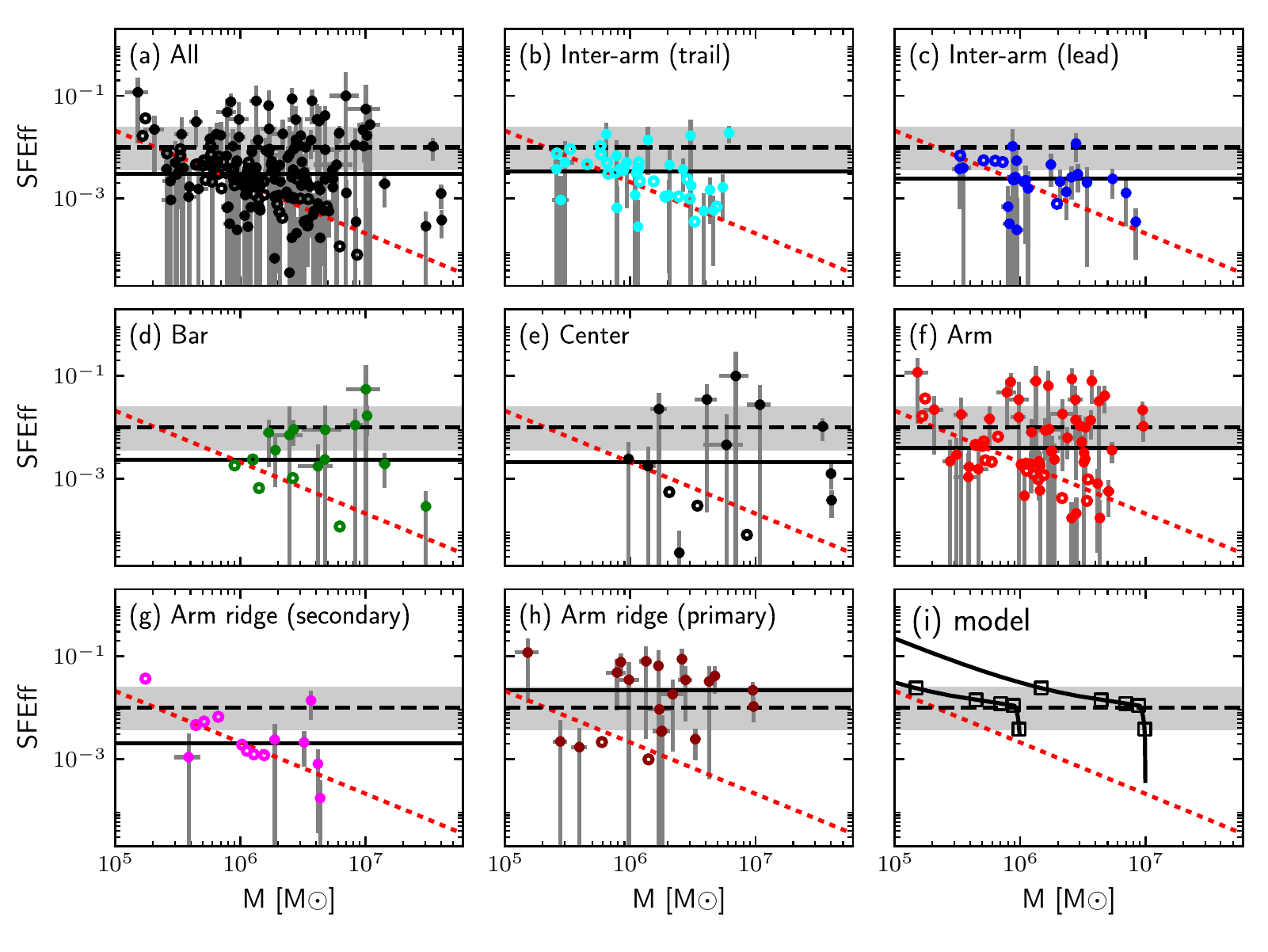}
\caption{(a-h)
Observationally derived 'apparent' value of $\SFEff$ as a function of cloud mass for the GMCs plotted for each regional mask.
The dotted red line in each plot indicates the nominal limit for the detection, obtained by assuming a constant SFR of 3.5 $\times$ 10$^{-4}$ $\MsunPerYear$ and a free-fall time of $\TauFF$ = 6.7 Myr.
(i)
Curves that show the evolution of $\SFEff$ for model GMCs calculated with the toy model of \cite{Feldmann2011}.
The left and right curves indicate the evolution of model GMCs with initial masses of 10$^6$ and 10$^7$ $\Msun$, respectively.
The intrinsic value efficiency, $\SFEffi$, is fixed as 0.01.
The feedback parameter is chosen such that the stellar feedback limits the lifetime of model clouds to 20 Myr.
Squares indicate the time steps for 10\$, 30\$, 50\$, 70\$, and 90\$ of the lifetime of model GMCs.
Gray shaded area in each plot indicates the range of $\SFEff$ covered by model GMCs within 10\% and 90\% of their lifetimes.
}
\label{FigSFEfftoM}
\end{center}
\end{figure*}
\subsubsection{Sampling effect of the stellar IMF}
If a cluster that illuminates an $\HII$ region contains a small number of stars, ionizing stars are likely less populated due to the stochastic sampling of the stellar IMF.
This sampling effect will lead to an underestimation of the SFR for small $\HII$ regions.
One might suspect this sampling effect could be responsible for the apparent variation in $\SFEff$, especially for the GMCs in inter-arm regions that exhibited an $\SFEff$ value that is systematically lower than the global mean of 0.01.
However, the variation in $\SFEff$ seen in the previous subsection was revealed with a conservative upper limit on the SFR of 3.5 $\times$ 10$^{-4}$ $\MsunPerYear$.
If this level of the SFR continues for 3--4 Myr, which is the nominal lifetime of ionizing stars, approximately 1100--1400 $\Msun$ of stellar mass is produced.
On the other hand, $\sim$1000 $\Msun$ of stellar mass is enough to avoid the severe effect of stochasticity of the stellar IMF sampling: for example, \citet{Calzetti2010} examined the $\HAlpha$ luminosity per stellar mass for clusters and indicated that the cluster-to-cluster variation is a factor of 2 for $\sim$1000 $\Msun$ clusters.
Thus, as long as the sampling effect of the stellar IMF alone is taken into account, the possible amount of the scatter for the apparent value of $\SFEff$ is at most a factor of 2 and is insufficient to explain the observed variation in $\SFEff$.
\subsubsection{Uncertainty of stellar ages}
As stated in \S\ref{SubsecHaToSFR}, the uncertainty of the star formation history for individual $\HII$ regions should make the SFR calibration factor, which relates the SFR and $\HAlpha$ luminosity, somewhat uncertain.
We consider here whether the observed regional variation in $\SFEff$ could be explained by variations in the SFR calibration factor.
\par
If stars in a cluster powering an $\HII$ region are formed instantaneously, the production rate of Lyman photons from the cluster decay quickly after 3--4 Myr because of the deaths of massive stars (e.g., Fig. 22 of \cite{Scoville2001M51}; Fig. 1 of \cite{Murray2011SFE}).
The strong time evolution of the Lyman continuum production rate implies there should be a time evolution of the SFR calibration factor.
To see the degree of the time variation of the SFR calibration factor, we here perform a simple calculation using Starburst99.
We assume the stellar IMF of \citet{Kroupa2001}, solar metallicity, and the burst mode of star formation, which produces all the stars in a cluster at the same time.
From the calculated results, we find that the $\HAlpha$ luminosity per stellar mass is almost constant for the first 2 Myr, and then starts to decline due to the deaths of the ionizing stars, being about 50\%, 20\%, and 10\% of the initial value at approximately 3.5, 4.5, and 5 Myr, respectively.
\par
The average value of $\SFEff$ in the arm region ($\sim$0.015) is approximately a factor of 3--5 larger than in the inter-arm regions ($\sim$0.0046 and $\sim$0.0027).
As far as the burst mode of star formation in clusters is concerned, the only way to produce this amount of variation in $\SFEff$ by just changing the SFR calibration factor is to assume that the $\HII$ regions in the inter-arm are preferentially older than those in the arm regions by a few Myr.
Referring to the values obtained with the Starburst99 calculation, if the $\HII$ regions in the arm region are younger than 2 Myr and the ones in the inter-arm regions are older than 4 Myr, it is possible to produce a factor of 3--5 variation in the apparent distribution of $\SFEff$.
However, considering the global galactic dynamics, the dwelling time of GMC in the inter-arm region should be a few to several tens of Myr for the galactocentric radii considered here (figure 17 in \cite{Hirota2014}).
The dwelling time is much longer than the time for the $\HAlpha$ luminosity of an $\HII$ region to decay.
Thus, the existence of $\HII$ regions in the inter-arm regions means that at least some of them are formed in situ.
Therefore, it is difficult to consider a process that produces just a few Myr difference in the age distributions of $\HII$ regions across the spiral arm and it is unlikely that the observed variation in $\SFEff$ is simply due to the uncertainty of the star formation history.
\par
The conclusion arrived at here will not change even if a cluster was created through several events of star formation that occurred over a few Myr (e.g., \cite{Venuti2018}).
If a cluster that powers an $\HII$ region was formed in a single instantaneous event, then the lifetime of the $\HII$ region is approximately 5 Myr, as we have seen above.
If a cluster was formed through several events spread over a few Myr, then the lifetime of the $\HII$ region is lengthened by a few Myr.
However, even if the extra few Myr has to be added to 5 Myr, the lifetime of $\HII$ regions is still likely below 10 Myr and well below the dwelling time in the inter-arm region.
\subsubsection{Rapid mass consumption of GMCs within the averaging time for the derivation of SFR}
The $\HAlpha$ emission from an $\HII$ region that is used to trace its SFR quickly decays after approximately 4 Myr, because of the deaths of massive stars.
Therefore, the SFR derived from the $\HAlpha$ luminosity of an $\HII$ region is essentially a time-averaged value with an averaging window length of approximately 4 Myr (e.g., \cite{Murray2011SFE}).
On the other hand, once massive stars are formed inside a GMC, they are considered to be capable of ionizing and disrupting a significant portion of the parental GMC
within a timescale of a few tens of Myr (e.g., \cite{Whitworth1979CloudDisruption}; \cite{WilliamsMcKee1997}), which is comparable with the estimated GMC lifetimes (15--40 Myr; \cite{Kawamura2009LMC}; \cite{Murray2011SFE}; \cite{Miura2012M33}).
The suggested timescale for cloud destruction is only a few times longer than the averaging window length of $\sim$4 Myr for the derivation of the SFR.
Therefore, a GMC associated with the $\HII$ regions could have lost a non-negligible fraction of its mass due to the stellar feedback within the averaging duration.
If this is the case, then the $\SFEff$ derived from the observed SFR and cloud mass should be biased upward.
\paragraph{Cloud mass evolution model}
\par
To investigate the impact of this time evolution effect, we employ a toy model introduced by \citet{Feldmann2011}, which solves a differential equation of mass evolution for a GMC:
\begin{equation}
\frac{d{\Mmol}}{dt}
= SFR(t) - \alpha_{\mathrm{fb}} M_{\mathrm{*}}(t),
\end{equation}
where $SFR(t)$ is the SFR for a GMC, $M_{\mathrm{*}}(t)$ is a the stellar mass associated with the GMC, and $\alpha_{\mathrm{fb}}$ is the feedback parameter.
SFR is further expressed as
\begin{equation}
SFR(t) = -{\SFEffi}\frac{{\Mmol(t)}}{\TauFF},
\end{equation}
where $\SFEffi$ is the {\it intrinsic} star formation efficiency per free-fall time.
We use the notation $\SFEffi$ to distinguish it from the observationally derived, $\SFEff$.
The stellar mass is obtained simply by integrating $SFR(t)$:
\begin{equation}
M_{\mathrm{*}}(t) = \int_{0}^{t'}|SFR(t)| dt'.
\end{equation}
Solving these equations, $\Mmol$ and $M_\mathrm{*}$ are described as a function of time.
\par
The observationally derived SFR just traces the stellar mass that is formed within a fixed time window.
Denoting the observed apparent value of the SFR as $\mathrm{SFR}_{a}$, it can be derived as
\begin{equation}
\mathrm{SFR}_{a}(t) = \frac{\int_{max(t - \tAVG,\ 0)}^{t} M_\mathrm{*}(t') dt'}{\tAVG},
\end{equation}
where $\tAVG$ is the time window traced by the SFR tracer.
As we use the $\HAlpha$ emission as a tracer of the SFR, we fix $\tAVG$ as 4 Myr.
If a cloud is younger than $\tAVG$, then $\mathrm{SFR}_{a}$ for the cloud underestimates the actual SFR because of the fixed averaging window used in the denominator.
Using the computed $\mathrm{SFR}_{a}(t)$, $\SFEff$ for a model cloud can be described as
\begin{equation}
\SFEffa(t) = \frac{\mathrm{SFR}_{a}(t)}{\Mmol(t)} \TauFF.
\end{equation}
\par
We adopt $\TauFF=6.7$ Myr and $\SFEffi=0.01$ as representative values for the GMCs in M83.
Although the actual value of $\alpha_{\mathrm{fb}}$ is not unknown, we can derive it by assuming the lifetime of GMC because with the preceding equation,
the lifetime of a model cloud is given as $\sim0.5\pi / \sqrt{\alpha_{\mathrm{fb}} \SFEffi / \TauFF}$ \citep{Feldmann2011}.
The estimated value of the GMC lifetime varies between 15 to 40 Myr (e.g., \cite{Kawamura2009LMC}; \cite{Murray2011SFE}; \cite{Miura2012M33}).
To check the maximum influence the feedback can exert, we here adopt the smallest value of 15 Myr: $\alpha_{\mathrm{fb}}$ is derived as 7.3 Myr$^{-1}$.
\par
Figure \ref{FigSFEfftoM} plots the observed $\SFEff$ as a function of the cloud mass for each region (a--h).
In addition, figure \ref{FigSFEfftoM}(i) shows the evolutionary tracks for model clouds with initial masses of 10$^6$ and 10$^7$ $\Msun$, respectively.
Along each of the two tracks, the squares indicate the time steps for 10\%, 30\%, 50\%, 70\%, and 90\% of the lifetime of the model GMC, from the bottom to the top.
\par
Initially, the model clouds move up almost straight on the plot as $\SFEffa$ increases, which is driven by the rapid growth of the observed SFR.
This rapid growth of the observed SFR is due to the underestimation of the observed SFR when the cloud's age is lower than $\tAVG$, as stated above.
Even after the initial rapid growth of $\SFEffa$, it continues to rise because the decrease in the cloud mass is always faster than the SFR decreases due to the averaging time imposed in the derivation of the SFR.
\par
With the adopted parameters, the range of the values for $\SFEffa$ is predicted to be within approximately 0.0038--0.024 for the 80\% of the GMC lifetime, which is approximately equivalent to a $\pm$0.4 dex variation.
The calculated range is indicated as the shaded area in figure \ref{FigSFEfftoM} to aid comparison with the actual $\SFEff$ of the GMCs in M83.
We note that this range is obtained by adopting the shortest estimate of the GMC lifetime available.
If a longer GMC lifetime is assumed, then the range covered by $\SFEffa$ becomes narrower.
For example, if the assumed lifetime is 30 Myr, the variation is approximately $\pm$0.2 dex.
\paragraph{Comparison with the observed distribution of $\SFEff$}
\par
Up to now, we have seen that the evolutionary effect of a cloud might cause a variation in the apparent efficiency, $\SFEffa$, even if the intrinsic efficiency is fixed as $\SFEffi=0.01$.
The predicted amount of the variation in $\SFEffa$ is up to approximately $\pm$0.4 dex over 80\% of the cloud lifetime, assuming the most intense role of the stellar feedback.
Therefore, if the observed variation in $\SFEff$ is over the predicted range of $\SFEffa$ obtained by assuming a fixed value of $\SFEffi$, then $\SFEffi$ is suggested to be nonuniform.
\par
Comparing the observed $\SFEff$ of GMCs with the predicted range of $\SFEffa$ in figure \ref{FigSFEfftoM}, we see that many GMCs deviate from the predicted $\pm$0.4 dex range of variations and also see signs of regional variations in the efficiency of star formation.
For example, in the inter-arm regions, most of the GMCs, especially GMCs more massive than 10$^6$ $\Msun$ are located below the lower end of the predicted range that corresponds to the 10\% of the model GMC lifetime.
Thus, if $\SFEffi$ is to be fixed as 0.01, then most of the GMCs in the inter-arm regions have to be in the very young stage of cloud evolution, younger than 10\% of the assumed cloud lifetime.
However, this appears implausible, because the dwelling time in the inter-arm regions observed here should be several tens of Myr, taking into consideration galactic dynamics (see figure 17 in \cite{Hirota2014}).
Therefore, the intrinsic efficiency, $\SFEffi$, should be significantly lower than the global average of 0.01 in the inter-arm regions.
\par
Another sign of regional variation in $\SFEff$ can be seen in the primary arm ridge, where a non-negligible number of GMCs exhibit an $\SFEff$ higher than the predicted range for it.
Applying the argument same as for the inter-arm regions, the upward deviation of the observed $\SFEff$ in the primary arm ridge suggests that GMCs form stars with an $\SFEffi$ that is higher than 0.01.
\par
Here, we have seen that the range of observed variation in $\SFEff$ is larger than its allowed range of variation with a fixed value of $\SFEffi$.
We also have seen regional variations in the distributions of $\SFEff$ which suggests variation in $\SFEffi$.
Therefore, we conclude that the observed variation in $\SFEff$ is not just an artifact caused by the rapid mass consumption of GMCs due to the stellar feedback.
\par

\subsection{Impact of stellar feedback in limiting the lifetime of GMCs}
\label{SectionFeedback}

\begin{figure} []
\begin{center}
\FigureFile(88mm,88mm){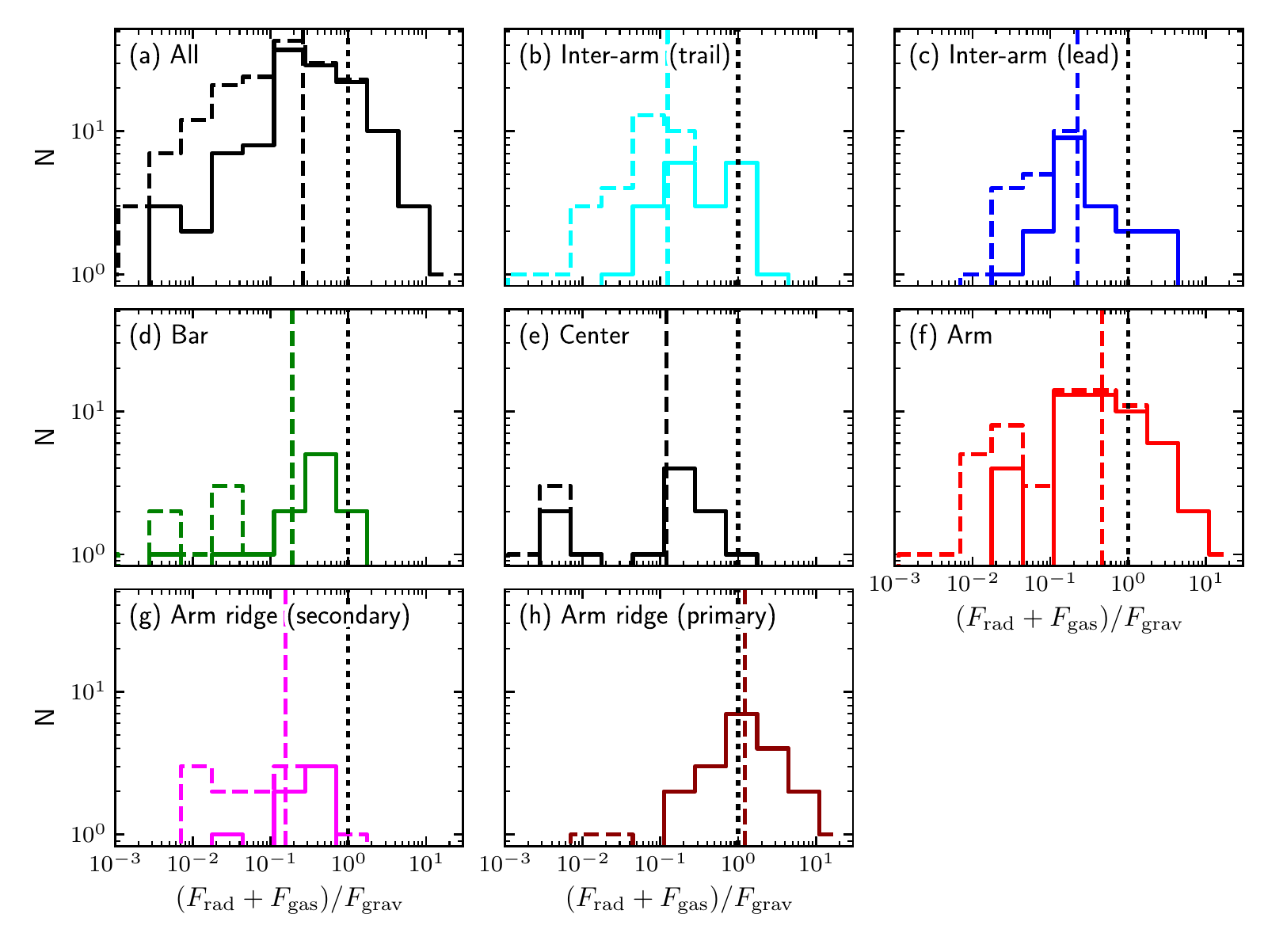}
\caption{
Histogram of the ratio of the outward force, which is a combination of radiation and gas pressure forces , to the self-gravity of GMC, $\Qfb$, for each region.
In each plot, the solid histogram shows the number distribution of the GMCs that have associated $\HII$ regions.
The dashed histogram in each plot includes both GMCs with and without associated $\HII$ regions.
Dotted and dashed vertical lines in each plot indicate $\Qfb$=1 and median value of $\Qfb$ for the samples.
}
\label {FigFradOverFgravHistogram}
\end{center}
\end{figure}
As we have seen in \S\ref{SubsecMassSpectrum}, the GMC mass function in the primary arm exhibited a steeper slope ($\sim-1.8$) compared to those in the other regions.
Disruption of GMCs due to the role of stellar feedback was argued as a possible mechanism for the steeper slope of the GMC function in the primary arm ridge.
To check the validity of the argument made in \S\ref{SubsecMassSpectrum}, we examine the impact of stellar feedback in this subsection.
\par
Stellar feedback takes place in various forms, such as momentum input by radiation pressure, gas pressure, stellar winds, and photo-ionization and photo-dissociation due to UV and FUV radiation.
At the scale of 10--100 pc which is relevant to GMCs, the radiation pressure and gas pressure associated with warm ionized gas are considered to play dominant roles in removing gas material from host GMC (e.g., \cite{Matzner2002}; \cite{Lopez2011}; \cite{Murray2011SFE}).
Here, we compare the outward force exerted by radiation pressure ($\Frad$) and ionized gas pressure ($\Fgas$) against the inward force of gravity for each GMC to see if stellar feedback is indeed strong enough to destroy the GMCs in M83.
\par
The force due to the radiation pressure from an $\HII$ region is calculated as
\begin{equation}
\Frad = L_{\mathrm{bol}}/c,
\end{equation}
where $L_{\mathrm{bol}}$ is the bolometric luminosity of an $\HII$ region and $c$ is the speed of light.
The derivation of $L_{\mathrm{bol}}$ for a GMC is made by converting back the SFR of the GMC (derived in \S\ref{SubsecCrossMatch}) into the extinction-corrected $\HAlpha$ luminosity using equation (\ref{EqSFRCalibration}), and then applying a bolometric-to-$\HAlpha$ luminosity ratio is calculated using Starburst99.
The parameters used for the calculation are taken to be as same as those used in \citet{Calzetti2007} to maintain consistency with the SFR calibration described in \S\ref{SubsecHaToSFR}.
We must note that the equation only considers the direct radiation emitted by massive stars.
In reality, emission absorbed by dust and re-emitted in infrared would boost the radiation pressure.
We omit this dust-processed emission from the calculation of radiation pressure for the following two reasons.
First, the dust-processed emission is not the most dominant source of the pressure for normal HII regions in a galactic disk (\cite{Lopez2011}; \cite{Lopez2014Feedback}).
Second, the calculation of the dust-processed emission requires computations of dust opacity that are far from trivial.
As the dust-processed radiation should have a non-negligible impact in the central region, it should be noted that the outward pressure estimated in this subsection is likely to be an underestimate for the central region.
\par
The force due to the pressure associated with warm ionized gas around an $\HII$ region is calculated as
\begin{equation}
\Fgas = 4\pi \RHII^2 (2 n_{e} k \THII),
\end{equation}
where $\RHII$ is the radius of the $\HII$ region, $n_{e}$ is the electron number density, $k$ is the Boltzmann constant, and $\THII$ is the ionized gas temperature.
We assume constant temperature of $\THII$ = 10$^4$ $K$.
The number density of electrons is estimated as $n_{e} = \sqrt{3 \QLyc / (4 \pi \RHII^3 \alphaRec)}$, where $\QLyc$ is the Lyman continuum production rate and $\alphaRec$ is the recombination coefficient.
By assuming Case B recombination, $\QLyc$ is derived from the $\HAlpha$ luminosity for each $\HII$ region.
\par
The force of gravity for a GMC is calculated as
\begin{equation}
F_{\mathrm{grav}} = G \frac{{\Mmol}^2}{{R}^2}.
\end{equation}
Figure \ref{FigFradOverFgravHistogram} shows the distribution of the ratio of the outward force to the inward force, which is $\Qfb$.
The radiation and gas pressure, $\Frad$ and $\Fgas$, calculated for each $\HII$ region is divided among the associated GMCs using the same method adopted for the SFR (\S\ref{SubsecCrossMatch}).
The median value of $\Qfb$ in the primary arm ridge is close to or above unity, suggesting that many GMCs in this region are certainly being disrupted by stellar feedback.
On the other hand, $\Qfb$ is mostly below 1 in other regions.
We again note that for the central region, $\Qfb$ is highly uncertain due to the omission of dust-processed radiation, and thus we do not discuss the central region here.
As far as subregions in the disk of M83 are concerned, stellar feedback is most efficiently disrupting GMCs in the primary arm.
This effectiveness of stellar feedback in the primary arm ridge is in agreement with the expectation that stellar feedback is responsible for forming the steeper slope of the mass function in this region.
\par
As a by-product of the analysis made here, an approximate estimate of the lifetime of the GMCs can be made.
First, we assume that GMCs evolve from a quiescent state in which they form a small amount of stars---and thus $\Qfb$ $<$ 1---to an active state in which they form many massive stars---and thus $\Qfb$ $>1$.
For the active GMCs with $\Qfb$ $>$ 1, it would be natural to assume further that they are about to be disrupted by the stellar feedback and thus the average remaining lifetime for them is comparable to the nominal lifetime of massive stars.
If stellar feedback is the only process that limits the lifetime of GMCs, then it is possible to gauge their lifetime from the number ratio of GMCs with $\Qfb$ above and below 1.
Applying a similar method, the lifetime of a GMC is estimated to be 15--40 Myr in the MW (\cite{WilliamsMcKee1997}; \cite{Murray2011SFE}), Large Magellanic Clouds \citep{Kawamura2009LMC} and M33 \citep{Miura2012M33}.
For the GMCs in M83, 37 out of 179 are found to have $\Qfb$ greater than 1.
Taking that average lifetime of massive stars as 4 Myr, the average lifetime of a GMC is estimated to be $\sim20$ Myr ($= 179 / 37 \times 4$ Myr), which is not far from the estimation made for other galaxies.
However, this estimation has to be treated with great care because it is merely an averaged value found in the limited area presented here, and it also does not take into consideration the role of shear that might be responsible for disrupting GMCs in the inter-arm regions \citep{Meidt2015Lifetime}.

\subsection{Implications of the spatial variation in SFE}
\label{SubsecImplications}
The cloud-scale examination of $\SFEff$ made in \S\ref{SubsecSFEff} indicated the following three points:
(1) mass-weighted mean value of $\SFEff$ is $\sim$0.93\% , which is in agreement with the average values found in other systems (e.g., \cite{Krumholz2012UniversalSF});
(2) however there is a large scatter in $\SFEff$ with a MAD of $\sim$0.6 dex and peak-to-peak variation of approximately three orders of magnitude;
and (3) there is a regional variation in $\SFEff$.
The most prominent characteristic of the regional variation in $\SFEff$ is the high median $\SFEff$ in the primary arm ridge ($\sim$0.027), compared to the inter-arm regions (4.6 $\times$ 10$^{-3}$ and 2.7 $\times$ $10^{-3}$) and the secondary arm ridge ($\sim$3.9 $\times$ 10$^{-3}$).
In this subsection, we discuss the implications of these findings.
\par
There is a group of theories that focus on the roles of turbulence in regulating star formation, and aim to provide a quantitative description which produces $\SFEff$ $\simeq$ 0.01 in a steady state (e.g., \cite{Krumholz2005TurbulenceSF}; \cite{Federrath2015}).
The goal of achieving $\SFEff$ $\simeq$ 0.01 in a steady state implies that GMCs are approximated as long-lived entities that form stars with a stable efficiency.
The first point, the mass-weighted mean of $\SFEff$ being approximated 0.01, agree with the expectation of the turbulence-regulated model.
However, the second point, the large scatter in the apparent distribution of $\SFEff$ does not agree with the assumption of stable efficiency in GMCs.
\par
Although the turbulence-regulated model of star formation is well applied to some studies made with mostly coarse resolutions that sample several clouds per beam (e.g., \cite{Krumholz2012UniversalSF}), it has been argued that the cloud-scale distribution of $\SFEff$ show some deviations from the expectation of turbulence-regulated models (\cite{Murray2011SFE}; \cite{Lee2016DynamicSF}).
In particular, \citet{Lee2016DynamicSF} observed a large scatter in $\SFEff$ in Galactic GMCs, and they argued that it is difficult to explain the large scatter in $\SFEff$ with the models of turbulence-regulated star formation, including \citet{Krumholz2005TurbulenceSF}, \citet{PadoanNordlun2011} and \citet{HennenbelleChabrier2011}.
Instead, they claimed that $\SFEff$ should be a time-dependent variable that dynamically increases during the lifetime of GMCs; at the later phase of their lifetime, GMCs produce stars with high $\SFEff$ and are disrupted by stellar feedback.
\par
The spread in $\SFEff$ observed in M83 is similarly huge as the one observed in Galactic GMCs by \citet{Lee2016DynamicSF}, and therefore the idea that $\SFEff$ increases with time might also hold in M83.
The $\SFEff$ observed in Galactic GMCs by \citet{Lee2016DynamicSF} is characterized by a median and scatter about the median of $\sim$1.8\% and 0.91 dex, respectively.
The scatter of 0.91 dex is comparable with the MAD of $\sim$0.6 dex observed in M83.\footnote{If the distribution in figure \ref{FigSFEffHistogram}(a) is approximated as a Gaussian, a MAD of 0.6 dex corresponds to a scatter of $\sim$0.9 dex}
We note that the median $\SFEff$ of $\sim$1.8\% obtained by \citet{Lee2016DynamicSF} is higher than the one obtained here in M83 (0.3\%), but it is likely due to the fact that \citet{Lee2016DynamicSF} has selected GMCs with active star formation.
It is also noteworthy that \citet{Leroy2017M51SFEff} also found the median value of $\SFEff$ in M51 to be $\sim$0.3\% with a resolution of 370 pc.
\par
If the notion that $\SFEff$ increases with time during the lifetime of a GMC is correct, then the regional variations of $\SFEff$ (the third point) may suggest that large-scale galactic structures exert an influence in organizing the life cycle of GMCs.
The GMCs in the primary arm ridge exhibit a higher $\SFEff$ compared to the inter-arm GMCs.
If $\SFEff$ increases over the lifetimes of GMCs, the GMCs in the primary arm ridge should be at a late stage of their evolution, producing stars with increased $\SFEff$, while GMCs in the inter-arm regions are at an early stage with low $\SFEff$.
The idea that GMCs in the primary arm ridge are at a late stage of their evolution is in agreement with the analysis made in \S\ref{SectionFeedback}, which suggested that the GMCs in the primary arm ridge are about to be disrupted by stellar feedback (\S\ref{SectionFeedback}).
\par
The discussion so far can be summarized as follows: $\SFEff$ increases with time during the lifetime of a GMC and galactic structures have a certain role in organizing the lifetimes of GMCs.
This scenario is in agreement with the suggestion that spiral arms can organize the buildup of massive GMCs (\cite{Egusa2011M51}; \cite{Colombo2014Env}).
A concern is the timescale of traversal across inter-arm regions which can be a factor of few longer than the suggested lifetime of GMCs, which is 15--40 Myr.
If all GMCs have the same lifetime and also increased $\SFEff$ in the same way, then at least a few GMCs in the inter-arm should exhibit $\SFEff$ values as high as those in the primary arm ridge.
However, the inter-arm GMCs do not exhibit such a high $\SFEff$.
Therefore, to explain the observed $\SFEff$, a mechanism that obstructs the evolution of GMCs in the inter-arm regions would be required.
The large-scale shear that acts to disrupt clouds \citep{Meidt2015Lifetime} might be a candidate mechanism.
If there exists a mechanism that disrupts GMCs in the inter-arm before increasing $\SFEff$, the lifetime of GMCs in the arm and inter-arm could be different from each other.
Complete mapping of a galaxy in CO and reliable star formation tracers will be required to fully reveal the life cycle of GMCs.

\section{Summary}
\label{SectionSummary}
Results of the mosaic $^{12}$CO (1--0) observations of the nearby barred galaxy M83 carried out with ALMA are presented.
The interferometric data are combined with the data obtained with the Nobeyama 45-m telescope to recover the total flux.
The mosaic observations cover a $\sim$13 kpc$^2$ region that includes the galactic center, eastern bar and spiral arm with a spatial resolution of \timeform{2''.03} $\times$ \timeform{1''.15} (44.3 pc $\times$ 25.1 pc), which is comparable to the typical sizes of GMCs.
The velocity resolution is $\sim$2.5 $\kmPerS$.
\begin{itemize}
\item{
With the GMC scale resolution, galactic structures including the spiral arm and the bar are resolved into narrow structures.
The bar appears as a continuous molecular ridge with a surface density as high as 200--800 $\MsunPerSqPC$, which exceeds the surface density of typical Galactic GMCs.
The spiral arm is resolved into two ridges, or chains of molecular clouds, that run parallel to each other.
The one at the leading side, referred to as primary arm ridge, is associated with numerous bright $\HII$ regions while the other one at the trailing side, referred to as secondary arm ridge, appears to be more quiescent.
Spurs are found at the leading side of the primary arm ridge and the bar.
}
\item{The distribution of the massive star-forming regions exhibits a higher degree of concentration, compared to that of CO emission.
Most of the $\HII$ regions are concentrated in particular regions, including the primary arm ridge and the bar, which suggests a spatial variation in the SFE.
}
\item{
We identify 179 GMCs from the CO data using the {\it astrodendro} software package.
When making the cloud identification, the data cube is smoothed in the spatial directions to a resolution of \timeform{2.1''} ($\sim$46 pc).
Assuming the Galactic CO-to-H$_2$ conversion factor, the median value of the cloud mass is found to be 1.6 $\times$ 10$^6$ $\Msun$.
The virial mass and CO luminosity are well correlated to each other, and the median value of the virial parameter for all the identified GMCs is found to be $\sim$1.4, suggesting that most of the GMCs are strongly influenced by their self-gravity.
The GMCs in the arm exhibited lower virial parameters with the median value of $\sim$1.0.
}
\item{
The mass spectrum for all the identified GMCs are fitted with a truncated power law with a slope of -1.58 $\pm$ 0.1, which is close to that of the Galactic GMCs.
The fitting is also performed for GMCs in each subregion, and the steepest slope is found in the primary arm ridge (-1.8).
We suggest that GMCs in the primary arm ridge are disrupted due to stellar feedback.
}
\item{
The identified GMCs are cross-matched with the catalog of $\HII$ regions to estimate the SFR for each GMC.
As the star formation history for individual $\HII$ regions is not constrained, there should be a factor-of-2 uncertainty in the calibration of the SFR.
Despite this weakness, the overall statistical distribution of the SFE for the GMCs in M83 are found to be in agreement with that of Galactic clouds.
The median SFE is $\sim$0.5 Gyr$^{-1}$ and the scatter is as large with a peak to peak variation of approximately two orders of magnitude.
}
\item{
The mass-weighted mean of $\SFEff$ is $\sim$9.4 $\times$ 10$^{-3}$, which is in agreement with the expectations of turbulence regulated star formation models.
However, its scatter is as large as $\sim$0.7 dex in MAD, which cannot be explained by a cloud evolution model with a constant $\SFEff$.
In addition, a regional variation in $\SFEff$ is also observed.
The median value of $\SFEff$ is highest in the primary arm ridge ($\sim$0.027) while it is more than a factor of 5 lower in the inter-arm regions and in the secondary arm ridge.
The large spread and significant spatial variation observed in $\SFEff$ support the idea that $\SFEff$ is not a steady time-invariant variable, but is a dynamic variable that increases as GMCs evolve.
In particular, the GMCs in the primary arm ridges are suggested to be reaching the last stage of their evolution with elevated $\SFEff$, because the feedback from massive stars appears to be large enough to disrupt them.

}

\end{itemize}

\section*{Acknowledgments}
We would like to thank the referee for carefully reading the paper and proving valuable comments.
This paper makes use of the following ALMA data: ADS/JAO.ALMA\#2011.0.00772.S. ALMA is a partnership of ESO (representing its member states), NSF (USA), and NINS (Japan), together with NRC (Canada), NSC and ASIAA (Taiwan), and KASI (Republic of Korea), in cooperation with the Republic of Chile. The Joint ALMA Observatory is operated by ESO, AUI/NRAO, and NAOJ.
This research made use of images provided by the Survey for Ionization in Neutral Gas Galaxies (Meurer et al. 2006) which is partially supported by the National Aeronautics and Space Administration (NASA).
This research has made use of the NASA/IPAC Extragalactic Database (NED) which is operated by the Jet Propulsion Laboratory, California Institute of Technology, under contract with the National Aeronautics and Space Administration.
This research made use of astrodendro, a Python package to compute dendrograms of Astronomical data (http://www.dendrograms.org/).
This research made use of APLpy, an open-source plotting package for Python hosted at http://aplpy.github.com.


\begin{thebibliography}{}



\bibitem[Ballesteros-Paredes et al.(2011)]{BallesterosParedes2011A} Ballesteros-Paredes, J., Hartmann, L.~W., V{\'a}zquez-Semadeni, E., Heitsch, F., \& Zamora-Avil{\'e}s, M.~A.\ 2011, \mnras, 411, 65

\bibitem[Bertoldi \& McKee(1992)]{BertoldiMcKee1992} Bertoldi, F., \& McKee, C.~F.\ 1992, \apj, 395, 140


\bibitem[Blitz \& Shu(1980)]{BlitzShu1980} Blitz, L., \& Shu, F.~H.\ 1980, \apj, 238, 148


\bibitem[Bolatto et al.(2013)]{Bolatto2013ConversionFactor} Bolatto, A.~D., Wolfire, M., \& Leroy, A.~K.\ 2013, \araa, 51, 207

\bibitem[Bresolin et al.(2016)]{Bresolin2016M83} Bresolin, F., Kudritzki, R.-P., Urbaneja, M.~A., et al.\ 2016, \apj, 830, 64

\bibitem[Calzetti et al.(2007)]{Calzetti2007} Calzetti, D., et al.\ 2007, \apj, 666, 870

\bibitem[Calzetti et al.(2010)]{Calzetti2010} Calzetti, D., Chandar, R., Lee, J.~C., et al.\ 2010, \apjl, 719, L158

\bibitem[Cardelli et al.(1989)]{Cardelli1989} Cardelli, J.~A., Clayton, G.~C., \& Mathis, J.~S.\ 1989, \apj, 345, 245

\bibitem[Chomiuk \& Povich(2011)]{ChomiukPovich2011} Chomiuk, L., \& Povich, M.~S.\ 2011, \aj, 142, 197

\bibitem[Colombo et al.(2014)]{Colombo2014Env} Colombo, D., Hughes, A., Schinnerer, E., et al.\ 2014, \apj, 784, 3

\bibitem[Comte(1981)]{Comte1981M83Parameters} Comte, G.\ 1981, \aaps, 44, 441

\bibitem[Crosthwaite et al.(2002)]{Crosthwaite2002M83} Crosthwaite, L.~P.,
Turner, J.~L., Buchholz, L., Ho, P.~T.~P.,
\& Martin, R.~N.\ 2002, \aj, 123, 1892



\bibitem[de Vaucouleurs et al.(1991)]{deVaucouleurs1991RC3} de Vaucouleurs,
G., de Vaucouleurs, A., Corwin, H.~G., Jr., Buta, R.~J., Paturel, G.,
\& Fouque, P.\ 1991, Volume 1-3, XII, 2069 pp.~7 figs..~ Springer-Verlag Berlin Heidelberg New York,

\bibitem[Dobbs(2008)]{Dobbs2008GMCFormation} Dobbs, C.~L.\ 2008, \mnras, 391, 844

\bibitem[Dobbs \& Bonnell(2006)]{DobbsBonnell2006Spurs} Dobbs, C.~L., \& Bonnell, I.~A.\ 2006, \mnras, 367, 873

\bibitem[Dopita \& Sutherland(2003)]{DopitaSutherland2003} Dopita, M.~A., \& Sutherland, R.~S.\ 2003, Astrophysics of the diffuse universe, Berlin, New York: Springer, 2003.~Astronomy and astrophysics library, ISBN 3540433627

\bibitem[Egusa et al.(2011)]{Egusa2011M51} Egusa, F., Koda, J., \& Scoville, N.\ 2011, \apj, 726, 85

\bibitem[Elmegreen(1980)]{ElmegreenDM1980Spurs} Elmegreen, D.~M.\ 1980, \apj, 242, 528


\bibitem[Elmegreen et al.(1998)]{Elmegreen1998M83DoubleRing} Elmegreen, D.~M., Chromey, F.~R., \& Warren, A.~R.\ 1998, \aj, 116, 2834


\bibitem[Evans et al.(2009)]{Evans2009c2d} Evans, N.~J., II, Dunham, M.~M., J{\o}rgensen, J.~K., et al.\ 2009, \apjs, 181, 321-350

\bibitem[Faesi et al.(2014)]{Faesi2014NGC300APEX} Faesi, C.~M., Lada, C.~J., Forbrich, J., Menten, K.~M., \& Bouy, H.\ 2014, \apj, 789, 81


\bibitem[Federrath(2015)]{Federrath2015} Federrath, C.\ 2015, \mnras, 450, 4035

\bibitem[Feldmann \& Gnedin(2011)]{Feldmann2011} Feldmann, R., \& Gnedin, N.~Y.\ 2011, \apjl, 727, L12

\bibitem[Freeman et al.(2017)]{Freeman2017} Freeman, P., Rosolowsky, E., Kruijssen, J.~M.~D., Bastian, N., \& Adamo, A.\ 2017, \mnras, 468, 1769

\bibitem[Fukui et al.(2008)]{Fukui2008} Fukui, Y., Kawamura, A., Minamidani, T., et al.\ 2008, \apjs, 178, 56-70

\bibitem[Goldreich \& Kwan(1974)]{GoldreichKwan1974} Goldreich, P., \& Kwan, J.\ 1974, \apj, 189, 441

\bibitem[Gratier et al.(2012)]{Gratier2012M33} Gratier, P., Braine, J., Rodriguez-Fernandez, N.~J., et al.\ 2012, \aap, 542, A108


\bibitem[Heald et al.(2016)]{Heald2016M83} Heald, G., de Blok, W.~J.~G., Lucero, D., et al.\ 2016, \mnras, 462, 1238

\bibitem[Hennebelle \& Chabrier(2011)]{HennenbelleChabrier2011} Hennebelle, P., \& Chabrier, G.\ 2011, \apjl, 743, L29

\bibitem[Heyer et al.(2001)]{Heyer2001OuterGalaxy} Heyer, M.~H., Carpenter, J.~M., \& Snell, R.~L.\ 2001, \apj, 551, 852

\bibitem[Heyer \& Brunt(2004)]{HeyerBrunt2004} Heyer, M.~H., \& Brunt, C.~M.\ 2004, \apjl, 615, L45

\bibitem[Heyer et al.(2009)]{Heyer2009} Heyer, M., Krawczyk, C., Duval, J., \& Jackson, J.~M.\ 2009, \apj, 699, 1092

\bibitem[Heyer \& Dame(2015)]{Heyer2015Review} Heyer, M., \& Dame, T.~M.\ 2015, \araa, 53, 583

\bibitem[Hirota et al.(2011)]{Hirota2011IC342} Hirota, A., Kuno, N., Sato, N., et al.\ 2011, \apj, 737, 40

\bibitem[Hirota et al.(2014)]{Hirota2014} Hirota, A., Kuno, N., Baba, J., et al.\ 2014, \pasj, 66, 46


\bibitem[Hughes et al.(2013)]{Hughes2013GMCs} Hughes, A., Meidt, S.~E., Colombo, D., et al.\ 2013, \apj, 779, 46

\bibitem[Inutsuka et al.(2015)]{Inutsuka2015} Inutsuka, S.-i., Inoue, T., Iwasaki, K., \& Hosokawa, T.\ 2015, \aap, 580, A49

\bibitem[Jarrett et al.(2013)]{Jarrett2013WISE} Jarrett, T.~H., Masci, F., Tsai, C.~W., et al.\ 2013, \aj, 145, 6


\bibitem[Kawamura et al.(2009)]{Kawamura2009LMC} Kawamura, A., Mizuno, Y., Minamidani, T., et al.\ 2009, \apjs, 184, 1

\bibitem[Kennicutt(1998)]{Kennicutt1998SchmidtLaw} Kennicutt, R.~C., Jr.\ 1998, \apj, 498, 541


\bibitem[Kobayashi et al.(2017)]{Kobayashi2017MF} Kobayashi, M.~I.~N., Inutsuka, S.-i., Kobayashi, H., \& Hasegawa, K.\ 2017, \apj, 836, 175

\bibitem[Koda et al.(2009)]{Koda2009} Koda, J., et al.\ 2009, \apjl, 700, L132

\bibitem[Koda et al.(2012)]{Koda2012} Koda, J., Yagi, M., Boissier, S., et al.\ 2012, \apj, 749, 20

\bibitem[Koda et al.(2016)]{Koda2016} Koda, J., Scoville, N., \& Heyer, M.\ 2016, \apj, 823, 76

\bibitem[Kroupa(2001)]{Kroupa2001} Kroupa, P.\ 2001, \mnras, 322, 231

\bibitem[Krumholz \& McKee(2005)]{Krumholz2005TurbulenceSF} Krumholz, M.~R., \& McKee, C.~F.\ 2005, \apj, 630, 250

\bibitem[Krumholz et al.(2012)]{Krumholz2012UniversalSF} Krumholz, M.~R., Dekel, A., \& McKee, C.~F.\ 2012, \apj, 745, 69

\bibitem[Kuno et al.(2007)]{Kuno2007Atlas} Kuno, N., et al.\ 2007, \pasj, 59, 117

\bibitem[Kurono et al.(2009)]{Kurono2009} Kurono, Y., Morita, K.-I., \& Kamazaki, T.\ 2009, \pasj, 61, 873

\bibitem[Lada et al.(2010)]{Lada2010} Lada, C.~J., Lombardi, M., \& Alves, J.~F.\ 2010, \apj, 724, 687

\bibitem[Larsen \& Richtler(1999)]{Larsen1999YMC} Larsen, S.~S., \& Richtler, T.\ 1999, \aap, 345, 59

\bibitem[Larson(1981)]{Larson1981} Larson, R.~B.\ 1981, \mnras, 194, 809

\bibitem[La Vigne et al.(2006)]{LaVigne2006} La Vigne, M.~A., Vogel, S.~N., \& Ostriker, E.~C.\ 2006, \apj, 650, 818

\bibitem[Lee et al.(1994)]{Lee1994G21625} Lee, Y., Snell, R.~L., \& Dickman, R.~L.\ 1994, \apj, 432, 167

\bibitem[Lee et al.(1996)]{Lee1996G21625} Lee, Y., Snell, R.~L., \& Dickman, R.~L.\ 1996, \apj, 472, 275

\bibitem[Lee et al.(2016)]{Lee2016DynamicSF} Lee, E.~J., Miville-Desch{\^e}nes, M.-A., \& Murray, N.~W.\ 2016, \apj, 833, 229

\bibitem[Leitherer et al.(1999)]{Leitherer1999} Leitherer, C., Schaerer, D., Goldader, J.~D., et al.\ 1999, \apjs, 123, 3



\bibitem[Leroy et al.(2013)]{Leroy2013ISMClumpiness} Leroy, A.~K., Lee, C., Schruba, A., et al.\ 2013, \apjl, 769, L12


\bibitem[Leroy et al.(2015)]{Leroy2015NGC253} Leroy, A.~K., Bolatto, A.~D., Ostriker, E.~C., et al.\ 2015, \apj, 801, 25

\bibitem[Leroy et al.(2017)]{Leroy2017M51SFEff} Leroy, A.~K., Schinnerer, E., Hughes, A., et al.\ 2017, \apj, 846, 71


\bibitem[Lopez et al.(2011)]{Lopez2011} Lopez, L.~A., Krumholz, M.~R., Bolatto, A.~D., Prochaska, J.~X., \& Ramirez-Ruiz, E.\ 2011, \apj, 731, 91

\bibitem[Lopez et al.(2014)]{Lopez2014Feedback} Lopez, L.~A., Krumholz, M.~R., Bolatto, A.~D., et al.\ 2014, \apj, 795, 121

\bibitem[Imara(2015)]{Imara2015G21625} Imara, N.\ 2015, \apj, 803, 38

\bibitem[Mac Low \& Klessen(2004)]{MacLow2004} Mac Low, M.-M., \& Klessen, R.~S.\ 2004, Reviews of Modern Physics, 76, 125


\bibitem[Maddalena \& Thaddeus(1985)]{Maddalena1985} Maddalena, R.~J., \& Thaddeus, P.\ 1985, \apj, 294, 231

\bibitem[Matzner(2002)]{Matzner2002} Matzner, C.~D.\ 2002, \apj, 566, 302

\bibitem[McMullin et al.(2007)]{McMullin2007CASA} McMullin, J.~P., Waters, B., Schiebel, D., Young, W., \& Golap, K.\ 2007, Astronomical Data Analysis Software and Systems XVI, 376, 127

\bibitem[Megeath et al.(2009)]{Megeath2009} Megeath, S.~T., Allgaier, E., Young, E., et al.\ 2009, \aj, 137, 4072

\bibitem[Meidt et al.(2015)]{Meidt2015Lifetime} Meidt, S.~E., Hughes, A., Dobbs, C.~L., et al.\ 2015, \apj, 806, 72

\bibitem[Meurer et al.(2006)]{Meurer2006SINGG} Meurer, G.~R., Hanish, D.~J., Ferguson, H.~C., et al.\ 2006, \apjs, 165, 307

\bibitem[Miura et al.(2012)]{Miura2012M33} Miura, R.~E., Kohno, K., Tosaki, T., et al.\ 2012, \apj, 761, 37

\bibitem[Mooney \& Solomon(1988)]{Mooney1988} Mooney, T.~J., \& Solomon, P.~M.\ 1988, \apjl, 334, L51


\bibitem[Murray(2011)]{Murray2011SFE} Murray, N.\ 2011, \apj, 729, 133

\bibitem[Orear(1982)]{Orear1982} Orear, J.\ 1982, American Journal of Physics, 50, 912

\bibitem[Padoan \& Nordlund(2011)]{PadoanNordlun2011} Padoan, P., \& Nordlund, {\AA}.\ 2011, \apj, 730, 40

\bibitem[Petry \& CASA Development Team(2012)]{Petry2012CASA} Petry, D., \& CASA Development Team 2012, Astronomical Data Analysis Software and Systems XXI, 461, 849

\bibitem[Rand et al.(1992)]{RandKulkarniRice1992} Rand, R.~J., Kulkarni, S.~R., \& Rice, W.\ 1992, \apj, 390, 66

\bibitem[Rand et al.(1999)]{RandLordHidgon1999M83} Rand, R.~J., Lord, S.~D., \& Higdon, J.~L.\ 1999, \apj, 513, 720

\bibitem[Renaud et al.(2013)]{Renaud2013Subpc} Renaud, F., Bournaud, F., Emsellem, E., et al.\ 2013, \mnras, 436, 1836




\bibitem[Rosolowsky(2005)]{Rosolowsky2005MassSpectra} Rosolowsky, E.\ 2005, \pasp, 117, 1403

\bibitem[Rosolowsky \& Leroy(2006)]{Rosolowsky2006CPROPS} Rosolowsky, E., \& Leroy, A.\ 2006, \pasp, 118, 590

\bibitem[Rosolowsky et al.(2007)]{Rosolowsky2007M33} Rosolowsky, E., Keto, E., Matsushita, S., \& Willner, S.~P.\ 2007, \apj, 661, 830


\bibitem[Sakamoto et al.(2004)]{Sakamoto2004} Sakamoto, K., Matsushita, S., Peck, A.~B., Wiedner, M.~C., \& Iono, D.\ 2004, \apjl, 616, L59

\bibitem[Sanders et al.(1985)]{Sanders1985} Sanders, D.~B., Scoville, N.~Z., \& Solomon, P.~M.\ 1985, \apj, 289, 373

\bibitem[Sault et al.(1995)]{SaultTeubenWright1995MIRIAD} Sault, R. J., Teuben, P. J., \& Wright, M. C. H.\ 1995, in ASP Conf. Ser. 77, Astronomical Data Analysis Software and Systems IV, ed. R. A. Shaw, H. E. Payne, \& J. J. E. Hayes (San Francisco, CA: ASP), 433

\bibitem[Sawada et al.(2008)]{Sawada2008} Sawada, T., et al.\ 2008, \pasj, 60, 445

\bibitem[Schinnerer et al.(2017)]{Schinnerer2017} Schinnerer, E., Meidt, S.~E., Colombo, D., et al.\ 2017, \apj, 836, 62

\bibitem[Schlegel et al.(1998)]{Schlegel1998DustMap} Schlegel, D.~J., Finkbeiner, D.~P., \& Davis, M.\ 1998, \apj, 500, 525

\bibitem[Scoville et al.(2001)]{Scoville2001M51} Scoville, N.~Z., Polletta, M., Ewald, S., et al.\ 2001, \aj, 122, 3017

\bibitem[Sheth et al.(2000)]{Sheth2000NGC5383} Sheth, K., Regan, M.~W., Vogel, S.~N., \& Teuben, P.~J.\ 2000, \apj, 532, 221

\bibitem[Solomon et al.(1987)]{Solomon1987Larson} Solomon, P.~M., Rivolo, A.~R., Barrett, J., \& Yahil, A.\ 1987, \apj, 319, 730


\bibitem[Thatte et al.(2000)]{Thatte2000} Thatte, N., Tecza, M., \& Genzel, R.\ 2000, \aap, 364, L47

\bibitem[Thilker et al.(2000)]{Thilker2000} Thilker, D.~A., Braun, R., \& Walterbos, R.~A.~M.\ 2000, \aj, 120, 3070

\bibitem[Thim et al.(2003)]{Thim2003M83Distance} Thim, F., Tammann, G.~A., Saha, A., Dolphin, A., Sandage, A., Tolstoy, E., \& Labhardt, L.\ 2003, \apj, 590, 256

\bibitem[Tilanus \& Allen(1993)]{TilanusAllen1993} Tilanus, R.~P.~J., \& Allen, R.~J.\ 1993, \aap, 274, 707

\bibitem[Tosaki et al.(2017)]{Tosaki2017} Tosaki, T., Kohno, K., Harada, N., et al.\ 2017, \pasj,

\bibitem[Utomo et al.(2015)]{Utomo2015} Utomo, D., Blitz, L., Davis, T., et al.\ 2015, \apj, 803, 16

\bibitem[Venuti et al.(2018)]{Venuti2018} Venuti, L., Prisinzano, L., Sacco, G.~G., et al.\ 2018, \aap, 609, A10

\bibitem[Wada \& Koda(2004)]{Wada2004Wiggle} Wada, K., \& Koda, J.\ 2004, \mnras, 349, 270

\bibitem[Walter et al.(2008)]{Walter2008THINGS} Walter, F., Brinks, E.,
de Blok, W.~J.~G., et al.\ 2008, \aj, 136, 2563

\bibitem[Whitworth(1979)]{Whitworth1979CloudDisruption} Whitworth, A.\ 1979, \mnras, 186, 59

\bibitem[Wiklind et al.(1990)]{Wiklind1990M83} Wiklind, T., Rydbeck, G., Hjalmarson, A., \& Bergman, P.\ 1990, \aap, 232, L11

\bibitem[Wilson \& Scoville(1990)]{Wilson1990M33} Wilson, C.~D., \& Scoville, N.\ 1990, \apj, 363, 435

\bibitem[Williams \& McKee(1997)]{WilliamsMcKee1997} Williams, J.~P., \& McKee, C.~F.\ 1997, \apj, 476, 166

\bibitem[Wong et al.(2011)]{Wong2011Magma} Wong, T., Hughes, A., Ott, J., et al.\ 2011, \apjs, 197, 16

\bibitem[Zuckerman \& Evans(1974)]{ZuckermanEvans1974} Zuckerman, B., \& Evans, N.~J., II 1974, \apjl, 192, L149

\bibitem[Zurita \& P{\'e}rez(2008)]{Zurita2008} Zurita, A., \& P{\'e}rez, I.\ 2008, \aap, 485, 5




\end{thebibliography}
\end{document}